\documentclass[12pt,preprint]{aastex}

\slugcomment{Last edited August 28, 2006}

\shorttitle{Infrared Nebulae Around YSOs}
\shortauthors{Connelley et al.}

\begin{document}


\title{Infrared Nebulae Around Young Stellar Objects}


\author{Michael S. Connelley\altaffilmark{1}}

\author{Bo Reipurth\altaffilmark{2}}

\and

\author{Alan T. Tokunaga\altaffilmark{3}}


\altaffiltext{1}{Institute for Astronomy, University of Hawai$'$i, 640 N. A'ohoku Pl., Hilo, HI 96720; msc@ifa.hawaii.edu}
\altaffiltext{2}{Institute for Astronomy, University of Hawai$'$i, 640 N. A'ohoku Pl., Hilo, HI 96720; reipurth@ifa.hawaii.edu}
\altaffiltext{3}{Institute for Astronomy, University of Hawai$'$i, 2680 Woodlawn Dr., Honolulu, HI 96822; tokunaga@ifa.hawaii.edu}


\begin{abstract}
  We present a K-band atlas of 106 reflection nebulae, 41 of which are new discoveries.  We observed these nebulae with the UH~2.2~m telescope in the course of an imaging survey of 197 objects that were selected to be nearby young Class I sources.  K-band images and flux calibrated surface brightness contour plots of each nebula are presented.  We found that the near-IR luminosities and physical sizes of the nebulae increase with the bolometric luminosity of the illuminating sources.  Only 22 nebulae, about 10\% of these candidate Class I sources, have indications of shocked H$_{2}$ emission.  The great variety of nebulae that we observed prevented us from classifying them based on morphology.  However, we note that as the spectral index decreases, the central star is more frequently visible at K-band and the flux from the central star tends to be dominant over the flux from the nebula.  For objects that have a higher spectral index, most of the K-band flux is from the reflection nebula, and the central star is less frequently visible.  The nebula around IRAS 05450+0019 has a unique morphology, and we speculate that it may be an example of a disk shadow being projected into the surrounding cloud.  We present J, H, and K-band images of this object with surface brightness contours, as well as its SED from 1.2~$\mu$m to 100~$\mu$m.

\end{abstract}




\section{Introduction}

   Reflection nebulae and star formation have been associated for a long time.  Hind's Variable Nebula (Hind 1864), near T Tauri, was the first reflection nebula associated with a young star, although T Tauri would not be identified as a young star for many more years.  The variable nebulae around the stars R Monocerotis (Hubble 1916) and R Corona Australis (Knox-Shaw 1916) were discovered half of a century later.  Cometary nebulae were subsequently associated with star formation (Ambartsumian 1954), but an extensive catalog of these nebulae (Gyulbudaghian \& Magakian 1977) was not published until the completion of the Palomar Sky Survey.  In 1979, Parsamian and Petrossian published a catalog of 106 cometary nebulae, classified into 4 groups, and a year later Cohen (1980) published a list of red and nebulous objects found on the Palomar plates.  Most recently, Magakian (2003) merged several of these catalogs of reflection nebulae.

   With the advent of large near-infrared arrays came the first major survey of infrared nebulae. Hodapp (1994) observed 164 CO outflow sources identified by Fukui (1989), and found many infrared reflection nebulae.  More recently, infrared reflection nebulae have been studied with HST and NICMOS (e.g. Padgett et al. 1999; Reipurth et al. 2000).  Technological developments lead to YSOs being studied from the ground out to 20~$\mu$m, as well as with the IRAS satellite.  Based on these data, new star forming regions were, and continue to be, found and new classes of YSOs identified.  Lada \& Wilking (1984) identified three broad types of YSOs.  Class I objects have an SED that is broader than a single temperature blackbody and rises beyond 2~$\mu$m, Class II objects have an SED that is broader than a single temperature blackbody and declines beyond 2~$\mu$m, and Class III objects have a stellar SED that is consistent with a single temperature blackbody.  More recently, Class 0 sources have been identified which have the SED of a cold single temperature blackbody (Andr\'{e} et al. 1993).  Class 0 and many Class I objects are believed to be true protostars, and Class II objects are now identified with the T Tauri stars. 

    Many studies of young stellar objects focus on a few objects at a time, which may become well known, while numerous interesting objects languish in obscurity.  We therefore developed our own sample of nearby Class I YSOs, that is not biased towards well known star forming regions or famous case objects, in the following manner.  First, we defined criteria for selecting Class I sources (described in the next section) by combining data from the IRAS Point and Faint Source Catalogs, the Digitized Sky Survey, and 2MASS.  Having selected our sample from these catalogs, we observed most of these objects at K-band with the University of Hawai$'$i 2.2~m telescope. We observed 197 of the 267 targets in our sample, omitting those sources that are too far south ($\delta<-40$) to be well observed from Mauna Kea, as well as very well known sources (e.g. R CrA).  Among the 197 objects that were observed, 106 were found to have an infrared reflection nebula.  These objects are presented in this paper. 

  During the course of our observations, we found that a significant fraction of our candidate Class I sources are associated with reflection nebulae, many with interesting morphology.  Judging by their citations in SIMBAD, we found that many of these candidate objects were not previously well known, including some of the largest and brightest nebulae in our survey. Among those sources we observed, about half showed some kind of nebula, and they are presented here.

\section{Source Selection Criteria and Distances}

  Our goal was to compose a sample that contains only Class I YSOs within $\sim$500~pc in order to be able to resolve close binary companions. We chose to define our own sample of Class I YSOs (versus drawing a sample from the literature) so that our sample would not be biased in favor of nor against the well known star forming regions, and so that we would have a clearly defined sample with well known properties.  

  We first selected all sources in the IRAS Point and Faint Source Catalogs that have increasing flux with wavelength from $12~\mu$m through $100~\mu$m, and that were detected in the $25~\mu$m, $60~\mu$m, and $100~\mu$m bands.  These selection criteria were chosen to select sources with cool ($\sim30~$K) dust and have allowed some objects into our sample that are classified as Class 0 sources. We then visually inspected $15' \times 15'$ images from the Digitized Sky Survey (DSS) for each of the $\sim$12,000 sources that satisfied the criteria described above, so that we could select sources that lie in the direction of nearby clouds with high optical extinction.  Since we wanted our sample of Class I objects to be nearby, we were looking for fields with a well defined dark cloud and few stars (if any) that appear to be in front of the cloud.  We also looked for light from gas emission or dust scattering since we sought to avoid more evolved objects in large HII regions or that have an optical reflection nebula.  We eliminated from further consideration fields centered on a galaxy, planetary nebula, bright star, a distant bright or dark cloud, or an empty star field (i.e. a star field without a dark cloud, bright nebula, or any of the previously mentioned objects).  This effectively eliminated galaxies, evolved stars, planetary nebulae, and other dusty objects from our sample.  In our review of the DSS images, choosing which dark nebulae appear nearby and which do not was inevitably a subjective process.  We have since found distance estimates to most of the objects in our sample.  The distribution of distances to these YSOs, which has a median of 470~pc, is presented in Figure 1 (right), and confirms that our expectations of the appearance of nearby dark clouds were generally correct.

   For all of the fields that satisfied the above criteria, 2MASS images in the J, H, and K-bands were then inspected to identify embedded sources.  We selected sources that are clearly brighter from J to H to K-bands and set aside targets where the only stars that were seen by 2MASS are also visible on the DSS plates.   Sources where no embedded stars were visible in 2MASS were not rejected because there are near-IR counterparts and nebulae that were undetected by 2MASS.  

    Our selection procedure yielded a sample of 267 objects that is almost entirely composed of Class I YSOs, as shown by the fact that the distribution of spectral indices (as defined by Lada 1990) for sources in this sample has a median value of +0.79 (Figure 1, left).  To calculate the spectral indices, we used the IRAS 12~$\mu$m to 100~$\mu$m measurements (or from 25 $\mu$m if the source was undetected at 12~$\mu$m).  We use the slope of the linear regression line through the data points as the spectral index. It must be noted that Lada (1990) defines the spectral index as $\alpha=-dlog\nu F_{\nu}/dlog\nu$, in which case a Class I protostar has a positive spectral index, Adams et al.(1987) define the spectral index without the minus sign, in which case a protostar would have a negative spectral index.  Kenyon \& Hartmann (1995) express the spectral index as $dlog\lambda F_{\lambda}/dlog\lambda$, in which case the spectral index of a protostar would be positive.  We adopt the convention where the spectral index of a protostar is positive, and use the equation given by Lada (1990).  There are a few targets that have a spectral index less than 0, which indicates that they may be Class II objects.  There are also a few objects in the sample that have an increasing SED with wavelength and no near-IR counterpart.  These are believed to be Class 0 sources.  These spectral indices contain flux from all sources within the broad IRAS beam.  Several sources have more than one YSO visible in the near-IR.  Higher angular resolution mid- and far-IR observations of individual sources may yield different spectral indices.  

\section{Observations and Data Reduction}

\subsection{UH~2.2~m}
   Our K-band observations of the reflection nebulae were conducted with QUIRC, the 1024$\times$1024 HgCdTe facility near-IR camera (Hodapp et al. 1996) on the University of Hawai$'$i 2.2~m telescope on Mauna Kea.  All observations were made through the MKO (Tokunaga et al. 2002; Tokunaga \& Vacca 2005) K-band filter (not K$'$ or K$_{s}$). The camera's plate scale is $0\farcs1886$ per pixel, yielding a field of view of $193\arcsec$.  Data were taken from January to September, 2003.  Details of when each object was observed, as well as the integration times, are listed in Table 1.  The mean image resolution (FWHM) for this data set is $0\farcs87$, with the best seeing being $0\farcs58$ and the worst $1\farcs56$.  Each source was observed using either a 5 or a 9 point dither pattern, the telescope being offset $\sim10\arcsec$ between exposures.  Exposure times were usually 60~s, although shorter exposure times were used on some very bright sources (e.g. IRAS~21004+7811).  In the case of IRAS~05450+0019, a very large nebula, separate sky frames were taken.

  The data were subsequently reduced using IRAF.  A dark image, itself the average of 10 individual darks and of the same exposure time as the target images, was subtracted from each on-source frame.  To make a flat, each dark subtracted frame was scaled to have the same average value.  These scaled images were then averaged together with a min-max rejection to make the flat, which was then normalized.  Each dark subtracted (but not scaled) image was divided by this flat, then a constant was subtracted from each image to set the average sky counts to zero.  Each processed frame was then registered and averaged together, using average sigma clipping rejection.  

   Selected UKIRT standard stars were observed and reduced in the same manner.  Photometry was performed using the PHOT routine in IRAF, using an aperture 20 pixels in radius with a 20 pixel wide buffer between the aperture and the 20 pixel wide sky annulus.  IDL was used to make the surface brightness contour plots.  Boxcar smoothing of 3 pixels for the brighter parts of the image (since this is slightly less than the average FWHM in pixels) and 6 pixels for the fainter areas was used to reveal the faint nebulosity in the contours.  The pixel counts in each reduced image was converted to a surface brightness through the relation:
\begin{equation}
\Sigma_{K} = K_{standard} - 2.5log(\frac{c_{nebula}-c_{sky}}{c_{standard}} \cdot \frac{t_{standard}}{t_{nebula}} \cdot A) - k(a_{standard}-a_{nebula})
\end{equation}

where $\Sigma_{K}$ is the K-band surface brightness of the nebula in magnitudes per square arcsecond, $K_{standard}$ is the standard star's K-band magnitude, $c_{nebula}$ is the number of counts for the pixel in question, $c_{sky}$ is the average background sky counts in the nebula image, $c_{standard}$ is the number of counts measured through aperture photometry from the standard star, $a_{standard}$ is the airmass at which the standard star was observed, $a_{nebula}$ is the airmass at which the nebula was observed, $k = 0.07$ is the coefficient for K-band atmospheric extinction per unit airmass (Krisciunas et al. 1987), $t_{nebula}$ is the exposure time used for the nebula (usually 60~s), $t_{standard}$ is the exposure time used for the standard star, and $A$ is the number of pixels that make a square arcsecond on the sky (in the case of QUIRC, $A = 28.114$).

  The dithering and flat fielding processes eliminated a majority of potential image artifacts.  The primary remaining artifact is a reduction of the background counts near a bright, spatially resolved object and in crowded fields.  This occurs when there is an object in the field that is similar in size to the offsets of the dither pattern, and thus the resulting flat contains some contribution from the objects in the field of view.  For most objects, this is not an issue.  Nevertheless, there are a number of nebulae where this continues to be a problem, such as for IRAS~16316$-$1540.  This does affect the calculated surface brightness of the nebula, particularly near the faint extremities of the nebula.  In such cases, we chose as our "sky" value the counts nearest the edge of the nebula to minimize this effect.  Another kind of artifact are the diffraction spikes from the secondary mirror support.  This is visible around very bright (usually saturated) stars, such as the near-IR counterpart to IRAS~20453+6746.  Saturated sources can be identified by a large inner surface brightness contour, whereas the contours for unsaturated sources become progressively smaller towards the center of the PSF.  Our images occasionally show elongated stellar PSFs caused by telescope wind shake.  Triangular PSFs are due to print-through of the primary mirror support structure.  One of the observing runs occurred immediately after the mirror was replaced in the telescope after recoating.  During this run, the primary mirror cell had not yet been precisely adjusted, resulting in uneven mirror support.

\subsection{IRTF}
   K-band and H$_{2}$ 2.122~$\mu$m narrow band observations of nebulae suspected of having shocked H$_{2}$ emission were conducted with the Spex 512$\times$512 pixel guider array on the NASA IRTF. We selected nebulae that had not yet been imaged with the 2.122~$\mu$m H$_{2}$ narrow band filter before. The guider array has a plate scale of $0\farcs12$ per pixel, yielding a field of view of $61\arcsec$. All of the data were taken on the night of 23 Sep. 2005 under nonphotometric conditions.  For each field, 7 K-band images were taken followed by (typically) 21 H$_{2}$ images.  Although we had K-band images of each nebula taken with QUIRC, we took comparison K-band images with Spex in order to identify the field and to have images with the same plate scale and seeing as the H$_{2}$ data.  Exposure times were typically 60~s, although some of the narrow band images were taken with a 120~s exposure time.  A 7 point dither pattern was used with $10\arcsec$ offsets.  The data reduction method was as described above.  Due to the observing conditions and the fact that we already had flux calibrated K-band images, standard stars were not observed. The nebula associated with IRAS 05399$-$0121 is the only case where H$_{2}$ was not confirmed.  For this object, there were no sources (either nebulous or stellar) in our H$_{2}$ images that we could use to register the images, so our data could not be reduced.

   K-band, K-continuum, and H$_{2}$ images are shown in these figures as appropriate, as well as a $15\arcsec$ scale bar.  The images labeled as "H$_{2}$ + K continuum" are broad K-band images.  We first identified H$_{2}$ emission regions by subtracting a scaled K-band image from a narrow band H$_{2}$ image, determining the amount of scaling by the relative flux of field stars.  The K-continuum images were made by subtracting a scaled narrow band H$_{2}$ image from the K-band image.  In this case, the scaling was determined by comparing the peak counts of H$_{2}$ features in the K-band and H$_{2}$ images in order to subtract out the H$_{2}$ features.  The images labeled as "H$_{2}$ only" were made by subtracting a scaled K-continuum image from a narrowband H$_{2}$ image; the scaling again being determined by the relative flux of field stars.




\section{The Atlas of Infrared Reflection Nebulae}

\subsection{Atlas of Nebulae}
   
   Figure 2 shows $60\arcsec\times60\arcsec$ K-band images of the majority of the nebulae in this paper.  In some cases a larger field of view is necessary to show the whole nebula, and these nebulae are presented in Figures 3 through 9.  Each image is centered on the K-band source, or occasionally offset to show two widely separated sources.  The (0,0) coordinate in each image is the IRAS Point Source or Faint Source Catalog location.  Tick marks designate a K-band source in each image (usually what we believe to be the near-IR counterpart to the IRAS source), and the 2MASS coordinate of that source is given in Table 2.  Under the name of each IRAS source is a 5,000 AU scale bar for those objects where we found a distance estimate.  The brightness and contrast of each image was chosen to most clearly show the morphological features of the nebulae.  K-band surface brightness contours, calibrated in the MKO photometric system, are overlaid at 1 magnitude per square arcsecond intervals.  The value of the first (outermost) K-band surface brightness contour is given in Table 1.  The inner contours are white in order to show them against the black areas of each image.  The J, H, and K$_s$ magnitudes presented in Table 2 are from 2MASS and are in the internal 2MASS photometric system.  When available, the magnitudes presented in Table 2 are from the 2MASS extended source catalog.  If the object is not included in the extended source catalog, then the photometry from the point source catalog was used.  Color conversion equations are presented in appendix A.  

\subsection{Notes on Sources}
   The following summarizes the results of a literature search on each source using the SIMBAD database.  The information presented here is intended to be representative of what has been reported on each source and is not comprehensive.  We hope that this information will be helpful in guiding the reader in future research.  Objects described as spatially resolved or non-stellar have a FWHM significantly larger than field stars in the same image.  Objects described as unresolved point sources have a FWHM consistent with field stars in the same image.  While the majority of sources have a near-IR counterpart at the IRAS coordinates, there are several cases where the suspected near-IR counterparts are up to tens of arcseconds away from the IRAS coordinates.  We believe that these cases can have a variety of causes, all rooted in the large size of the IRAS beam.  The hot near-IR counterpart may not be centered in the cooler region detected by IRAS at 60~$\mu$m and 100~$\mu$m.  The centroiding of a faint IRAS source may be affected by a much brighter nearby source.  Finally, there are cases such as IRAS 16288$-$2450 where the IRAS position is right between two bright near-IR counterparts, both of which have reflection nebulosity.  In such a case IRAS detected both sources, and they are too close to be distinguished by IRAS.  

  We believe that a nebula is newly discovered if there are no citations in SIMBAD that include near-IR images.  Among the 41 newly discovered nebulae, we noticed that there are two regions in the sky with a large number of previously unreported nebulae.  One group with 10 nebulae is centered around 6\fh0 $-10^{\circ}$, near the optical reflection nebula NGC~2149 in SW Monoceros.  Being near Orion, it is likely to be associated with that star forming region.  The other group is centered near 18\fh5 $-1^{\circ}$ in Serpens and has 5 nebulae.  These unexplored star forming clouds may harbor more young stars.

\textbf{IRAS 00182+6223.}
  This is a previously unreported nebula whose cloud has been detected in CO emission (Wouterloot \& Brand 1989).  This is the first near-IR image published of this source.  We observed a $\sim10\arcsec$ diffuse nebula extending to the north, as well as two stars near the nebula to the ESE.

\textbf{IRAS 00465+5028.}
  This nebula is associated with RNO~3, and the IRAS source has been extensively observed in the mm and radio.  Yun \& Clemens (1994a) were the first to publish near-IR images of this nebula.  We observed a monopolar nebula extending to the north of an unresolved point source.

\textbf{IRAS 01166+6635.}
  This is a previously unreported nebula whose illuminating source has been associated with H$_{2}$O maser (Codella et al. 1995) and CO emission (Wouterloot \& Brand 1989).  This is the first near-IR image published on this object, showing a small diffuse nebula to the SE of an unresolved point source.

\textbf{IRAS 02086+7600.}
  This YSO in L1333 was first observed in the near-IR by Fujii et al. (2002), who conclude that this is likely to be a YSO rather than a post-ABG star. It has been detected in C$^{18}$O, including a feature that may be an outflow (Obayashi et al. 1998). Our image shows what appears to be a small bipolar nebula.  There is also a large, faint nebula to the west.  

\textbf{IRAS 03225+3034.}
  This source is associated with the well known, highly embedded protostellar binary star L1448N~IRS~3.  VLA observations have shown this source to be a 7\farcs3 binary (Reipurth et al. 2002).  Cardi et al. (2003) were only able to see one component of this binary at mid-IR wavelengths, concluding that the binary consists of a Class I source and a Class 0 source. This is the origin of HH~193A-C and HH~194A and B (Bally et al. 1997).  We observed a $\sim90\arcsec$ long jet-like nebula that may be mostly H$_{2}$ emission based on its morphology.  Our image is presented in Figure 3.

\textbf{IRAS 03245+3002.} 
  This nebula is associated with RNO~15~FIR in the L1455 cloud.   This source is the origin of a CO outflow (Levreault 1988), water maser emission (Cesaroni et al. 1988), and the HH~318 flow  (Bally et al. 1997).  Davis et al. (1997) present H$_{2}$ S(1) imaging of the molecular jet that drives the CO outflow.  There is no observed near-IR counterpart to this IRAS source, only a jet-like nebula $\sim6\arcsec$ long to the SW ending in a small bow shock. 

\textbf{IRAS 03260+3111.}
  A source in NGC~1333 that is well known for studies of silicate (Noguchi et al. 1993) and other molecular spectral features.  It was first studied at near-IR wavelengths by Haisch et al. (2004).  IRAS~8 in the study of NGC~1333 by Jennings et al. (1987) corresponds to this object.  There is a large diffuse nebula around the bright stars to the east in Figure 4.  To the west is a reflection nebula, with a fainter surrounding nebula elongated to the east and west.
  
\textbf{IRAS 03271+3013.}
  This object was first observed in the near-IR by Aspin (1992), who found shocked H$_{2}$ emission and estimated that there are about 50 magnitudes of extinction towards the H$_{2}$ line emission region.  Our image shows a central nebula, with faint nebulosity arcing to the NE and to the SW.

\textbf{IRAS 03301+3057.} 
  Also known as Barnard~1~IRS, this source is believed to be the origin of a CO outflow (Hirano et al. 1997).  HH~431 was found near this source by Yan et al. (1998).  This may be the first near-IR data published on this nebula.  Our image shows a small nebula with a jet-like feature extending $\sim5\farcs5$ to the WSW.

\textbf{IRAS 03331+6256.}
  This is a previously unreported near-IR nebula, and these are the first near-IR data published on this object.  CO emission was detected by Wouterloot \& Brand (1989).  Our image shows a small faint nebula to the north of a point source that is likely the near-IR counterpart of the IRAS source.  A faint companion $\sim2\farcs5$ to the SSW of the source was observed to significantly brighten in the near-IR in mid-2004.

\textbf{IRAS 03445+3242.}
  Also known as Barnard~5~IRS~1, this object is the well known source of a CO outflow (Fuller et al. 1991; Yu et al., 1999) and the parsec scale jet HH~366 (Bally et al. 1996).  Moore \& Emerson (1994) found the nebula to show a steady decrease in luminosity of 0.27 magnitudes per year with constant near-IR color.  The near-IR spectrum shows H$_{2}$ and weak CO emission (Reipurth \& Aspin 1997).  In addition to a faint reflection nebula, our K-band image also shows what appear to be a few small knots of H$_{2}$ emission symmetrically placed to the ENE and WSW of the central star.

\textbf{IRAS 03507+3801.}
  This previously unreported near-IR reflection nebula is near HH~462, which was discovered by Aspin \& Reipurth (2000).  Our image shows a bright point source, a faint monopolar nebula to the SW, and what appears to be a faint loop to the NE.  

\textbf{IRAS 04016+2610.}
  A well known YSO in L1489, this is the origin of a molecular outflow (Terebey et al. 1989; Hogerheijde et al. 1998) and several HH objects (G\'{o}mez et al. 1997b).  The mass of the associated core has been estimated to be 2~M$_{\odot}$ based on NH$_{3}$ emission (Benson \& Myers 1989).  This Class I source has been detected as an unresolved 6~cm source by the VLA (Rodr\'\i guez et al. 1989), and was observed with HST by Padgett et al. (1999) who observed a 600~AU dust lane.  Lucas et al. (2000) describe this system as having two perpendicular bipolar jets, based on VLA, MERLIN, and near-IR H$_{2}$ images.  Our image shows a bipolar reflection nebula, brighter to the south, bisected by a dark lane.

\textbf{IRAS 04067+3954.}
  This is a previously unreported, large, and morphologically interesting reflection nebula.  Although most of the K-band flux is scattered light from the bright spatially resolved nebula, there is a faint point source 5\arcsec\ to the west of the brightest part of the nebula.  The faint point source is the brightest object in the field at L$'$, and probably represents the near-IR counterpart to the IRAS source.

\textbf{IRAS 04073+3800.}   
  Also known as PP13S*, Sandell \& Aspin (1998) found that this object is an FU Orionis pre-main sequence star and present near-IR imaging, spectroscopy, and sub-mm data.  Aspin \& Reipurth (2000) showed that this is the origin of the HH~463 outflow.  Our image shows a bright point source with a bright loop of nebulosity to the SW.

\textbf{IRAS 04169+2702.}
  This well studied object is described in Park \& Kenyon (2002) as a Class I object, and is the origin of HH~391 (G\'{o}mez et al. 1997b). Our image shows a bright point source with a faint monopolar nebula to the SW.  There also appears to be a few small knots of H$_{2}$ emission to the SW.

\textbf{IRAS 04189+2650.}
  This is FS~Tau, also known as Haro~6-5.  FS~Tau A (to the east) is a close binary and FS~Tau~B is the origin of the HH~157 outflow (Mundt \& Raga 1991).  FS~Tau B is bisected by a dust lane $\sim$600 AU wide (Padgett et al. 1999).  Our image shows two bright unresolved point sources, with FS~Tau~B being surrounded by a faint reflection nebula.
 
\textbf{IRAS 04191+1523.}
  This nebula around a Class I source was imaged by Hodapp (1994), and is located $1\arcmin$ NE of the Class 0 source IRAM~04191+1522 (Andr\'{e} et al. 1999).  The IRAM source shows evidence for outflow (Moriarty-Schieven et al. 1992), infall, and rotation in its 28,000~AU diameter envelope (Belloche et al. 2002).  Our image shows a close pair of objects.  The SW object is surrounded by an elongated nebula with a faint tail to the east.

\textbf{IRAS 04223+3700.}
  This source has a bright near-IR counterpart and is not visible optically .  Wouterloot et al. (1993) were not able to detect CO emission from the associated cloud.  We find that this object is an E-W oriented 1\farcs0 binary.  There is also a faint reflection nebula $\sim9\arcsec$ to the ENE.

\textbf{IRAS 04239+2436.}
  This is a well studied Class I protostar shown to be a 0\farcs3 binary by Reipurth et al. (2000) using NICMOS aboard HST.  This is the origin of a CO outflow (Moriarty-Schieven et al. 1992) and the parsec scale HH~300 jet (Reipurth et al. 1997).  Our image shows a bright central point source with a faint nebula to the north and east that may represent a wide cavity.  

\textbf{IRAS 04248+2612.}
  Also known as HH~31~IRS~2, this is a well studied Class I source and jet, with near-IR extended emission described in Park \& Kenyon (2002).  Despite its impressive nebula, Kenyon \& Hartmann (1995) estimated that the luminosity of this source is only 0.4~L$_{\odot}$.  HST imaging by Padgett et al. (1999) showed that this object is a 0\farcs16 binary.  Our image shows a central point source with a large, bipolar-like reflection nebula.
   
\textbf{IRAS 04275+3531.}
  This is a previously unreported near-IR reflection nebula.  CO emission has been detected from the associated cloud by Wouterloot \& Brand (1989).  Our image shows a spatially resolved central star with a faint monopolar nebula to the south.

\textbf{IRAS 04287+1801.}
  Well known as L1551~IRS~5, this is the origin of the HH~154 outflow.  This is a binary radio source  separated by 50~AU (Bieging \& Cohen 1985), and is believed to have two aligned protoplanetary disks (Rodr\'\i guez et al. 1998).  This FU Orionis type object shows deep CO bands in its near-IR spectrum that are characteristic of this kind of YSO (Reipurth \& Aspin 1997).  Our image shows a bright, large reflection nebula with a tail arcing to the SW.

\textbf{IRAS 04302+2247.}
  Known as the "Butterfly Star", this is a low luminosity source (estimated to have a bolometric luminosity of 0.34~L$_{\odot}$ by Kenyon \& Hartmann 1995) with a large, bright nebula.  The associated cloud has been detected in CO emission (Bontemps et al. 1996) and this source is the origin of the HH~394 outflow (G\'{o}mez et al. 1997b).  This nebula was observed with HST by Padgett et al. (1999), who describes a spectacular bipolar nebula bisected by a $\sim$900~AU dust lane that completely obscures the central star.  Wolf et al. (2003) present a model of this disk that calls for larger dust grains in the disk and smaller dust grains (similar to those in the ISM) in the envelope.  Our image shows an edge-on bipolar nebula with an obvious dark lane, as well as fainter absorption features on either side of the dark lane.

\textbf{IRAS 04325+2402.}
  Also known as L1535~IRS, this is a well known and complex object seen entirely in scattered light in the near-IR.  Hartmann et al. (1999) used HST to image the complex morphology of this bipolar nebula.  They comment on the low luminosity ($10^{-2}~L_{\odot}$) companion to the north that could be a young brown dwarf.  Both stars appear to be bisected by nearly edge-on disks.  Wang et al. (2001) found this object to be the driving source of three Herbig-Haro objects (HH~434-6).  Our image shows two spatially resolved objects, with the companion superimposed on the bipolar nebula of the primary source.

\textbf{IRAS 04327+5432.}
   This is a previously unreported near-IR nebula associated with HH~378 and L1400.  Wouterloot et al. (1993) detected CO emission.  Our image shows a point source with a faint nebula $\sim4\arcsec$ to the SW.  Faint nebulosity $\sim1\arcmin$ to the NW was found to be H$_{2}$ emission. 

\textbf{IRAS 04365+2535.}
  Also known as TMC-1A, this is a very well studied Class I source, with its near-IR extended emission described in Park \& Kenyon (2002).  Through CO observations,  Brown \& Chandler (1999) used the Keplerian rotation of the molecular envelope of this object to estimate a mass of 0.35 to 0.7~M$_{\odot}$.  Terebey et al. (1989) observed high velocity CO indicative of an outflow.  Windshake of the telescope has elongated our image in the east-west direction by about $0\farcs5$.  We observed a bright point source with a faint bipolar nebula, slightly brighter to the south.

\textbf{IRAS 04381+2540.}
  Associated with TMC-1, this is a well studied Class I source. Its near-IR extended emission is described in Park \& Kenyon (2002).  Terebey et al. (1989) observed high velocity CO indicative of an outflow. Through CO observations, Brown \& Chandler (1999) used the Keplerian rotation of the molecular envelope of this object to estimate a mass of 0.2 to 0.4~M$_{\odot}$.  Our image shows a bright central source with a very faint bipolar reflection nebula.  Windshake of the telescope has elongated our image in the east-west direction by about $0\farcs5$.

\textbf{IRAS 04530+5126.}
  This source illuminates a previously unreported near-IR reflection nebula associated with RNO~33.  V347 Aurigae, the near-IR and optical counterpart of the IRAS source, is a flat-spectrum pre-main sequence H$\alpha$ emission line star (Weintraub 1990) associated with the L1438 dark cloud.  We observed a very bright point source with a faint arc-shaped nebula to the south.

\textbf{IRAS 04591$-$0856.}
  This is a previously unreported near-IR reflection nebula.  This source of a Herbig-Haro object was found by Persi et al. (1988) to possibly be in transition from Class I to Class II.  Tapia et al. (1997) report that the colors are consistent with a Class II source, and their nondetection of nebulosity prompts them to suggest it may be a Class III source.  Our image shows a bright central star with a faint reflection nebula that may be bipolar.  Windshake of the telescope has elongated our image in the east-west direction by about $0\farcs5$.
 
\textbf{IRAS 05155+0707.}
  This source is in the $\lambda$ Ori molecular shell. It is between the parsec scale HH~114 and HH~115 flows, and there is a chain of faint HH knots that appear to emanate from this source (Reipurth et al. 1997). The cold Class 0 source HH~114~MMS is 1\farcm5 to the NW (Chini et al. 1997), and may be the source of the HH~328/329 flows. Lee et al. (2002) find that the CO morphology and kinematics of the HH~114 and HH~115 outflows are consistent with a single driving source.  Our image shows a spatially resolved central object in a bright, complex nebula with a possible companion 6\farcs5 to the east.  We also see H$_{2}$ emission extending $\sim75\arcsec$ to the west, in the direction of HH~114.

\textbf{IRAS 05302$-$0537.}
  Also known as Haro~4-145 and associated with the Orion A-west molecular outflow, this object has been extensively studied at radio and mm wavelengths. CO emission from the bipolar outflow associated with this source was discovered by Fukui et al. (1986).  Felli et al. (1992) found this object to be a source of H$_{2}$O maser emission.  This object appears to be in the southern part of the Orion cloud, so we assume a distance of 470~pc.  We observed a faint, diffuse nebula extending to the north from the bright star to the NW, as well as a small nebula to its SE.

\textbf{IRAS 05311$-$0631.}
  Also known as HH~83~VLA~1. The nebula associated with this source was imaged with HST by Reipurth et al. (2000).  As in the HST image, we see a reflection nebula characteristic of a cavity.  Optical spectroscopy shows that this is an emission line T Tauri star, and the infrared spectrum shows Br $\gamma$ emission and no CO absorption or emission (Reipurth \& Aspin 1997).
 
\textbf{IRAS 05327$-$0457.}
  This is a near-IR reflection nebula in NGC~1977 (just north of the Great Orion Nebula), associated with a maser observed by Henning et al.(1992).  Mookerjea et al. (2000) present $143~\mu$m and $185~\mu$m observations, as well as HIRES processed IRAS maps of this source.  Tsujimoto et al (2003) present near-IR colors and found that this X-ray source is a protostar.  Our image shows a bright star, a diffuse nebula to the south, and three close pairs of stars nearby.  Windshake of the telescope has elongated our image in the east-west direction by about $0\farcs5$.
  
\textbf{IRAS 05378$-$0750.}
This object in the L1641 cloud is associated with object \#146 in Chen \& Tokunaga's (1994) H and K imaging data.  The nebulous object we observed is the faint source to the SSW of star \#7 in their image of object \#146.  Morgan et al. (1991) did not detect a molecular outflow from this IRAS source.  We observed a faint bipolar nebula around an unresolved point source.

\textbf{IRAS 05379$-$0758.}
  This source is associated with object \#163 in Chen \& Tokunaga (1994).  They note that star \#7 (the star to the NE of the three in our image) is the brightest source at M band in object \#163 and is thus likely to be the near-IR counterpart of the IRAS source.  We observed faint diffuse nebulosity to the NE of this star and to the south of the southern star.

\textbf{IRAS 05384$-$0808.}
  Also known as L1641~S4.   The cloud associated with this source was observed from $800~\mu$m to 1.3~mm by Zavagno et al. (1997), who found a dust temperature of 32~K and mass of $0.4~M_{\odot}$.  This nebula is \#132 in the Magakian (2003) catalog of reflection nebulae.  We observed diffuse nebulosity around the bright star to the south and west.  

\textbf{IRAS 05391$-$0841.}
  This source is also in the L1641 cloud.  Strom et al. (1989) and Chen \& Tokunaga (1994) present near-IR images of this region.  We observed three point sources, and a small nebula to the east that may be illuminated by a separate source.

\textbf{IRAS 05399$-$0121.}
  This source in the LBS~30 cloud core is a Class I YSO that is the origin of the large HH~92/93 outflow (Bally et al. 2002a).  We observed knots of emission (possibly from H$_{2}$) extending $1\arcmin$ to the NW. There is no identifiable near-IR point source.

\textbf{IRAS 05403$-$0818.}
  Also known as L1641~S2, this source of a CO outflow (Chen et al. 1992) was studied by Chen \& Tokunaga (1994) and was found to have an SED consistent with a Class I object.  We observed a nebula extending to the east, towards knots of H$_{2}$ emission $\sim30\arcsec$ away.

\textbf{IRAS 05404$-$0948.}
  This is a previously unreported near-IR reflection nebula.  The cloud associated with this source has been mapped in ammonia (Jijina et al. 1999; Harju et al. 1993).  We observed a bright point source with a diffuse nebula to the north and a companion to the SW.  There is also a monopolar nebula, with no central point source visible in the near-IR, $28\arcsec$ to the SE.

\textbf{IRAS 05405$-$0117.}
  This is a previously unreported near-IR reflection nebula.  The cloud associated with this source has been mapped in ammonia (Jijina et al. 1999; Harju et al. 1993).  We observed a bright point source with a faint diffuse nebula to the NW.

\textbf{IRAS 05413$-$0104.}
  This source is the origin of the highly symmetrical HH~212 molecular outflow, discovered by Zinnecker et al. (1998).  Water maser emission has been observed by the VLA (Furuya et al. 1999) from this Class 0 source.  Wiseman et al. (2001) found that the 0.2~M$_{\odot}$ core is elongated perpendicular to the jet axis, presumably due to rotational flattening.  We observed several knots and bow shocks of what appears to be H$_{2}$ emission linearly and symmetrically arranged about a central source, which is not seen.  The nebula extends beyond the $3\arcmin$ field of view of our image.  Our image is presented in Figure 3.

\textbf{IRAS 05417+0907.}
  This source is in the L1594 ($\lambda$ Orionis) dark cloud, which has been the source of mm and sub-mm continuum and line studies.  This object was found to be a source of water maser emission (Terebey et al. 1992) and is the origin of a CO outflow (Terebey et al. 1989).  Hodapp (1994) was the first to present near-IR data on this nebula.  Our image shows a diffuse nebula to the north of the central star, which is a 1\farcs2 binary.  There is also a faint nebula to the NE of a star to the south.

\textbf{IRAS 05450+0019.}
  The associated dense core has been extensively surveyed for ammonia emission (e.g., Jijina et al. 1999; Harju et al. 1993), and has been mapped at $450~\mu$m with SCUBA by Motte et al. (2001).  This nebula is covered in more detail in section 6 of this paper.  Images are presented in Figure 9, and the SED in Figure 10.

\textbf{IRAS 05510$-$1018.}
  This is a previously unreported nebula.  We observed a point source to the NE and a spatially resolved object to the SW with a faint nebula to its east.

\textbf{IRAS 05513$-$1024.}
  A candidate Herbig Ae/Be star suspected to have a spectral type of B7 (Vieira et al. 2003).  J through M band photometry was given by Fouqu\'{e} et al. (1992).  The nearby reflection nebula is the 158$^{th}$ entry in Magakian's Merged Catalogue of Reflection Nebulae (2003).  In addition to a very bright star, we also observed an arc-shaped nebula $\sim8\arcsec$ to the NE.
  
\textbf{IRAS 05548$-$0935.}
  This is a previously unreported nebula.  Of the three stars in our image near the IRAS source, the middle one has a faint nebula to the east.

\textbf{IRAS 05564$-$1329.}
  This is a previously unreported nebula whose source has one prior citation (Codella et al. 1995, a non-detection of H$_{2}$O maser emission).  Our image shows a 4\farcs5 binary star, each with nebulosity to the SE.

\textbf{IRAS 05581$-$1026.}
  This is a previously unreported nebula associated with a source with no prior citations.  We observed a faint monopolar nebula to the SW.  The source appears to be bisected by an edge-on disk.

\textbf{IRAS 05582$-$0950.}
  This is a previously unreported infrared nebula that is associated with RNO~60 in Cohen's (1980) catalog of red nebulous objects.  Our observation shows several stars and nebulae.  There are two nebulae in our field, and the IRAS source position lies between them.  To the east is a bipolar nebula, with a very bright half to the east and a much fainter half to the west.  To the west of the IRAS source lies a faint nebula near a faint point source.  Our image is presented in Figure 5.

\textbf{IRAS 05596$-$0903.}
  This is a previously unreported nebula associated with a source with no prior citations.  We observed a small clump of nebulosity with no source visible at K-band.  There is another small patch of nebulosity, $88\arcsec$ to the SE, that may be related.

\textbf{IRAS 05598$-$0906.}
  This source has been found to be an H$_{2}$O maser source (Codella et al. 1995; Persi et al. 1994), and is near the X-ray source 1RXS J060216.7$-$090700.  This object is 1\farcm2 north of the nebulous star GGD~10 (Gyulbudaghian et al. 1978).  Follow up observations have shown that the bright star to the north, as well as the eastern and southern stars of the four to the south, are sub-arcsecond binaries.  Our image shows diffuse nebulosity near each of the bright stars except for the one farthest west.

\textbf{IRAS 06027$-$0714.}
  This is a previously unreported reflection nebula, with no other publications with near-IR data.  Similar in morphology to IRAS~05581$-$1026, we observe a monopolar nebula to the NNW of a spatially resolved source.  

\textbf{IRAS 06033$-$0710.}
  This is a previously unreported reflection nebula, with no other publications with near-IR data.  SIMBAD classifies this source as a galaxy.  We observe an elongated nebula with a tail to the ENE.

\textbf{IRAS 06047$-$1117.}
  Yun et al. (2001) recently discovered this 0.4~pc jet originating from this candidate Class I source, and present continuum and narrow band H$_{2}$ images.  Our observations through a narrow-band filter show shocked H$_{2}$ emission from a jet 2\farcm9 from end to end.  Our image is presented in Figure 6.
   
\textbf{IRAS 06057$-$0923.}
  This is a previously unreported reflection nebula, with no publications with near-IR data.  SIMBAD classifies this source as a galaxy although no galaxy is apparent in the DSS.  Our image shows what could be a nebula with an edge-on disk or an embedded binary star.

\textbf{IRAS 06249$-$0953.}
  This is a previously unreported near-IR reflection nebula associated with a source with no prior citations.  We observed a spatially resolved source surrounded by a triangular nebula.  Our follow-up L$'$ observations have found a companion star embedded in the nebula to the west.

\textbf{IRAS 06368+0938.}
  This is a previously unreported near-IR nebula in NGC~2264, associated with object \#10 in Margulis et al. (1989) in the Monoceros OB1 cloud.  They derive a luminosity of 5.7~L$_{\odot}$ and a spectral index of +0.8 for this source via a different method. Our image shows several stars near the IRAS source.  Two of the fainter ones to the west have faint nebulae, one of them bipolar.

\textbf{IRAS 06381+1039.}
  This object is near object \#27 in Margulis et al. (1989) in the Monoceros OB1 cloud, and they derive a luminosity of 87~L$_{\odot}$ and a spectral index of +1.7 via a different method.  Based on near-IR imaging by Reipurth et al. (2004), this is the source of the HH~576/7 outflow.  This nebula in NGC~2264 has a spatially resolved central star and a curving nebula extending 20\arcsec\ to the east.  The structure of the nebula suggests it is from H$_{2}$ emission.  

\textbf{IRAS 06382+1017.}
  This nebula in the Monoceros OB1 star forming region is associated with HH~124 VLA and HH~571/2 (Reipurth et al. 2004).  The VLA observations show a compact  group of 6 sources near the IRAS source.  This nebula was first imaged in the near-IR by Moneti \&  Reipurth (1995), and was identified as a Class I source by Margulis et al. (1989).  Our K-band image shows a complex nebula extending to the NE of a spatially resolved source.  Our follow up L$'$ observations show that a tripple star (likely to be the near-IR counterpart to the IRAS source) lies in the resolved K-band peak 10/arcsec to the west of the IRAS position, and another fainter L$'$ source lies at the tip of the nebulosity to the east.  H$_{2}$ and [FeII] images by Reipurth et al. (2004) show several regions of shocked emission from multiple jets.  

\textbf{IRAS 06393+0913.}
  This is a previously unreported near-IR nebula in the Monoceros OB1 star forming region.  The H$_{2}$O maser source (Codella et al. 1995) was detected by IRAS, and the associated cloud has been detected in CO (Oliver et al. 1996; Wouterloot \& Brand 1989).  We observed a bright pair of stars at the location of the IRAS source.  There is a fainter spatially resolved object, $23\arcsec$ to the WSW, with nebulosity on either side.

\textbf{IRAS 07180$-$2356.}
  This source is also known as HH~72~IRS, and is in the Bok globule L1660.  Davis et al. (1997) presents $2.12~\mu$m H$_{2}$ and narrow-band continuum images, showing a string of H$_{2}$ knots $4\arcmin$ long.  This source was not detected as a water maser source by Furuya et al. (2003), but was detected by Felli et al. (1992).  Reipurth \& Aspin (1997) published a near-IR spectrum of this object, showing H$_{2}$ emission lines and CO bands in emission.  They give the distance to this source as 1500~pc.  Although the IRAS position is $\sim23\arcsec$ west of a bright (K = 11.6) red (H$-$K = 2.5) star, MSX observed an $8.3~\mu$m source at the location of this star.  Thus, we believe that this star is the near-IR counterpart to the IRAS source.  Our image shows a bright star surrounded by a small, faint nebula, with knots of H$_{2}$ emission to the east and west extending beyond our field of view.

\textbf{IRAS 15398$-$3359.}
  This is a previously unreported near-IR nebula associated with HH~185, in the Lupus 1 (Barnard 228) cloud.  The envelope of this source was detected at 1.3~mm with a flux of 310~mJy (Reipurth et al. 1996).  We observed a spatially resolved source with a broad, jet-like nebula extending 12\farcs5 to the ENE.

\textbf{IRAS 16288$-$2450.}
  This source is better known as $\rho$ Oph South.  The IRAS source is between the binary T Tauri star to the west and a smaller nebula to the east (itself near the location of IRAS 16289-2450). Since the two near-IR counterparts are widely separated, they are presented in two separate panels.  L1689 IRS~5 is the wide T Tauri binary to the west.  This is the origin of a high velocity molecular outflow (Wu et al. 2004).  Our images show a faint, diffuse nebula around the binary T Tauri star.  The nebula to the east has a higher surface brightness, is more elongated, and has a bright point source in the center.

\textbf{IRAS 16295$-$4452.}
  Moreira et al. (2000) were the first to publish near-IR data of this nebula in Norma, which they conclude is likely to be a Class I protostar.  We observed a point source with faint, diffuse nebulosity in poor seeing (FWHM=1\farcs2).

\textbf{IRAS 16316$-$1540.}
  Also known as L43~VLA~2 and RNO~91, this is a T Tauri star known to have a CO outflow (Mathieu et al. 1988), a bolometric luminosity of 4.3~L$_{\odot}$ (Terebey et al. 1993), a mass of 0.5~M$_{\odot}$ (Levreault 1988), and a disk 1700~AU in diameter (Weintraub et al. 1994).  We observed a bright central star surrounded by a large, bright nebula.  This in turn is surrounded by a large semi-circular nebula, roughly centered on the star, with a radius of $\sim40\arcsec$.
  
\textbf{IRAS 16442$-$0930.}
  This is a previously unreported near-IR reflection nebula in L260.  This object was detected at $850~\mu$m by SCUBA (Visser, Richer, \& Chandler 2002).  Myers et al. (1987) present J through L$'$ photometry, and derive a bolometric luminosity of 0.97~L$_{\odot}$.  We observed a point source in poor seeing (FWHM~=~1\farcs3) with faint, diffuse nebulosity to the south.

\textbf{IRAS 17364$-$1946.}
  This is a previously unreported near-IR nebula in L219 (Barnard 276) that is associated with an outflow (Wu et al. 2004).  Our image shows a N-S elongated, spatially resolved near-IR source with faint nebulosity to the south.

\textbf{IRAS 17441$-$0433.}
  This is a previously unreported near-IR reflection nebula near L425.  Although Carballo et al. (1992) could not determine if this is a galaxy or a YSO based on IRAS photometry and appearance on photographic plates, we are confident that this is a YSO due to its K-band morphology and its coincidence with a nearby dark cloud.  We observed a spatially resolved near-IR source with a small nebula to the north with a smaller, faint nebula to the south.

\textbf{IRAS 18148$-$0440.}
  Also known as L483~FIR, this source was found to be in transition between Class 0 and Class I by Tafalla et al. (2000), whereas Pezzuto et al. (2002) classify this as a Class I object based on its ISO colors.  Visser et al. (2002) find a total mass of 10~M$_{\odot}$, extinction of A$_{v}\approx182$, and detected the associated envelope at $450~\mu$m and $850~\mu$m.  This is also the origin of a CO outflow and a jet with shocked H$_{2}$ emission (Buckle et al. 1999), as well as water maser emission (Furuya et al. 2003).  Although no near-IR point source was observed, we did notice that the morphology of the nebula has changed significantly compared to the images published by Fuller et al. (1995).  We observed a faint nebula divided into two main lobes.  The western lobe has two K-band peaks.  There is also a faint patch of nebula $55\arcsec$ west of the center of the western lobe.

\textbf{IRAS 18264$-$0143.}
  This is a previously unreported near-IR nebula, for which this is the first near-IR data published.  We observed a spatially resolved source with a jet-like, but crooked, nebula extending $14\arcsec$ to the west.

\textbf{IRAS 18270$-$0153.}
  This is a previously unreported near-IR reflection nebula.  The nebula is found $\sim30\arcsec$ east of a star located at the IRAS coordinates, and engulfs a pair of point sources.

\textbf{IRAS 18274$-$0212.}
  This is a previously unreported near-IR nebula.  Our image shows an unresolved near-IR point source with a small, faint nebula to the north.  There is also a faint nebulous ribbon of emission $70\arcsec$ to the ESE that may be related to the IRAS source.

\textbf{IRAS 18278$-$0212.}
  This is a previously unreported near-IR nebula.  MSX observed a mid-IR source at the location of the very red (H-K~$\approx$~3.2) wide binary $\sim10\arcsec$ south of the IRAS source location, and we believe this is the near-IR counterpart to the IRAS source.  In addition to these stars, we observed a small but bright arc-shaped nebula $25\arcsec$ to the NW.

\textbf{IRAS 18331$-$0035.}
  This source is also known as HH~108~VLA 1.  The associated HH flow was discovered by Reipurth \& Eiroa (1992) with optical imaging.  Siebenmorgen \& Kr\"{u}gel (2000) found this to be a triple source, along with HH~108~MMS and a second core.  Whereas the IRAS object is a source of emission at $14~\mu$m, the other two cores are seen in absorption against a diffuse background.  Our image shows a spatially resolved nebula at the location of the IRAS source, with faint nebulosity extending to the south and north.

\textbf{IRAS 18341$-$0113.}
  This is a previously unreported near-IR reflection nebula, that 2MASS data shows is significantly redder than the bright star to the north.  Our image also shows a small jet-like nebula, possibly associated with a faint nearby star, $27\arcsec$ to the NW of the bright star.

\textbf{IRAS 18383+0059.}
  This is a previously unreported near-IR reflection nebula. The PSF in this image is triangular because the primary mirror support was out of adjustment.  Our image shows a wide binary star at the center of a faint monopolar nebula.

\textbf{IRAS 18595$-$3712.}  
  This is a previously unreported near-IR reflection nebula in the VV~CrA area.  This source is also known as ISO-CrA~182 and MMS~23 in Chini et al. (2003).  Olofsson et al. (1999) identify this object as a YSO candidate based on its mid-IR excess observed by ISO.  To better show the bipolar morphology of this nebula, we employed additional smoothing over 12 pixels to show surface brightness contours  down to K~=~21.
    
\textbf{IRAS 19266+0932.}
  Also known as HBC~687 and Parsamian~21, and associated with HH~221, this is an FU Orionis type star (Staude et al. 1992).  Polomski et al. (2005) found no mid-IR companions to this source, and present a detailed SED from 1~$\mu$m to 100~$\mu$m.  Unique among the objects in this paper, the appearance of this nebula at K-band is very similar to its appearance in the optical.

\textbf{IRAS 19411+2306.}
  This is a previously unreported and complex reflection nebula and group of young stars. This object has not been detected as a source of methanol maser emission (Szymczak et al. 2000; Slysh et al. 1999), but it is the origin of a CO outflow (Beuther et al. 2002).  The distance was estimated to be 4.3~kpc by Watson et al. (2003), whereas Guetter (1992) estimated a distance of 2.1~kpc to NGC~6823, which apparently hosts this source.  Our image shows small patches of nebulosity near stars surrounded by a larger diffuse nebula.

\textbf{IRAS 20353+6742.}
  This nebula in L1152 was first observed in the infrared by Myers et al. (1987), who presents J through L-band photometry and calculate a bolometric luminosity of 3.3~L$_{\odot}$ and a spectral slope of +2.06.  This object may be the source of HH~376, and it is the origin of a bipolar high velocity CO outflow (Bontemps et al. 1996).  We observed a spatially resolved near-IR source which appears to be bisected by a dust lane at the center of a faint bipolar nebula. It appears that the NE half is obscured just NE of the source.

\textbf{IRAS 20361+5733.}
  This is a previously unreported near-IR nebula that was observed to be a source of OH maser emission (Slysh et al. 1997).  Our image shows three faint nebulae.  The two to the west appear to be parts of a bipolar nebula.

\textbf{IRAS 20377+5658.}
  This faint reflection nebula in L1036 was first imaged in the near-IR by Hodapp (1994).  This is the origin  of a high velocity bipolar molecular outflow (Wu et al. 2004).  We observed a bright point source with a faint loop of nebulosity to the NW.  
  
\textbf{IRAS 20386+6751.}
  This source is also known as L1157~IRS and HH~375~VLA. This is the origin of a precessing bipolar molecular outflow (Gueth et al. 1996), an H$_{2}$O maser (Furuya et al. 2003), a CO outflow (Umemoto et al. 1992), shocked SiO emission (Zhang et al. 1995) and H$_{2}$ (Hodapp 1994), as well as emission from other molecules. Rodr\'\i guez \& Reipurth (1998) observed this Class 0 source at the VLA to have a flux of 0.26~mJy at 3.6~mm.  Molecular abundance gradients in the shocked gas show that this is a chemically active outflow (Bachiller et al. 2001).  Our image shows two regions of shocked emission $\sim1\arcmin$ to the north and south of an unseen source.  Our image is presented in Figure 6.

\textbf{IRAS 20453+6746.}
  Better known as PV~Cephei or RNO~125, this Herbig Ae/Be star is the origin of the parsec scale outflow HH~215.  ISO spectra show no PAH emission, but do show $10~\mu$m silicate absorption (Acke \& van den Ancker 2004).  The star and associated nebula were brightest from 1977-79, having atomic emission lines with P Cygni profiles, but since then the emission lines have weakened (Magakian \& Movsessian 2001).  Optical [S II] images show knots in a precessing outflow, suggesting intermittent eruptions every $\sim2000$ years (G\'{o}mez et al. 1997a).  Our K-band image shows a very bright star, with an arc-like nebula to the east extending to the NE.  

\textbf{IRAS 20568+5217.}
  This source is also known as HH~381~IRS.  This is a previously unreported near-IR nebula.  Near-IR spectroscopy shows a red continuum with deep CO absorption (Reipurth \& Aspin 1997).  Our image shows a bright central star at the center of a large bipolar nebula.

\textbf{IRAS 20582+7724.}
  This object in L1228 is the source of the HH~199 outflow.  Reipurth et al. (2000) used HST/NICMOS and found a one armed spiral of circumstellar material similar to that seen around FU Ori stars.  Bally et al. (1995) found that the axis of the CO outflow and HH knots differs from the axis of H$_{2}$ knots closer to the source by $\sim40^{\circ}$.  The narrow "jet" to the west in the contour plot is an electronic artifact from the detector array.  In addition to a bright central star and monopolar nebula, we also observe faint knots of emission to the east and west that appear to be the above mentioned H$_{2}$ knots.

\textbf{IRAS 21004+7811.}
  This source is associated with the nebulae GY~21 and RNO~129.  Persi et al. (1988) classified this object as a T Tauri star based on their J through L photometry.  Our image shows a bright central binary with a nebulous loop to the NW.  There is another larger and fainter loop-like nebula $\sim50\arcsec$ to the west.
  
\textbf{IRAS 21007+4951.}
  This source in L988 has been observed in the near-IR by Hodapp (1994).  It is the origin of a bipolar outflow (Clark 1986; Wu et al. 1996) as well as an H$_{2}$O maser (Brand et al. 1994).  We observed a spatially resolved near-IR source with a faint, broad, jet-like nebula extending $\sim25\arcsec$ to the WNW.

\textbf{IRAS 21017+6742.}
  This is a previously unreported near-IR reflection nebula associated with a YSO in the L1172D core that has been extensively studied at mm and radio wavelengths.  It has been found to be the origin of a CO outflow (Myers et al. 1988).  The envelope associated with this source has also been detected with SCUBA at $850~\mu$m (Visser et al. 2002).  Our image shows a bright star surrounded by a faint, diffuse nebula.  An elongated K-band source is $18\arcsec$ to its NW.  To the west, near the IRAS coordinates, is an unresolved star surrounded by a faint, diffuse nebula.

\textbf{IRAS 21169+6804.}
  Also known as CB~230, this object in L1177 was found by Yun (1996) to be a wide binary infrared source.  N$_{2}$H$^{+}$ maps show emission from both components, with the primary source having a thick disk with radius $\sim$1600~AU.  There are also two aligned CO outflows, one for each component (Launhardt 2001).  Only a single object is detected by SCUBA at $450~\mu$m and $850~\mu$m (Huard et al. 1999).  Our image shows a large, bright monopolar nebula with a fainter spatially resolved companion $\sim10\arcsec$ to the east.

\textbf{IRAS 21352+4307.}
 This is a previously unreported near-IR reflection nebula.  Also known as CB~232, in the B~158 cloud, this is the origin of a CO outflow (Yun \& Clemens 1994b). Yun \& Clemens (1995) classify this source as a Class I object and present near-IR photometry and colors.  A strong $850~\mu$m source was detected near CB232:SMM2, $\sim6\arcsec$ to the west of the IRAS source (Huard et al. 1999). Our image shows an unresolved point source with a small, faint nebula to the north.

\textbf{IRAS 21388+5622.}
  This is the source of HH~588 in a bright rimmed cloud in IC~1396, which has been detected in  CS emission (Bronfman et al. 1996) and H$_{2}$O maser emission (Brand et al. 1994).  Our image shows a bright near-IR source in a monopolar nebula extending $\sim10\arcsec$ to the ENE.  There is also a spatially resolved object at the SW end of a line of stars to the north of the monopolar nebula.

\textbf{IRAS 21391+5802.}
  This source is in L1121, in the IC~1396N core.  Beltr\'{a}n et al. (2002, 2004) used CS and methanol emission to probe the outflow, and estimate that there is about 5~M$_{\odot}$ of circumstellar material.  They conclude that the source and outflow properties are consistent with this being a low mass Class 0 source.  This object is also a source of H$_{2}$O maser emission (Furuya et al. 2003).  Our image shows a bright, jet-like nebula near the IRAS coordinate as well as several large patches of nebulosity around the field.

\textbf{IRAS 21445+5712.}
  This is a previously unreported near-IR reflection nebula in IC~1396.  It has been detected in CO emission (Wouterloot \& Brand 1989).  Our image shows an E-W elongated nebula, as well as several stars embedded in a nebula $\sim85\arcsec$ to the NW.

\textbf{IRAS 21454+4718.}
  Better known as V1735 Cygni, this is in L1031 in the IC~5146 star cluster, itself the subject of a study by Herbig \& Dahm (2002).  Although this is a well known FU Orionis star, Pezzuto et al. (2002) classified this as a T Tauri star based on its ISO colors.  \'{A}braham et al. (2004) found this star faded by $\sim$40\% in the near-IR from about 1975 to 2000.  This source was observed to be the origin of a CO outflow by Levreault (1983).  The reflection nebula to the ENE is at the location of the submillimeter source V1735 Cyg SM1.

\textbf{IRAS 21569+5842.} 
  This source in L1143 is likely to be a YSO in transition between Class I and II based on its IRAS SED and optical/near-IR imaging (Magnier et al. 1999).  Our image shows a bright non-stellar object with a faint nebula to the north.
  
\textbf{IRAS 22051+5848.} 
  Also known as HH~354~IRS, this object is in L1165 (Reipurth et al. 1997).  Near-IR spectra show a red continuum with deep CO absorption that is very similar to the spectrum of L1551~IRS~5 (Reipurth \& Aspin 1997).  The cold envelope associated with this IRAS source was detected at $850~\mu$m by Visser et al. (2002), who derive a total mass of 29.4~M$_{\odot}$, and whose $^{12}$CO maps show a bipolar outflow.  Persi et al. (1988) reported near-IR photometry of this nebula and found it to be blueward of the main sequence reddening vector in a J-H vs. H-K color-color diagram, suggesting that scattered light dominates the near-IR flux from this nebula.  Our image shows a spatially resolved near-IR source with a nebula extending to the SSW.  The bright E-W elongated bar of nebulosity $\sim13\arcsec$ to the SSW of the source, as well as a faint streak of nebulosity extending farther to the SSW, are optically visible in the DSS plates.

\textbf{IRAS 22176+6303.} 
  This is the well known source Sharpless 140 IRS~1, also known as RAFGL 2884, in the L1240 dark cloud.  This is the origin of a high velocity CO outflow (Bally \& Lada 1983), HH~615-8 (Bally et al. 2002b), as well as methanol and H$_{2}$O masers.  Beichman et al. (1979) found three 20~$\mu$m sources associated with this nebula.  Radio observations have shown that IRS~2 and IRS~3 are both binary (Schwartz 1989).  Megeath et al. (2004) analyses color-color data from the Spitzer Space Telescope to classify cluster members in the field around S140~IRS~1.  At K-band we observed IRS~1 as the bright central star in a large, complex nebula.  IRS~2 is a fainter object embedded in the nebula to the east.  IRS~3, to the north, was not detected.  Our image is presented in Figure 7.

\textbf{IRAS 22266+6845.}
  Umemoto et al. (1991) found this source in L1221 to drive an unusual U-shaped CO outflow.  Lee et al. (2002) report that the CO emission is complicated by multiple outflows with multiple driving sources, and state that there may be another outflow source 25\arcsec\ to the east of the IRAS source.  This object is likely to be the origin of the HH~363 flow, and it has been detected as a source of water maser emission by Furuya et al. (2003).  Our image shows a wide binary star with nebulosity to the north and east.
  
\textbf{IRAS 22267+6244.}
  This nebula in L1203 was first observed in the near-IR by Hodapp (1994), and was found to be the origin of a CO outflow (Wouterloot \& Brand 1989).  MSX shows a double source separated by $39\arcsec$ at 8.28~$\mu$m that is too embedded to be seen in our K-band data.  We observed a large, faint nebula with a broad, dark band across the center.  There is a star, with a bright nebula to its west, on the southern edge of the large nebula.  Our image is presented in Figure 8.

\textbf{IRAS 22272+6358.}
  This source is in L1206.  Our image shows a bright point source to the east (IRAS 22272+6358B) surrounded by a diffuse nebula to the NE.  IRAS 22272+6358A, about 45\arcsec{~} to the west, is seen only as an elongated nebula.  Ressler \& Shure (1991) found that IRAS 22272+6358A is a Class I object with a nearly edge on disk, based on their J through M photometry as well as IRAS and radio data.  It is seen only in scattered light in the near-IR and is the origin of a CO outflow (Sugitani et al. 1989). IRAS 22272+6358B is seen directly in the near-IR and is either a late Class I or an early Class II object.  Pezzuto et al. (2002) classify IRAS 22272+6358B as a Class I source, although it nearly has the ISO colors of a Class II source.

\textbf{IRAS 22376+7455.}
  These are previously unreported near-IR reflection nebulae in L1251B.  An optically visible nebula and HH outflows were discovered by Eiroa et al. (1994), although this IRAS source does not seem to be the origin of the HH~189 outflow.  Sato \& Fukui (1989) observed two CO outflows in L1251, one from IRAS 22343+7501, and another from this source.  Nikolic et al. (2003) found this source, which they identify with core \#4 (H2b) in their paper, to be coincident with  emission from H$^{13}$CN, HN$^{13}$C, $^{13}$CO, C$^{18}$O, C$^{34}$S,  H$^{13}$CO$^{+}$, and SO.  This was also observed to be the source of water maser emission by Furuya et al. (2003).  Our image shows several stars in a small group.  At the SE end is a large bipolar nebula.  In the middle of the group is a small nebula arcing north of a faint, spatially resolved source.  To the west is a faint star with an elongated nebula to its south.  Faint nebulosity engulfs the whole group.
  
\textbf{IRAS 23037+6213.}
  This source, situated in the Cepheus C cloud, has been observed in the near-IR by Hodapp (1994).  It has been found to be the origin of H$_{2}$O maser emission  (Wouterloot \& Walmsley 1986) and CO emission (Wouterloot \& Brand 1989).  Megeath et al. (2004) uses color-color data from the Spitzer Space Telescope to classify cluster members over a 3~pc wide region.

\section{General Properties of Sources}

\subsection{Luminosity Distribution}

  In order to calculate the bolometric luminosity of our sources from the IRAS fluxes, we used the following relation: 
\begin{equation}
L_{bol}(L_{\odot}) = [4.16\times10^{-6}\cdot F_{12} + 3.54\times10^{-6}\cdot F_{25} + 9.59\times10^{-7}\cdot F_{60} + 6.57\times10^{-7}\cdot F_{100}] \cdot D^{2}
\end{equation}
where F is the IRAS flux density in Jy at the wavelength (in microns) indicated by the subscript, and D is the distance to the source in parsecs.  This formula, described by Reipurth et al. (1993), includes an estimate of flux emitted at sub-millimeter wavelengths based on their finding that the material responsible for the flux beyond $100~\mu$m tends to have a characteristic temperature of 36~K.  The IRAS $100~\mu$m flux has been used to roughly estimate the submillimeter contribution to the bolometric luminosity by fitting a 36~K blackbody to the IRAS $100~\mu$m data point.  The luminosity distribution presented in Figure 11 has a median value of $L_{bol}~=~5.8~L_{\odot}$, with most of the sources having a luminosity between $0.5~L_{\odot}$ and $30~L_{\odot}$.

\subsection{Nebula Evolution and Morphology}
   
  As a star begins to form, an outflow from the young star starts to carve a bipolar cavity out of the surrounding material.  The cavity gets larger and wider as the star evolves and the outflow disperses the surrounding cloud (Arce \& Sargent (2006) and citations therein).  Once enough of the cloud has been dispersed, the young star becomes optically visible in the T Tauri phase.  Having produced this atlas of near-IR reflection nebulae, we sought to classify them into morphological groups in an attempt to derive an evolutionary sequence and test this model. 

  We sought to do this through a comparison of the average spectral index for various morphological groups, under the assumption that a higher spectral index indicates a younger object.  After careful inspection of all of the nebulae, it became apparent that a clear, one dimensional classification scheme was not practical because we were unable to unambiguously place each nebula in a morphological group.  Although there are a number of cases where it is easy to classify the morphology of an individual nebula, there is also a large number of nebulae that defy simple morphological classification.  If an evolutionary trend were to be found, another method was required. Assuming that the spectral index decreases with time, we examined the nebulae in this atlas in the order of their spectral indices.  After doing this, we noticed the following trends.
 
   The clearest trend, and the most quantifiable, is that the central star of the nebula tends to become more visible as the spectral index decreases.  Among the one-third of the nebulae with the highest spectral index, there are 4 cases where a star, believed to be the central source, is visible as an unresolved point source.  For these objects, most of the flux is scattered light from the nebula or what appears to be shocked emission. Among the one-third of the nebulae with the lowest spectral index, there are 5 cases where we cannot see a central unresolved point source.  Light from the central star dominates the K-band flux for these objects.

   Other trends we noticed are less obvious and less quantifiable.  In order to understand if younger sources are more likely to have shocked H$_{2}$ emission, we considered the one-third of these 106 nebulae with the highest spectral index versus the one-third of sources with the lowest spectral index. Among the one-third of the nebulae with the highest spectral index, 25\% (9/35) show what appears to be shocked H$_{2}$ emission.  Among the one-third of the nebulae with the lowest spectral index, we found that 9\% (3/35) of cases appear to have shocked H$_{2}$ emission.  In these cases, the apparent H$_{2}$ emission is confined to a few isolated knots, as opposed to the long chains of emission often found around the younger sources.  This is consistent with the expectation that young stars experience accretion events, leading to bursts of outflow activity.

  We also noticed that the nebulae around the older objects tend to be round and less elongated, diffuse and closer to the central star.  For the younger objects, structured nebulosity could more often be found farther from the IRAS source, often as part of shocked emission or as a bipolar nebula.  Comparatively less of the scattered light is very close to the  central source.  Finally, we noticed that we tend to find diffuse, arc shaped reflection nebulae (e.g. IRAS 03445+3242) only around sources with a low spectral index.  

   Overall, these general trends that we noticed are consistent with the concept of nebula evolution described above.  Younger objects are surrounded by more dust, and thus are rarely seen directly at K-band.  Shocked H$_{2}$ more likely to be found around younger objects, especially as large jets.  As the cloud is dissipated, the star becomes directly visible and dominates the near-IR flux.  We interpret nebulae such as those around IRAS 03445+3242 and HH~83~VLA~1 as light scattering off of the walls of a cavity that has been almost completely dispersed.  

\subsection{Nebula Brightness and Size}

  The nebulae in this paper range in size from being just larger than a stellar point to over 3$'$ long, and display a similarly broad range of morphologies. The amount, distribution, and structure of the surrounding dust, as well as our point of view, can lead to nebulae appearing very different although they are at the same evolutionary stage. This limits the accuracy of our comparison of the sizes and near-IR brightness of the nebulae to the bolometric luminosities of these objects.  Also, in cases where the central star is very bright at K-band, such as IRAS 21004+7811, the extent of the surface brightness contour may be dominated by the scattered light from the telescope rather than from the nebula.  Cases where the flux is clearly dominated by light from the star rather than the nebula or where scattered light from the star dominates the apparent size of the nebula were not considered in the following analysis. Since our goal was to determine if there is a correlation between the nebula brightness and size with the source bolometric luminosity, we only considered the part of the nebula believed to be illuminated by the IRAS source.  For example, we found two small nebulae near IRAS 06368+0938, but they are far enough from the IRAS source position that it is unclear which nebula, if any, is associated with the IRAS source.  Such cases were not considered in this analysis.

  The K-band magnitudes for the nebulae are from the 2MASS extended source catalog for the cases where the nebula is included in this catalog.  If not, the magnitude from the point source catalog was used.  These magnitudes were converted to absolute magnitudes for the sources whose distances are known.  This was then divided by 2.5 to directly compare the log of the bolometric luminosity to the log of the K-band flux.  Figure 12 includes the 73 sources that have both 2MASS fluxes and distance estimates.  The bolometric luminosities are as described above.  Although we excluded objects where flux from the central star dominated over the flux from the nebula, in most cases there is still some near-IR flux from associated stars.  The extent of this contamination varies from case to case.

  We found a correlation between the near-IR brightness of the nebulae and the bolometric luminosity of the IRAS source as shown in Figure 12.  The best fit power law relation has an index of $0.75\pm0.09$, as this is the slope of the linear regression line on the log-log plot.  The standard deviation from the regression line is 1.63 magnitudes.  Since this analysis did not include nebulae where the near-IR luminosity is dominated by the central star, and since the central star tends to be visible on older objects, this correlation is biased against older stars.  A significant amount of scatter is expected due to the diversity of objects in this sample, i.e. some sources are clearly seen through little extinction while others are nearly completely obscured at $2.2~\mu$m.

   A mathematical model developed by Pendleton et al. (1990) shows that, for the case of a star illuminating a homogeneous medium, the luminosity of a reflection nebula should be proportional to the luminosity of the illuminating star.  The nebulae illuminated by high bolometric luminosity sources appear to be underluminous in the near-IR because, as we have found, sources with higher bolometric luminosity tend to have a higher spectral index.  Objects with a higher spectral index tend to be younger and cooler, and thus emit less near-IR flux than sources with a lower index.  Furthermore, we expect that younger sources will tend to be surrounded by more dust, increasing the amount of extinction.  Thus a nebula around such a source would appear to be underluminous in the near-IR given its bolometric luminosity.

  To estimate the size of a nebula, we determined the total area contained within the K~=~19 magnitudes per square arcsecond surface brightness contour.  This contour was chosen as it is the faintest contour in nearly all of the figures.  We took the square root of the area within the contour to determine the size of a box that would contain the same area, and this is the value we used as the apparent size of the nebula.  We find a correlation between the bolometric luminosity of the IRAS source and the nebula's linear size for those nebulae where we have a distance estimate, and this is presented in Figure 13.  Those cases where the measured size of the nebula is dominated by scattered light from a bright central star are noted by an asterisk in the last column of Table 2.  On a log-log plot, the slope of the regression line is $0.41\pm0.04$, with the data points having a standard deviation from the regression line of 0.25 (a factor of 1.76).  The sizes \emph{d} of the nebulae tend to follow the power law relation over several decades of luminosity.  We find it interesting that the two sources with the lowest luminosity (IRAS 04302+2247, $L_{bol}~=~0.28~L_{\odot}$, and IRAS 04248+2612, $L_{bol}~=~0.33~L_{\odot}$) are both associated with bright and well known reflection nebula.  In the case of IRAS 04302+2247, the central star cannot be seen at near-IR wavelengths because the circumstellar disk is seen edge on, whereas the central source of IRAS 04248+2612 has been resolved as a 0\farcs3 binary by HST (Reipurth et al. 2000).

  We believe that the scatter in the plot is due to the various ages, the various morphologies, the amount of extinction and the uncertainties in the distance to the nebula.  We do not believe that uncertainties in the distance to these sources is the origin of these correlations between bolometric luminosity, near-IR luminosity, and size.  Distance uncertainties are typically no greater than a factor of 2, whereas these relationships of near-IR brightness and size vs. bolometric luminosity cover several orders of magnitude in scale.

\subsection{Shocked H$_{2}$ Emission}

    Having observed 197 sources at K-band, we found 22 nebulae that appear to be the source of shocked H$_{2}$ emission.  We suspected a nebula to be a source of shocked H$_{2}$ emission if we found multiple distinct clumps of emission, linearly or symmetrically arranged about the central star. In comparison, reflection nebulae due to dust scattered light tend to be smooth, with diffuse edges.  We performed follow-up observations using a 2.122~$\mu$m H$_{2}$ filter for 9 nebulae suspected of having shocked H$_{2}$ emission but not yet observed through a H$_{2}$ filter.   We confirmed the presence of H$_{2}$ emission for all of them, thus we are confident that what we believe looks like H$_{2}$ emission actually is.  These nebulae are listed in Table 3, which also includes the coordinates of a marked star or clump in each field.

  Although $61\%$ of the objects presented in this paper are known sources of molecular outflows (either through millimeter line emission, or they are the source of an HH object), relatively few show shocked H$_{2}$ emission.  For example, IRAS 05311$-$0631 is the source of a well known Herbig-Haro outflow (HH~83), yet shows no signs of shocked emission within 1\farcm5. It is possible that there is shocked emission that we did not see, either because it is too faint or outside of our $3\arcmin$ field of view with QUIRC.  At any given time, it appears that $\sim10\%$ (22/197) of Class I objects are sources of shocked H$_{2}$ emission.

   Figures 14 through 22 show, as necessary, broad K-band, K-band $-$ H$_{2}$, H$_{2}$ $-$ K-band, and H$_{2}$ emission of the nebulae we observed in H$_{2}$.  A $15\arcsec$ scale bar is under the label in each figure.  The morphology of this emission varies from a few knots, such as to the east of IRAS~03445+3242, to a series of bow shocks such as to the west of IRAS~05155+0707.  In several cases, the morphology of the emission is in the form of a few small bow shocks.  IRAS~21391+5802 has a series of shocks to the NE and a few knots to the NW of the source, which itself is only seen in scattered continuum light.  There are many cases where reflection nebulae and H$_{2}$ emission are coincident.  In such cases, narrow band imaging is very useful to confidently identify the nature of the observed nebulae.  Aside from a small knot near IRAS 05403-0818, the H$_{2}$ emission we observed tends to be found quite far from the YSO.  It is not uncommon for shocked emission to be found $80\arcsec$ away.  In the case of IRAS~00182+6223, the H$_{2}$ emission $53\arcsec$ away has a projected separation of 1.2~pc.  

\subsection{Binary Frequency}

  The observations presented in this paper were made to explore the circumstellar environment around our sample of candidate Class I sources.  As such, these data have limited use for constraining the binary frequency of these sources.  An exposure time of 60~s was used for all observations to obtain a homogeneous data set, which resulted in many of the central stars being saturated.  A program is underway to determine the binary frequency of a large number of Class I YSOs.  We have tailored an observing program for this purpose, being careful not to saturate and observing through H, K, and L$'$ filters.  A forthcoming publication will present the results of this program.

\section{IRAS 05450+0019}

   Although it does not have an unusually high bolometric luminosity, IRAS 05450+0019 is the source of  one of the brightest reflection nebulae in this sample.  This nebula is also one of the largest, being over 2$'$ (0.3~pc) in extent.  Only large collimated outflows such as HH~212 and IRAS 06047$-$1117, and nebulae around very luminous sources such as IRAS 22176+6303, are larger.  This suggests that a special set of criteria must be met (e.g. a luminous source, enough dust distributed over a large area to scatter light but not enough to obscure the nebula) for such a bright and large nebula to be visible.  

   Although it is not well known as a bright nebula, IRAS 05450+0019 has been observed several times in the radio and mm.  It has been observed in the course of surveys for emission from H$_{2}$O (Wouterloot \& Walmsley 1986) and methanol (Slysh et al. 1999) masers, CO (Likkel et al. 1991), ammonia (e.g., Jijina et al. 1999), H$^{13}$CO, and CO$^{18}$O (Aoyama et al. 2001).  This object is 7\farcm6 ESE of NGC~2071, a center of high mass star formation (e.g., Aspin et al. 1992), in the Orion B molecular cloud.  It is associated with NGC~2071e and is at a distance of $\sim$500~pc (Clark 1991).  Harju et al. (1993) estimated the bolometric luminosity to be 47~L$_{\odot}$, which is considerably greater than our estimate of 27.6~L$_{\odot}$.  Having measured the ammonia mass, they infer a total mass of 86~M$_{\odot}$ for a clump 40\arcsec\ E of the IRAS source (near the K-band peak), plus 12~M$_{\odot}$ for a clump 2$'$ WNW of the IRAS source.  They found that this pair of clumps has about 20 times as much potential energy as kinetic energy, and thus it is possible that they are coalescing.  Launhardt et al. (1996) observed dust emission from the associated dust envelope at 1.3~mm but found no compact source, suggesting this object has an optically thin dust envelope.  They estimate this object has about $10~M_{\odot}$ in gas with 2~M$_{\odot}$ in the dust envelope, and is internally heated.  The 1.3~mm peak corresponds to the bright component of the nebula to the east of the IRAS source.  They also observed this nebula at K-band using the La Silla 2.2~m telescope, and present a small near-IR image overlaid with the millimeter map.

   J, H, and K-band images with surface brightness contours are presented in Figure 9.  In addition to IRAS observations at 12, 25, 60, and $100~\mu$m, IRAS 05450+0019 was observed by ISO at $6.7~\mu$m and $14.3~\mu$m.  Figure 10 shows the SED of this source, combining these data with data from IRTF at L$'$-band and the 2MASS extended source catalog for the J, H, and K$_{s}$ data points.  The nebula appears to be bipolar, with a broad dark band across the center.  The eastern side is much brighter than the western, suggesting that the eastern side is tilted towards our line of sight.  If the central dark band  is from a disk seen nearly edge on, then it would be much larger ($\sim$15,000~AU or 0.07~pc in diameter) than would be expected ($\sim$100~AU for a protoplanetary or debris disk, $\sim$1000~AU for a CTTS disk).  Finally, this dark band appears tilted relative to the axis of the reflection nebula by $\sim20^{\circ}$, which is unique among the nebulae in this paper.

  It is possible that the dark band is the shadow of a much smaller disk being projected into the surrounding dust cloud.  This concept is proposed and examined by Hodapp et al. (2004) to explain the unusually large apparent size ($\sim3000$AU) of the disk around ASR41 in NGC~1333.  They argue that if what they see is caused by an obscuring band in front of a nebula, then the band should be more prominent at shorter wavelengths, which is contrary to their observations.  We note that in the case of our observations, the dark band does not seem more apparent at shorter wavelengths.   The nebula seen by Hodapp et al. (2004) is almost equally bright on both sides of the dark band and thus is seen nearly edge-on, which helps to make the disk shadow easy to see.  The nebula around IRAS 05450+0019 does not seem to be seen edge-on from our point of view, rather the eastern side is tilted towards us. Pontoppidan \& Dullemond (2005), who present more detailed modeling of disk shadows, show that a dark band can be visible for inclinations as low as $60^{\circ}$ ($90^{\circ}$ being edge-on).  Thus, even for cases where we do not see the disk edge-on, as it seems for the case of IRAS 05450+0019, a disk shadow may still be visible.
   
\section{Conclusions}
   We have conducted a K-band imaging survey of 197 nearby candidate Class I sources.  Most of our sources are between 150~pc and 500~pc away, and have a spectral index from 0 to 2.  We found 106 near-IR counterparts that are associated with nebulae, 41 of which have not been previously reported.  Among these are two areas, one in Orion and another in Serpens, that have a significant number of new nebulae.  Based on our observations and archival data, we have come to the following conclusions:

1)  Sources with higher bolometric luminosities tend to have larger, brighter nebulae.  However, the near-IR brightness of the nebulae do not scale directly with the bolometric luminosity of the source.  We believe this is because the more luminous sources have a tendency to be younger.

2)  We found that the near-IR flux from younger objects tends to be mostly from the nebula, and the central star is rarely visible at K-band.  For older YSOs, the central star is almost always visible at K-band.  Most of the near-IR flux comes from the central star, and the much fainter nebula does not tend to be visible as far from the star.

3) Although many Class I sources are known to have molecular outflows, only $\sim10\%$ show what appears to be shocked H$_{2}$ emission.  This suggests that, at any given time, shocked H$_{2}$ emission is rare within $\sim0.15$~pc the central star.  Also, Class I YSOs that appear to be less evolved are more likely to be the source of H$_{2}$ emission than ones that appear more evolved.

4) Among all of the nebulae observed in this sample, IRAS 05450+0019 deserves special attention.  Although it is relatively unknown, it is among the largest and brightest nebulae in this sample.  It appears to be a nearly edge-on bipolar reflection nebula with a very large dark band that may be the shadow of a smaller disk projected into the surrounding envelope.




\acknowledgments

\emph{Acknowledgments}  
We would like to thank our referee for a thorough review and constructive comments, which significantly improved this paper.  This research has made use of the SIMBAD database, operated at CDS, Strasbourg, France.  This research has made use of NASA's Astrophysics Data System.  The Digitized Sky Surveys were produced at the Space Telescope Science Institute under U.S. Government grant NAG W-2166.  The images of these surveys are based on photographic data obtained using the Oschin Schmidt Telescope on Palomar Mountain and the UK Schmidt Telescope. The plates were processed into the present compressed digital form with the permission of these institutions.  The National Geographic Society - Palomar Observatory Sky Atlas (POSS-I) was made by the California Institute of Technology with grants from the National Geographic Society.  The Second Palomar Observatory Sky Survey (POSS-II) was made by the California Institute of Technology with funds from the National Science Foundation, the National Geographic Society, the Sloan Foundation, the Samuel Oschin Foundation, and the Eastman Kodak Corporation.  The Oschin Schmidt Telescope is operated by the California Institute of Technology and Palomar Observatory.  The UK Schmidt Telescope was operated by the Royal Observatory Edinburgh, with funding from the UK Science and Engineering Research Council (later the UK Particle Physics and Astronomy Research Council), until 1988 June, and thereafter by the Anglo-Australian Observatory. The blue plates of the southern Sky Atlas and its Equatorial Extension (together known as the SERC-J), as well as the Equatorial Red (ER), and the Second Epoch [red] Survey (SES) were all taken with the UK Schmidt.  Supplemental funding for sky-survey work at the ST ScI is provided by the European Southern Observatory.  This research has made use of the NASA/ IPAC Infrared Science Archive, which is operated by the Jet Propulsion Laboratory, California Institute of Technology, under contract with the National Aeronautics and Space Administration.  This publication makes use of data products from the Two Micron All Sky Survey, which is a joint project of the University of Massachusetts and the Infrared Processing and Analysis Center/California Institute of Technology, funded by the National Aeronautics and Space Administration and the National Science Foundation.



\appendix

\section{2MASS to MKO Color Transformations}
   In this paper, we use both the Mauna Kea Observatories (MKO) and 2MASS photometric systems.  The surface brightness contours in Figure 2 through 9 are in the MKO system whereas the photometry in Table 2 is in the 2MASS system.  We have derived the following equations to convert between these two systems.  These equations were derived by comparing the UKIRT faint standard stars, observed by UKIRT in the MKO system, to the 2MASS magnitudes of the same stars.  The values and one sigma uncertainties given are based on linear fits to the data.  These relations are limited by the fact that there are few red UKIRT standard stars.  Also, using the (J$-$H) and (H$-$K) colors to derive an H magnitude (eqs. 3 - 6) may not yield the same result.  To do so would require that (J$-$H)/(H$-$K) is a constant, which is generally not true.  However, in the case of the UKIRT standard stars, this is approximately true.

\begin{equation}
 J_{2MASS} = (0.015 \pm 0.005) + (0.050 \pm 0.010)(J-H)_{MKO} +  J_{MKO}
\end{equation}

\begin{equation}
 J_{MKO} = (-0.012 \pm 0.005) + (-0.055 \pm 0.008)(J-H)_{2MASS} +  J_{2MASS}
\end{equation}

\begin{equation}
 H_{2MASS} = (0.025 \pm 0.004) + (-0.028 \pm 0.007)(J-H)_{MKO} +  H_{MKO}
\end{equation}

\begin{equation}
 H_{2MASS} = (0.021 \pm 0.003) + (-0.044 \pm 0.010)(H-K)_{MKO} +  H_{MKO}
\end{equation}

\begin{equation}
 H_{MKO} = (-0.026 \pm 0.003) + (0.029 \pm 0.006)(J-H)_{2MASS} +  H_{2MASS}
\end{equation}

\begin{equation}
 H_{MKO} = (-0.021 \pm 0.003) + (0.041 \pm 0.011)(H-K_{s})_{2MASS} +  H_{2MASS}
\end{equation}

\begin{equation}
 K_{s,2MASS} = (0.014 \pm 0.004) + (0.042 \pm 0.013)(H-K)_{MKO} +  K_{MKO}
\end{equation}

\begin{equation}
 K_{MKO} = (-0.016 \pm 0.004) + (-0.030 \pm 0.015)(H-K_{s})_{2MASS} +  K_{s,2MASS}
\end{equation}

\begin{equation}
(J-H)_{2MASS} = (-0.010 \pm 0.006) + (1.078 \pm 0.011)(J-H)_{MKO}
\end{equation}

\begin{equation}
(J-K_{s})_{2MASS} = (0.004 \pm 0.006) + (1.018 \pm 0.008)(J-K)_{MKO}
\end{equation}

\begin{equation}
(H-K_{s})_{2MASS} = (0.007 \pm 0.004) + (0.914 \pm 0.013)(H-K)_{MKO}
\end{equation}





\clearpage 

\begin{figure}
\plottwo{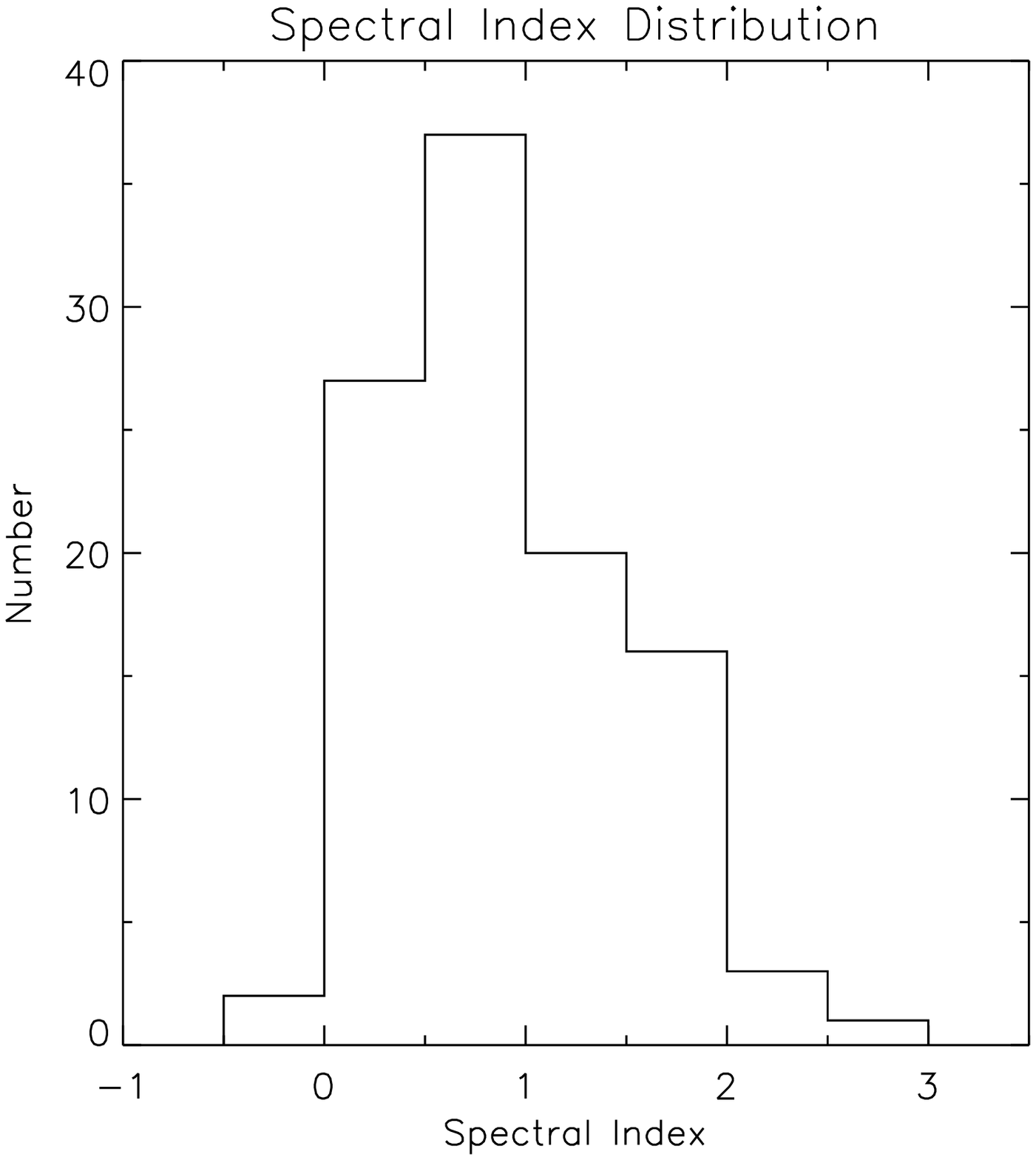}{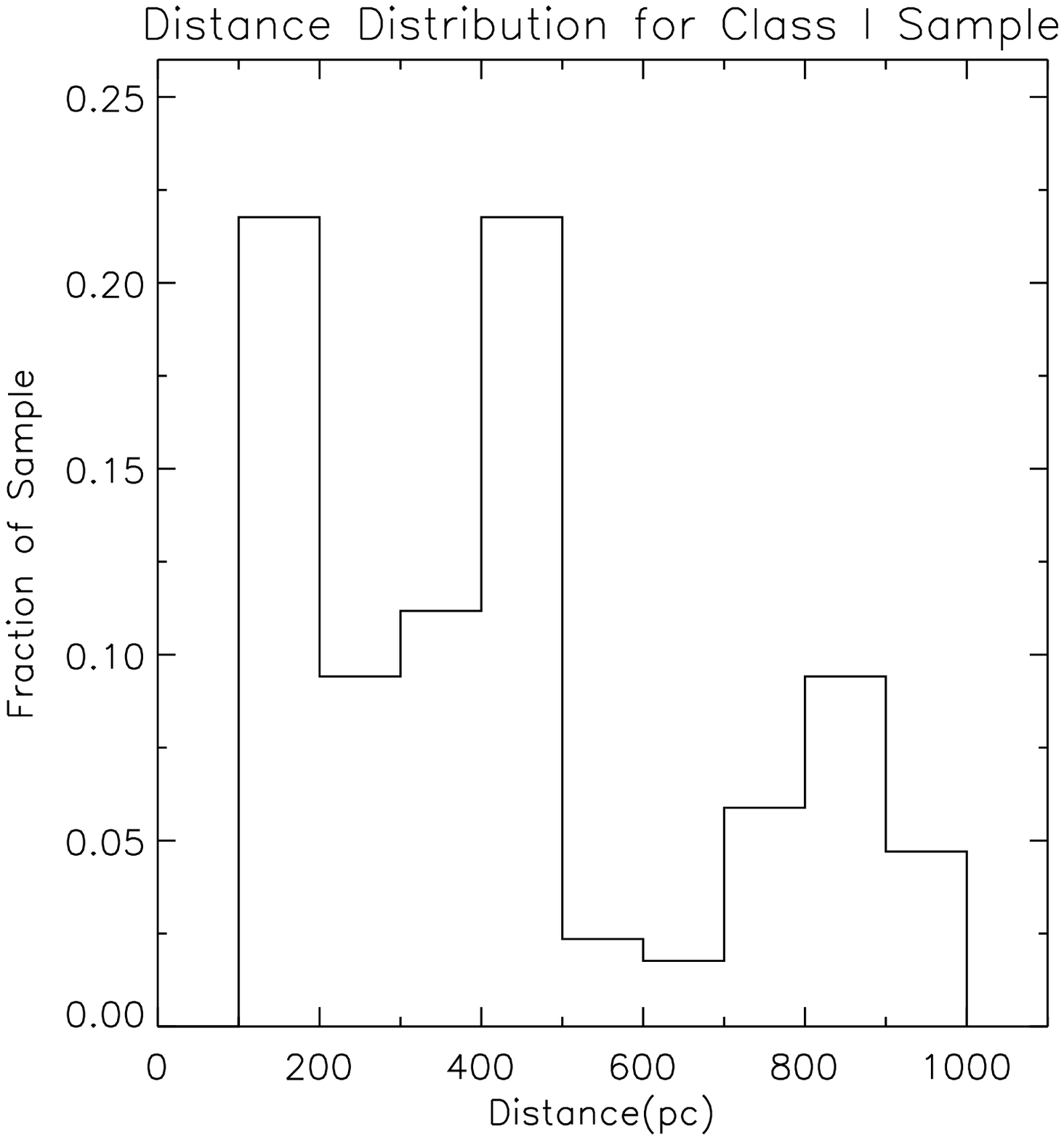}
\caption{Left: Our sample's spectral index distribution, with a median value of +0.79.  This plot shows that only a few of our sources are T Tauri stars (index $< 0$), and the majority are transitional (i.e. between Class II and I) or Class I YSOs.  Right: Distribution of distances to our sample objects, which has a median value of 470~pc.   \label{fig1}}
\end{figure}

\clearpage

\begin{figure}
\plotone{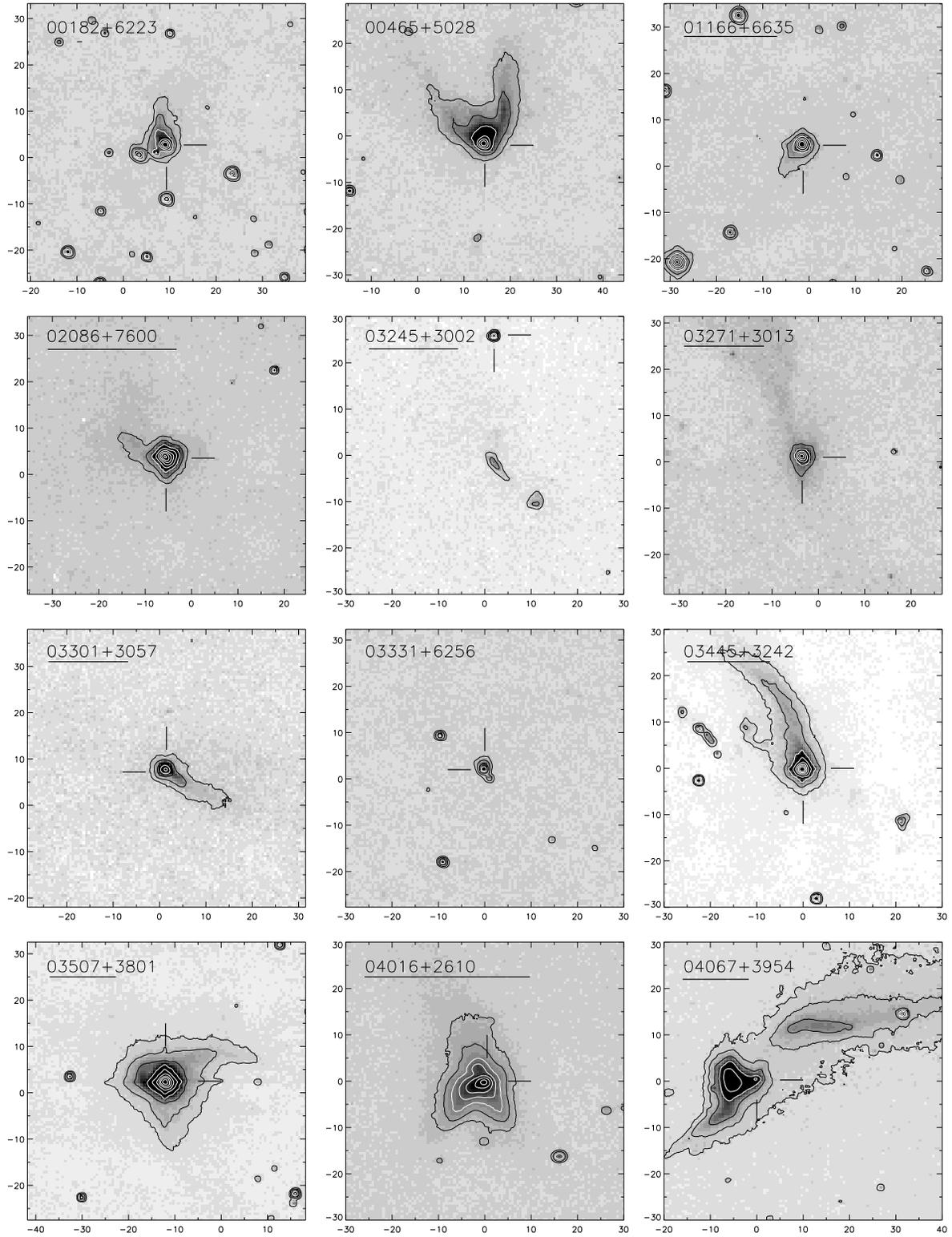}
\caption{$60\arcsec \times 60\arcsec$ images of each source, with the IRAS position at (0,0).  The 2MASS coordinates for the star indicated with tick marks is given in Table 2.  A 5,000~AU scale bar is under the name of each object for which we have a distance estimate.}
\end{figure}

\clearpage

\begin{figure}
\addtocounter{figure}{-1}
\plotone{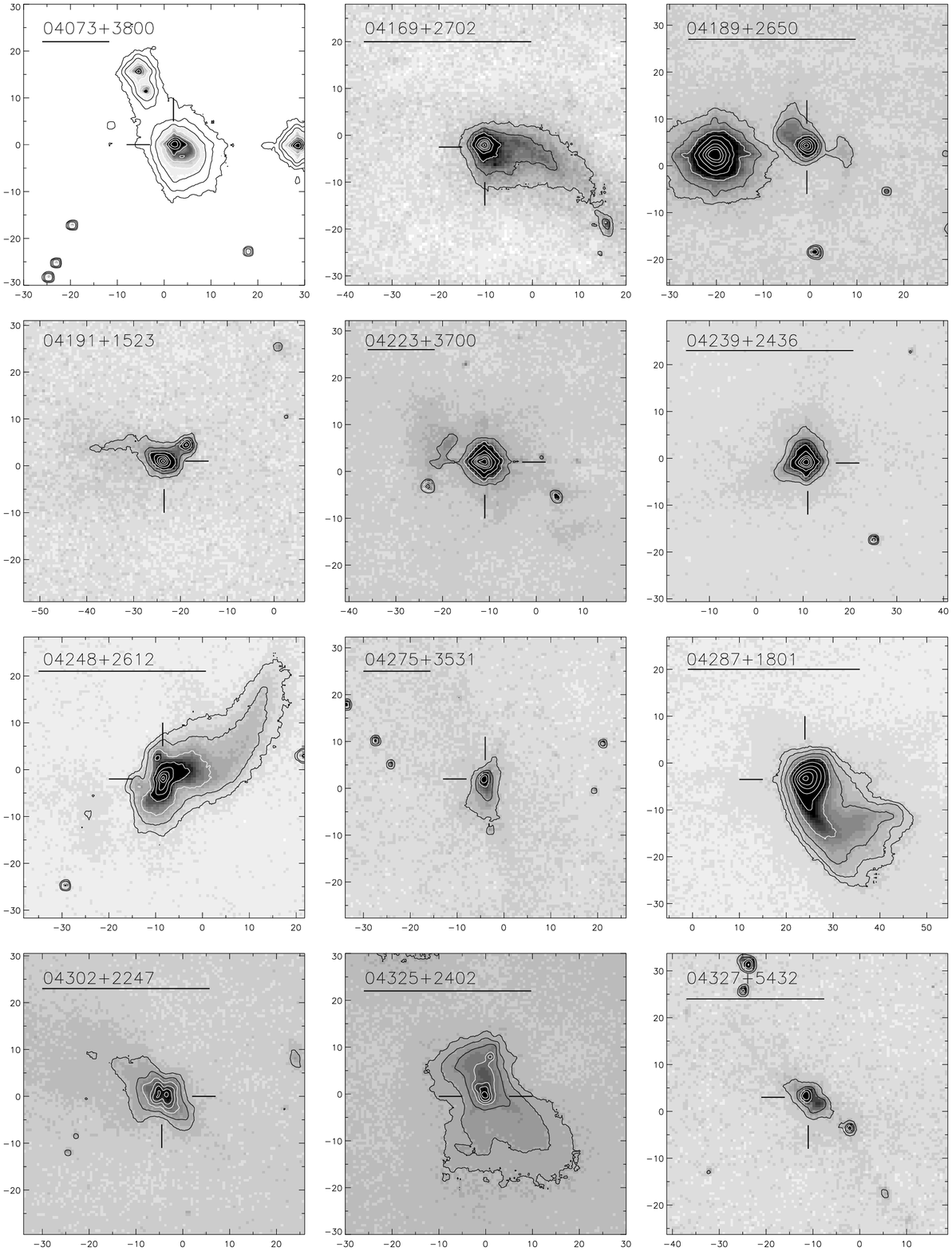}
\caption{}
\end{figure}

\clearpage

\begin{figure}
\addtocounter{figure}{-1}
\plotone{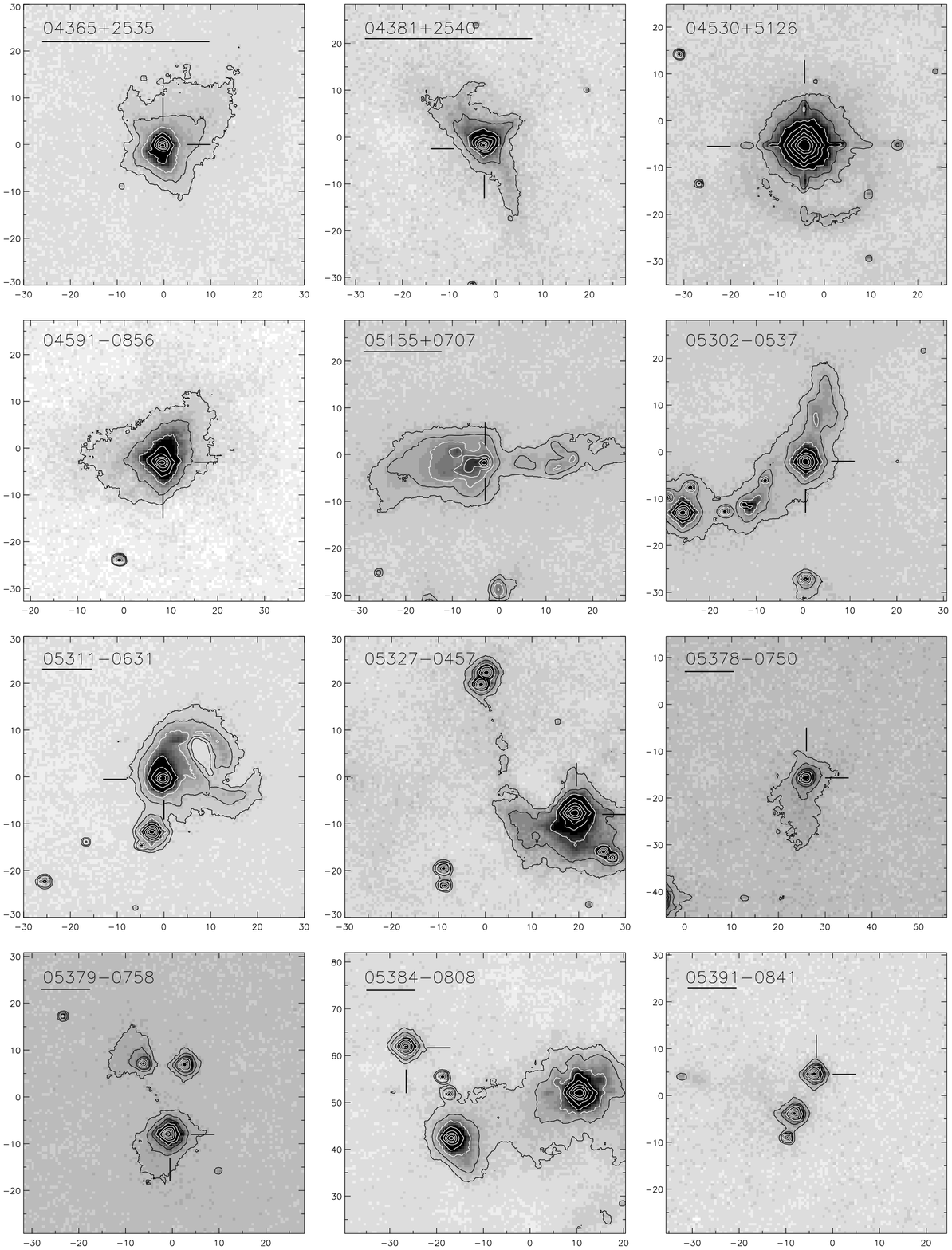}
\caption{}
\end{figure}

\clearpage

\begin{figure}
\addtocounter{figure}{-1}
\plotone{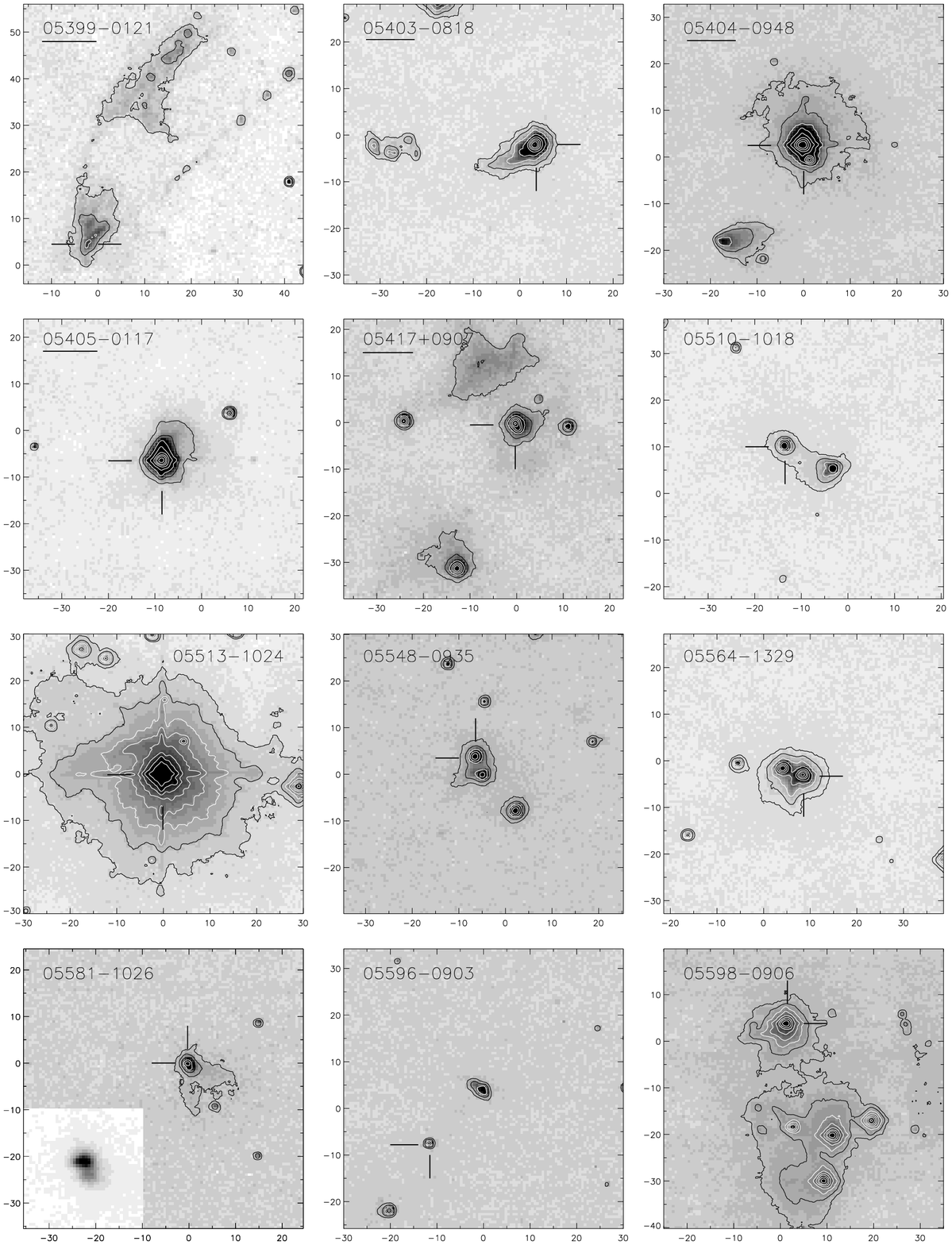}
\caption{}
\end{figure}

\clearpage

\begin{figure}
\addtocounter{figure}{-1}
\plotone{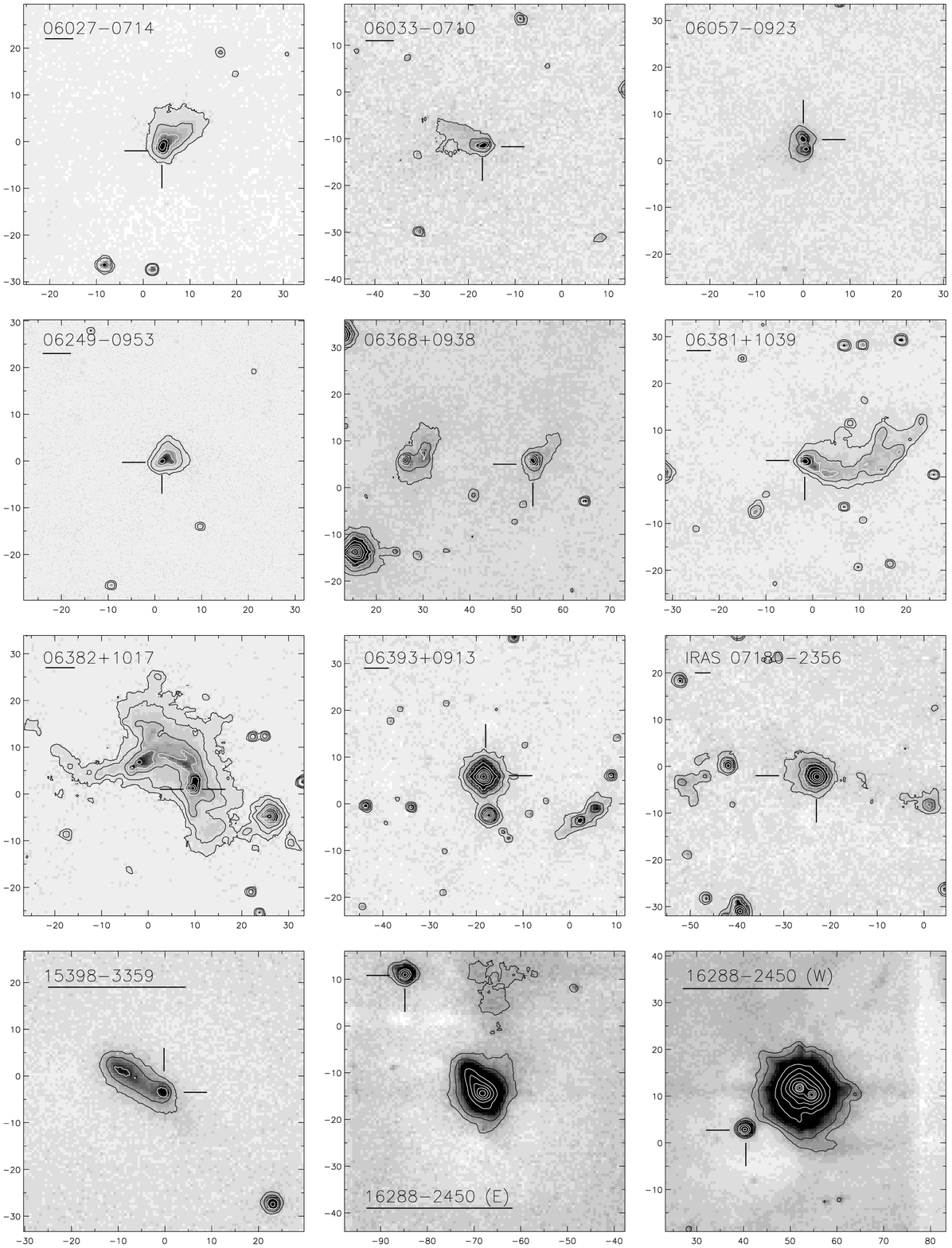}
\caption{}
\end{figure}

\clearpage

\begin{figure}
\addtocounter{figure}{-1}
\plotone{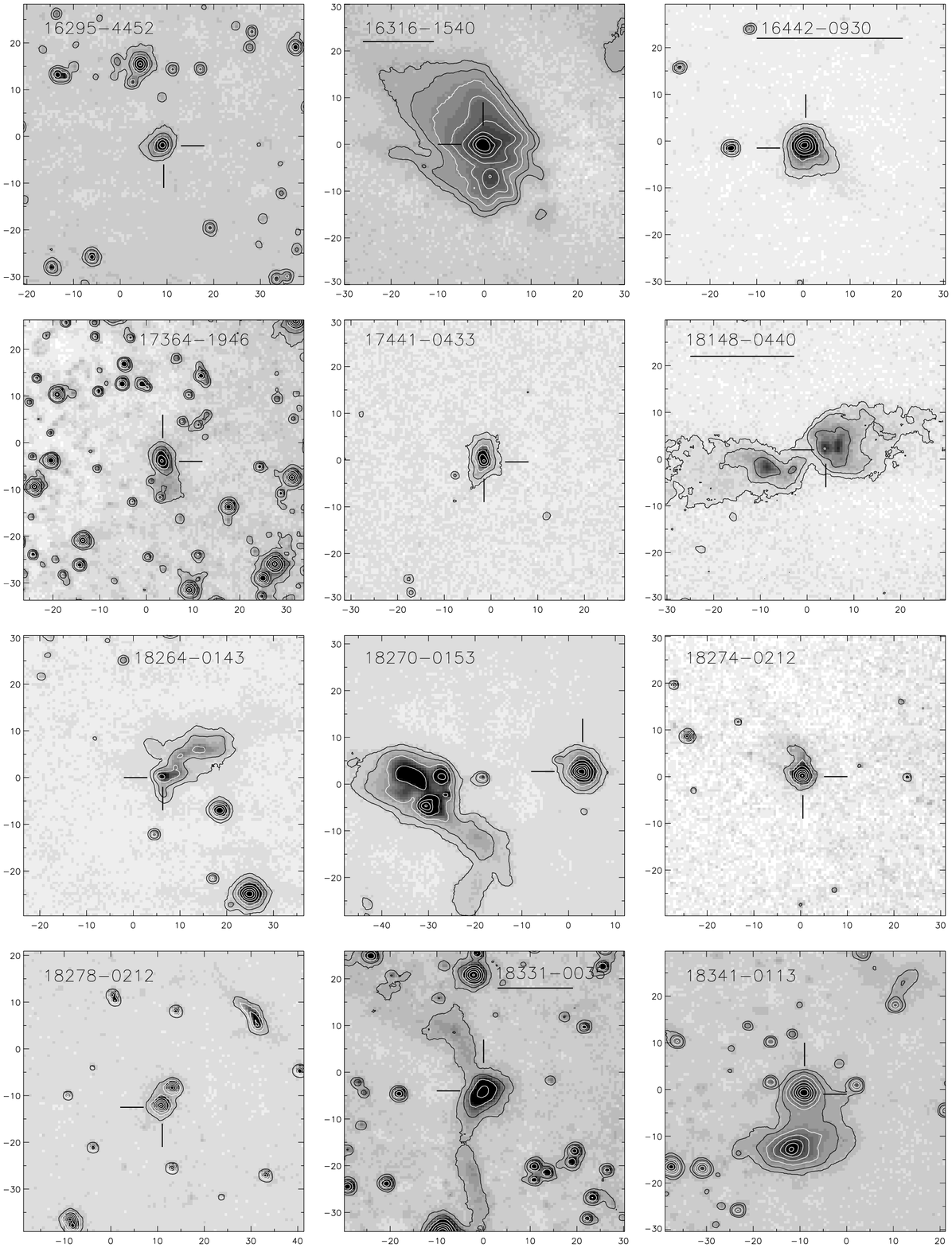}
\caption{}
\end{figure}

\clearpage

\begin{figure}
\addtocounter{figure}{-1}
\plotone{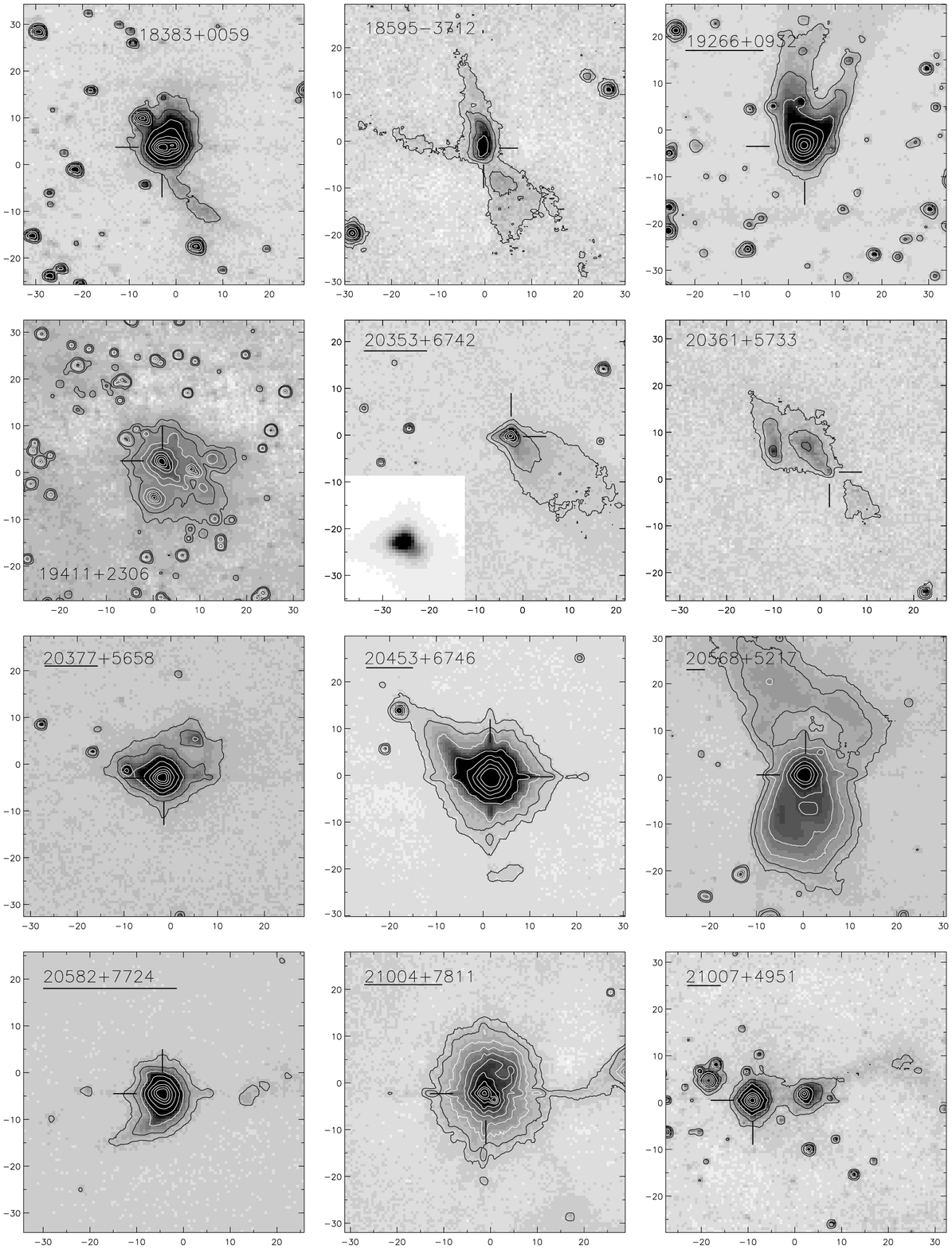}
\caption{}
\end{figure}

\clearpage

\begin{figure}
\addtocounter{figure}{-1}
\plotone{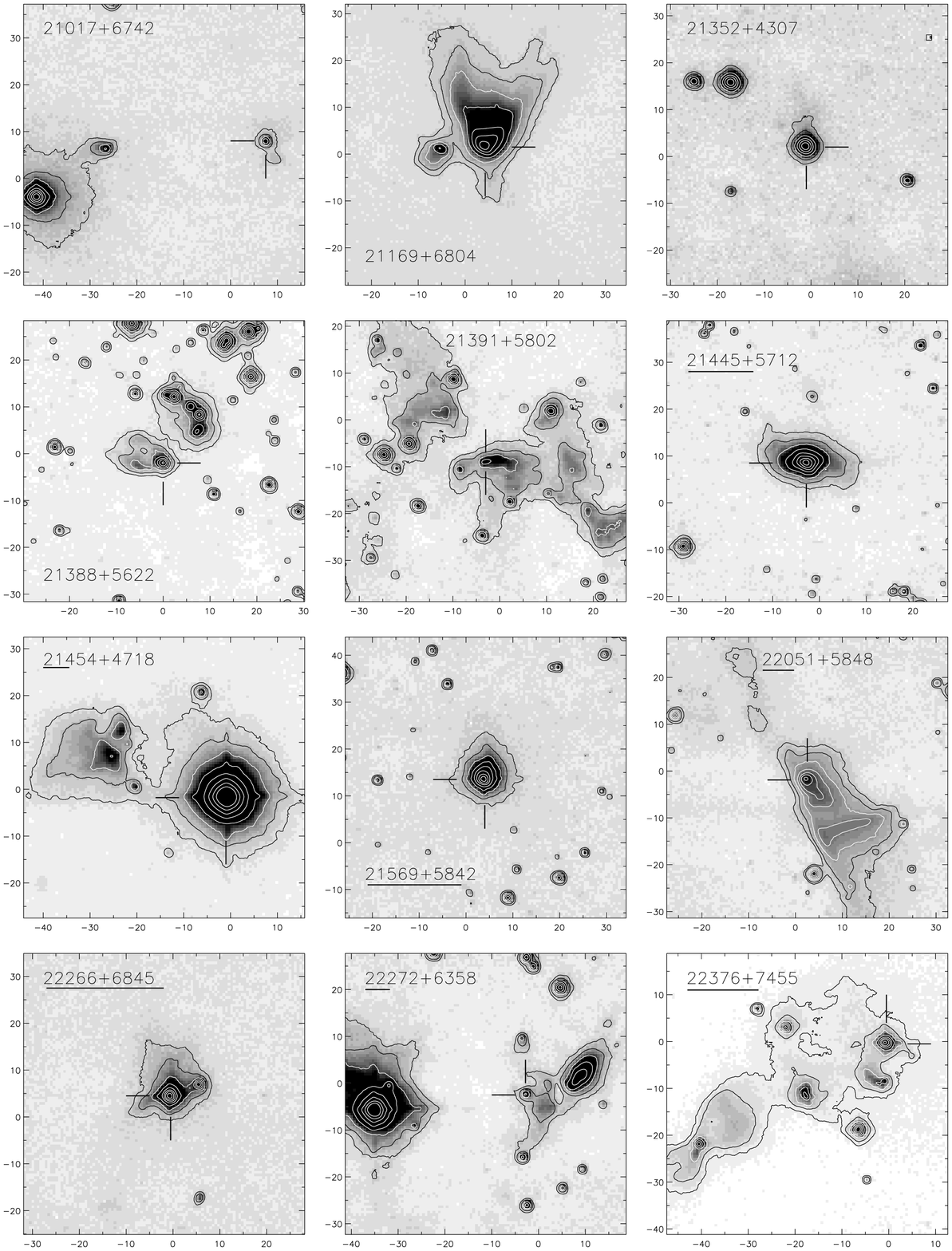}
\caption{}
\end{figure}

\clearpage

\begin{figure}
\addtocounter{figure}{-1}
\plotone{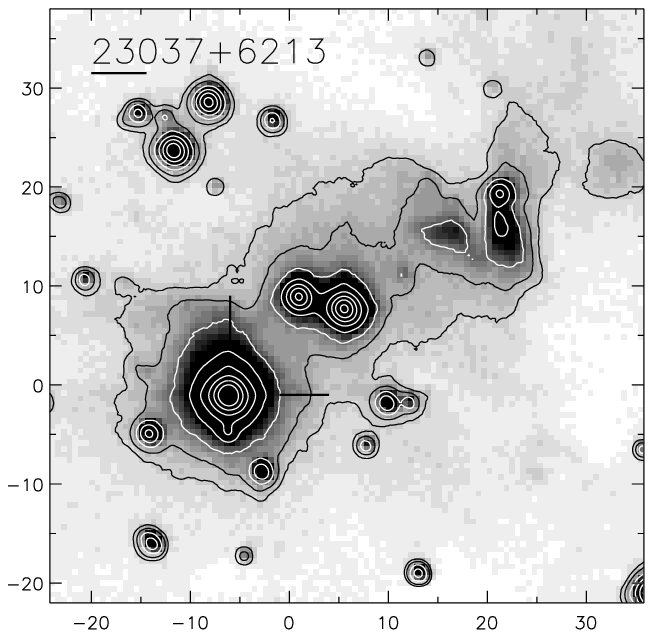}
\caption{}
\end{figure}

\clearpage 

\begin{figure}
\plottwo{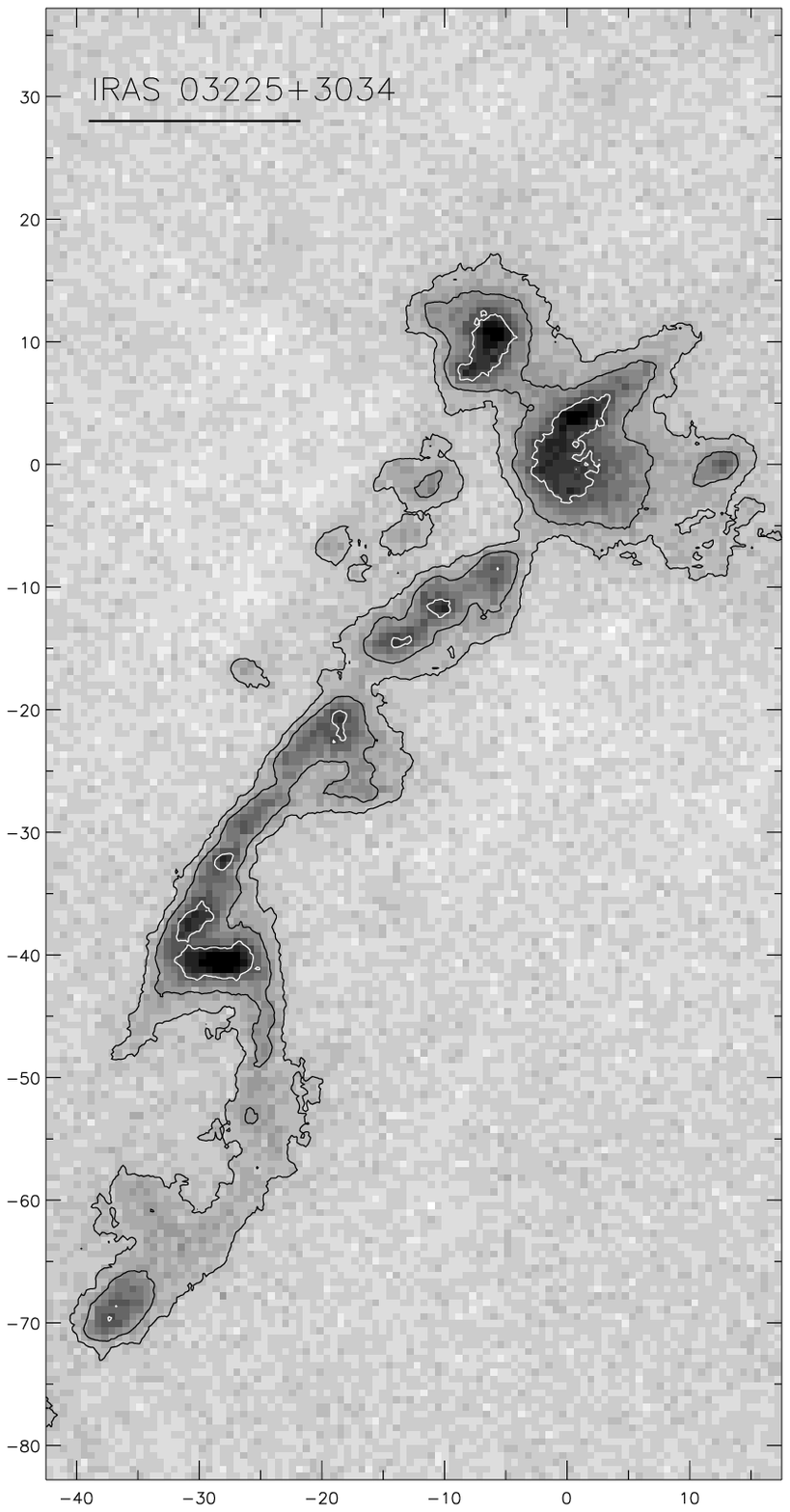}{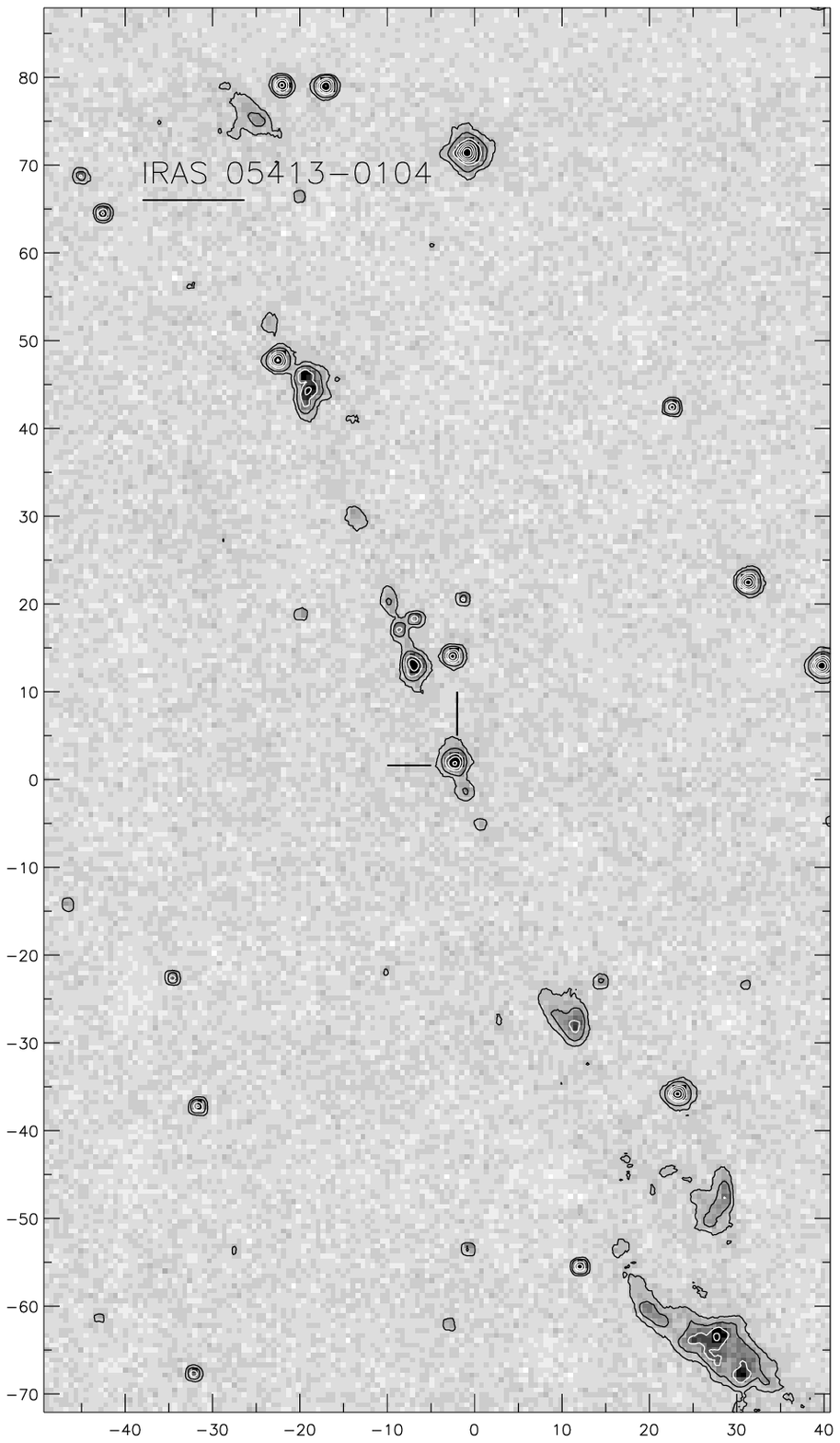}
\caption{Left: IRAS 03225+3034, a Class 0 7\farcs3 binary detected at 3.6~cm at the VLA by Reipurth et al. (2002).  Right: HH~212. Most of the K-band flux is from shocked H$_{2}$ emission.  \label{fig9}}
\end{figure}

\clearpage

\begin{figure}
\plotone{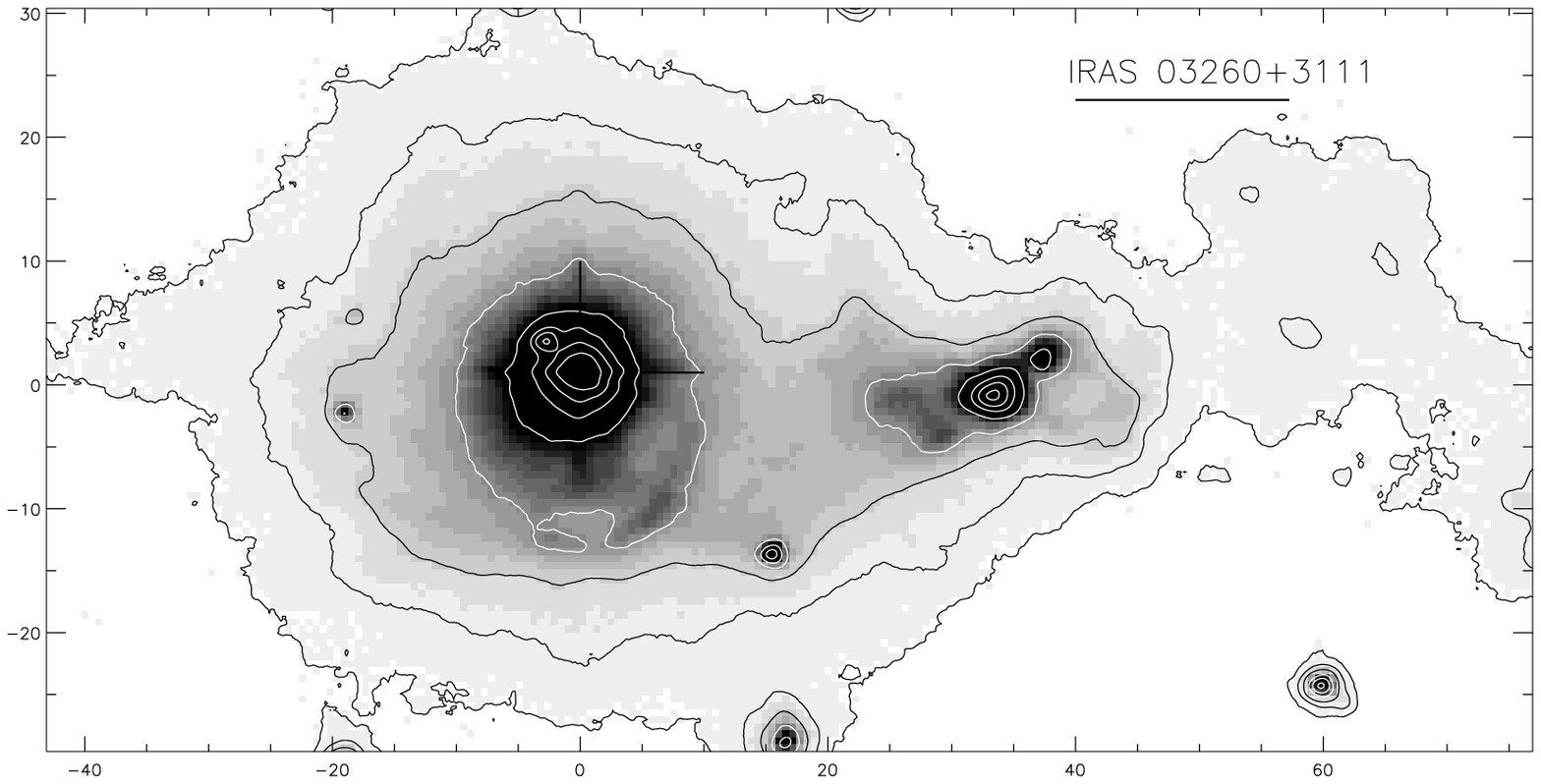}
\caption{IRAS 03260+3111  \label{fig9}}
\end{figure}

\clearpage

\begin{figure}
\plotone{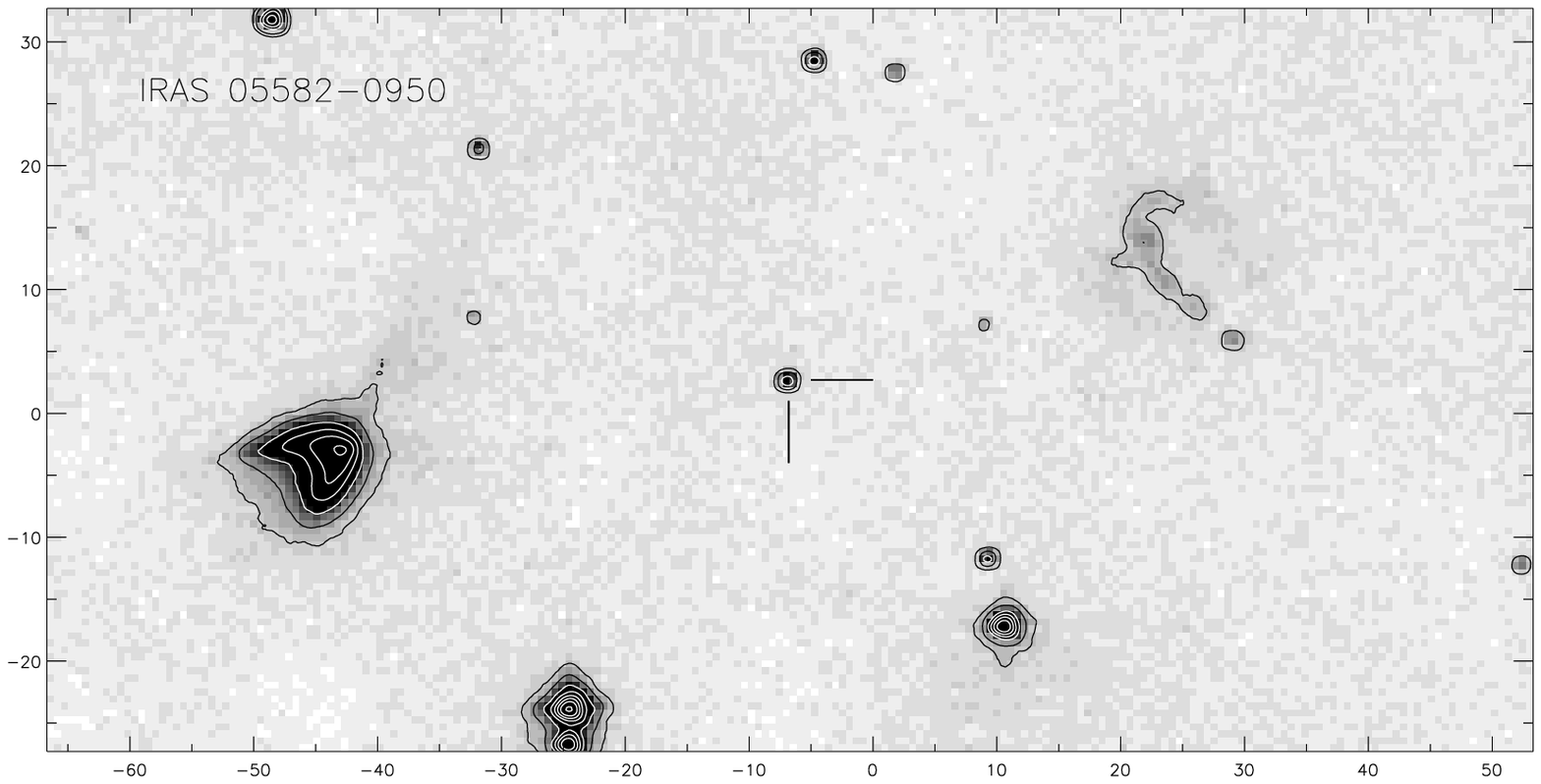}
\caption{IRAS 05582-0950  \label{fig9}}
\end{figure}

\clearpage 

\begin{figure}
\plottwo{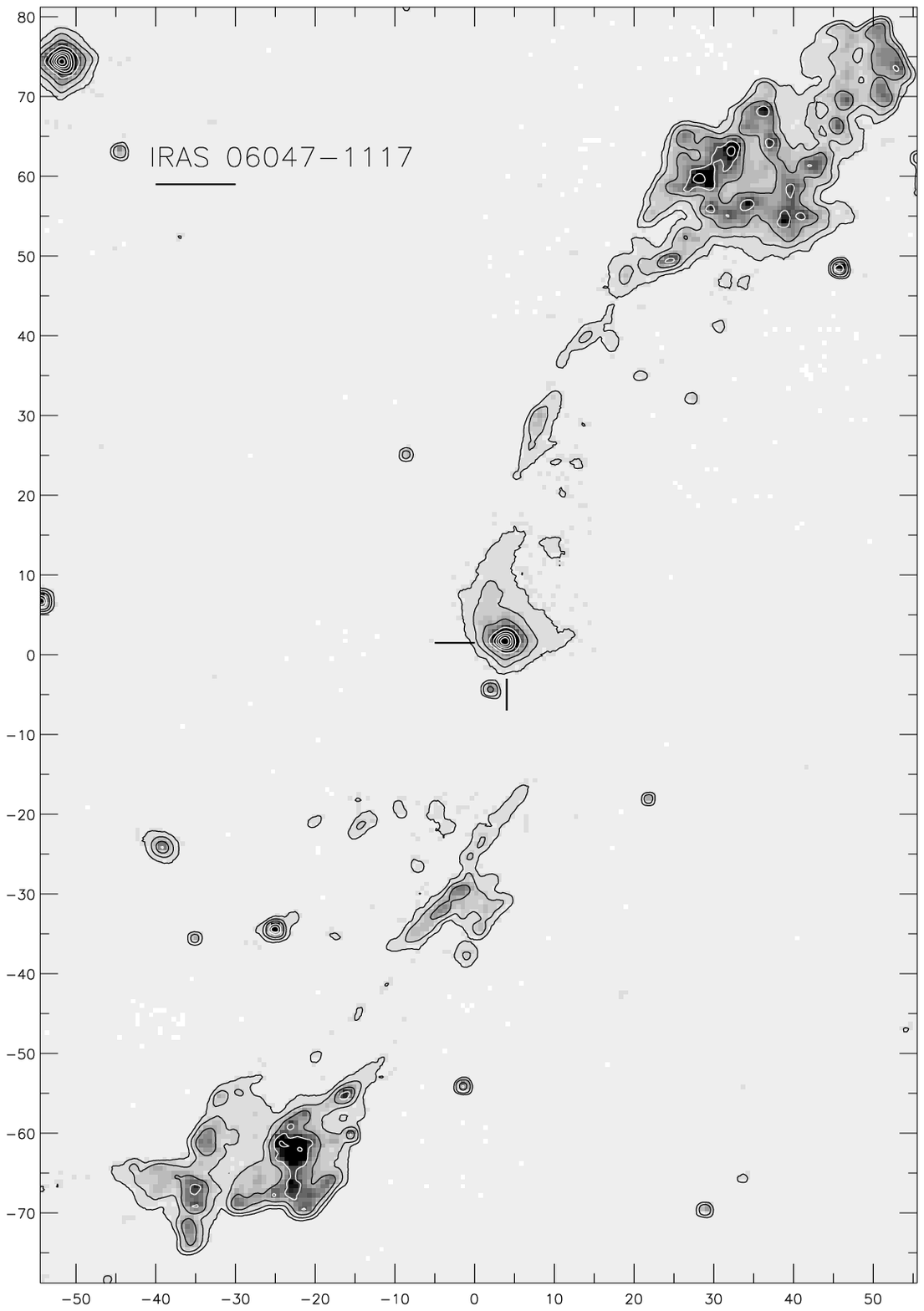}{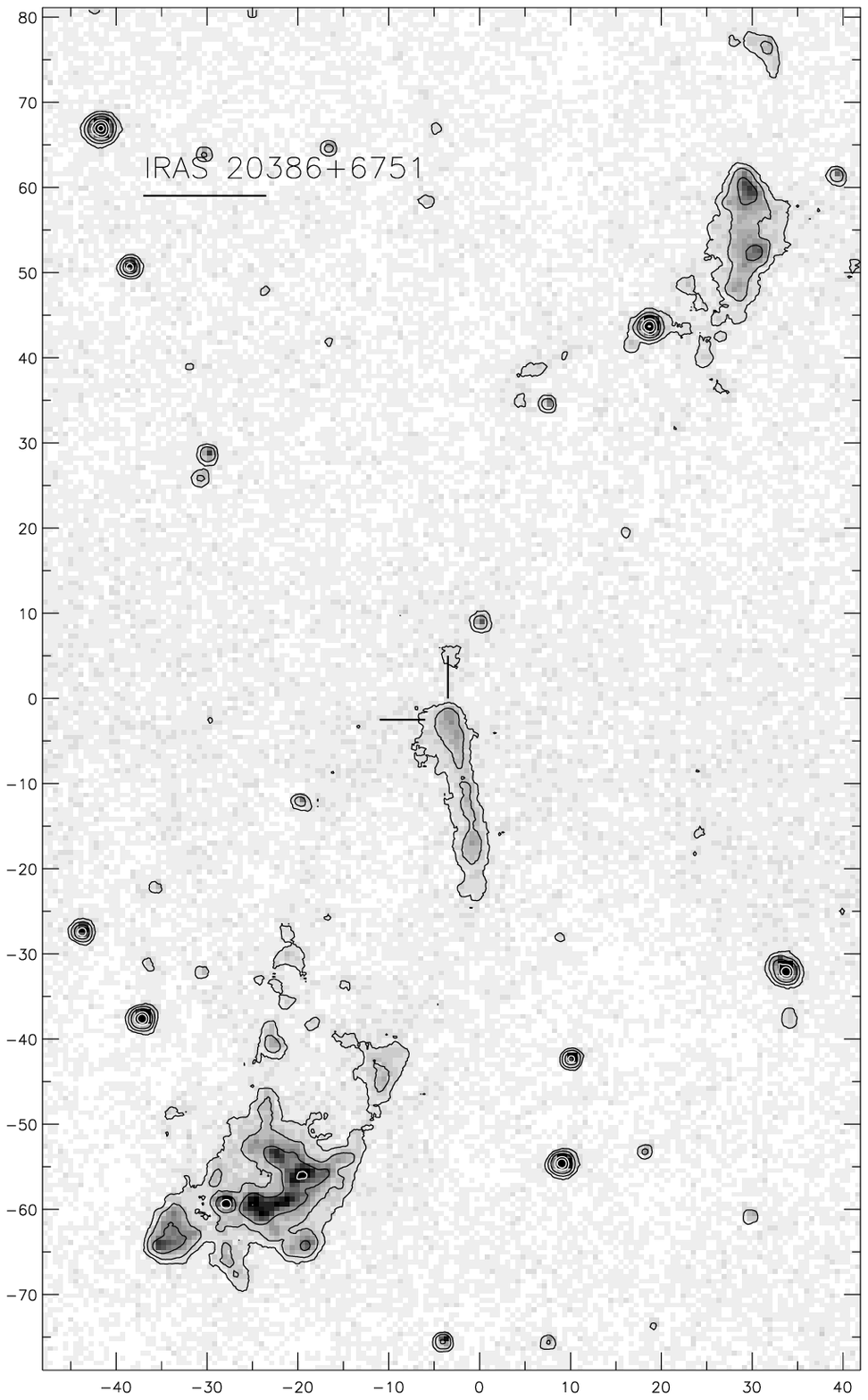}
\caption{Left: IRAS 06047-1117, a large recently discovered outflow.  Right: IRAS 20386+6751.  \label{fig9}}
\end{figure}

\clearpage 

\begin{figure}
\plotone{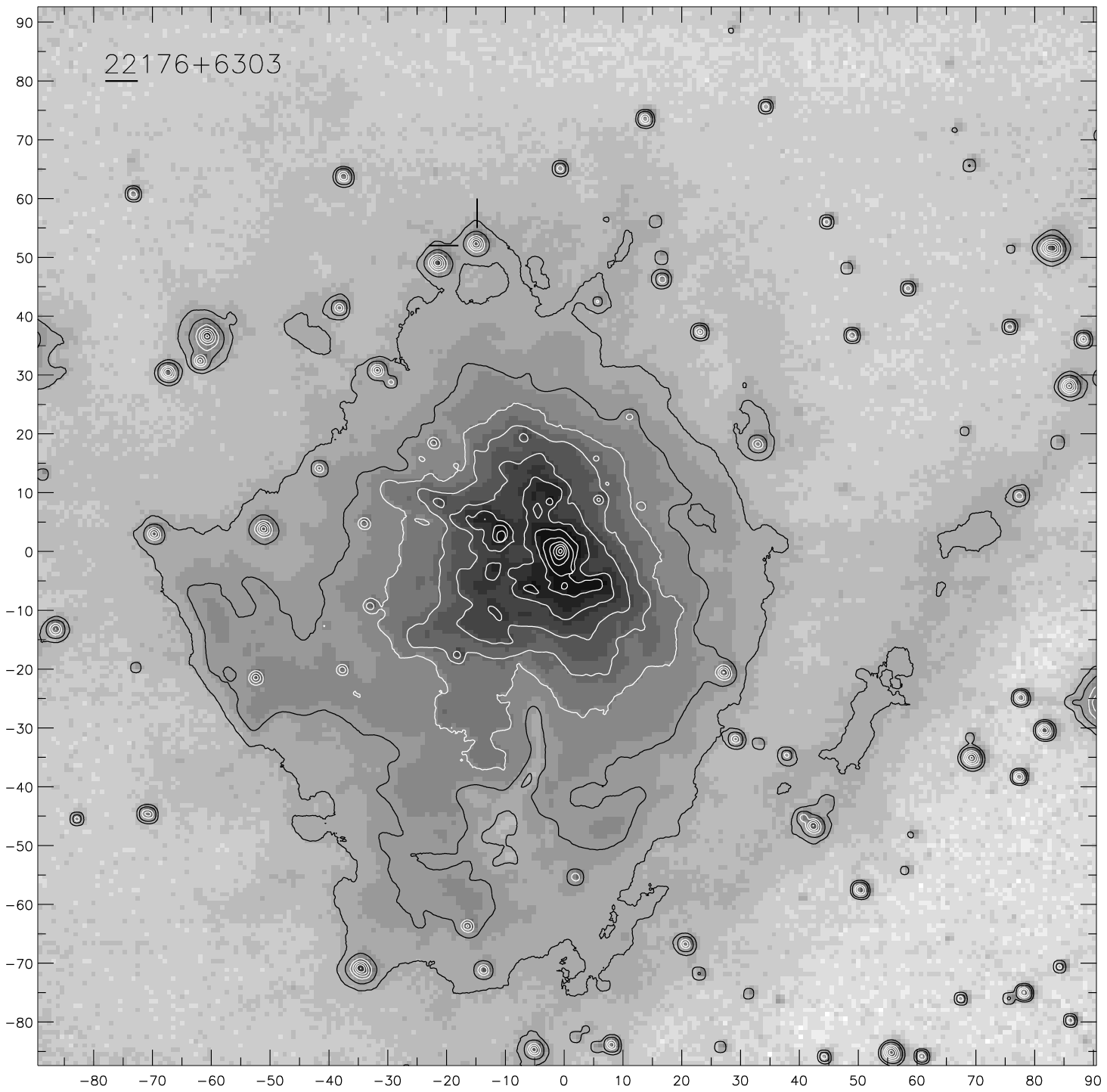}
\caption{IRAS 22176+6303. \label{fig9}}
\end{figure}

\clearpage 

\begin{figure}
\plotone{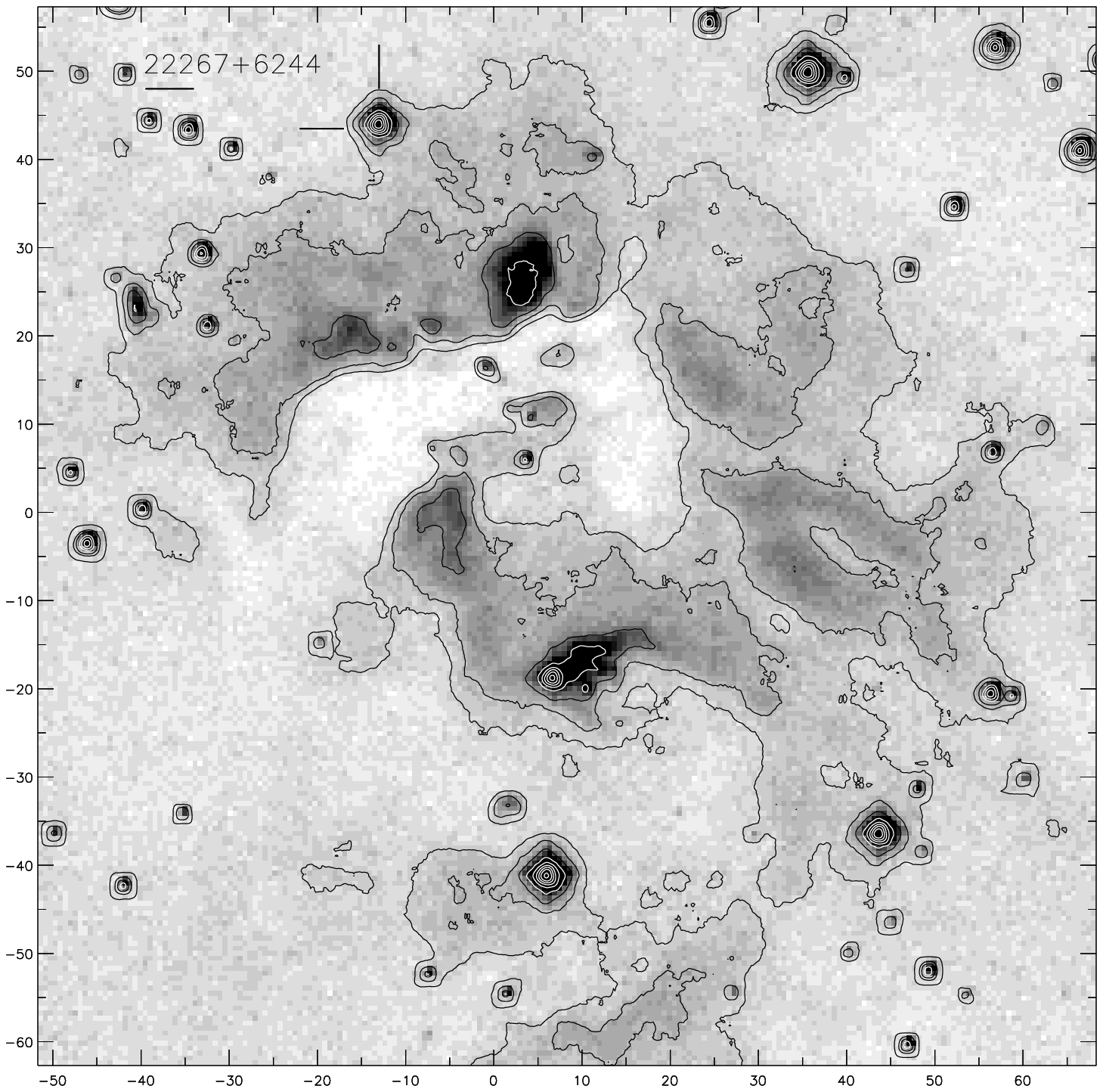}
\caption{IRAS 22267+6244.  \label{fig9}}
\end{figure}

\clearpage

\begin{figure}
\plotone{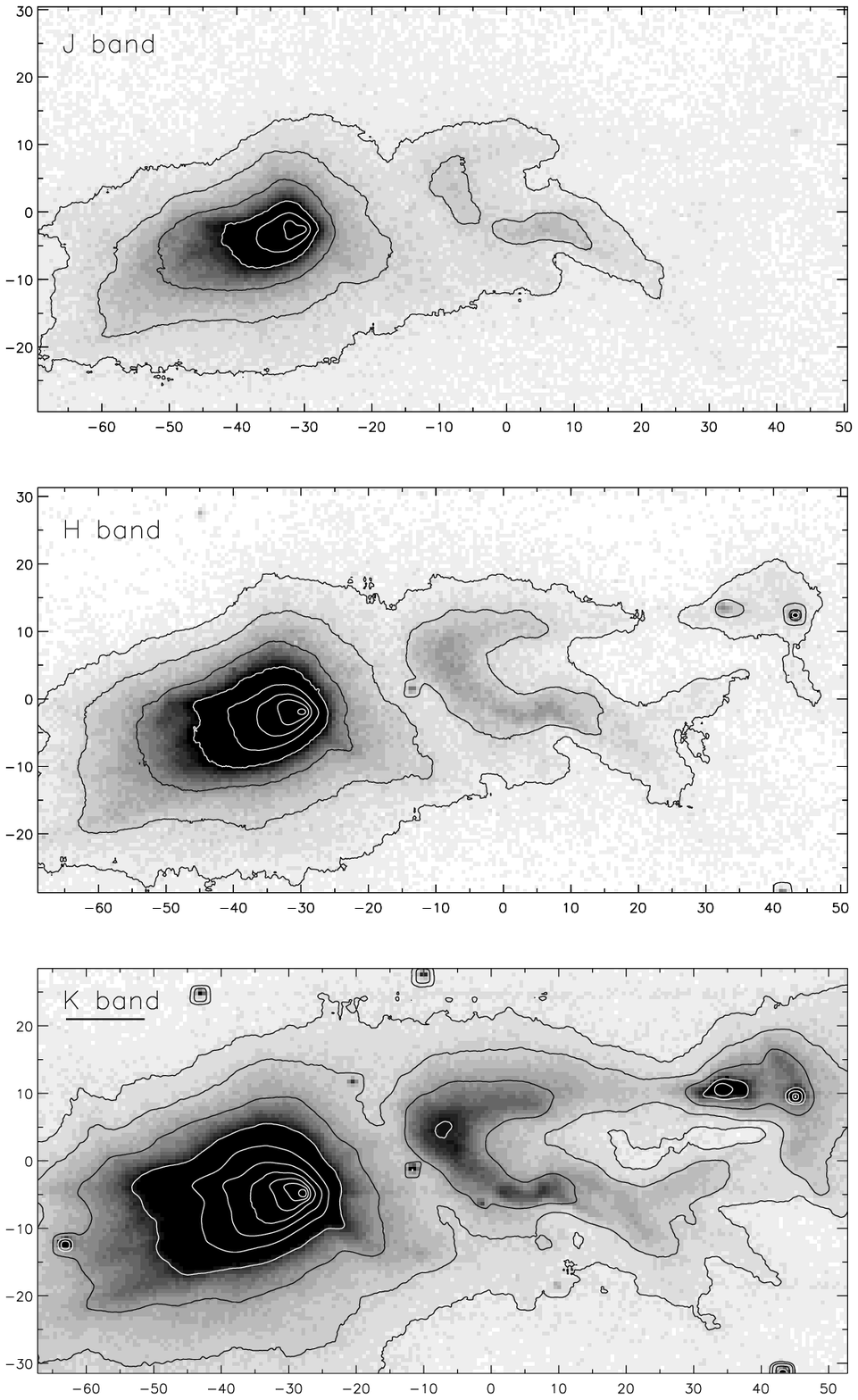}
\caption{J, H, and K contour plots of IRAS 05450+0019.  The outer most contours are J~=~21, H~=~20, and K~=~20 magnitudes per square arcsecond, and the contours are at 1 magnitude intervals.  As in the other plots, (0,0) is at the IRAS coordinate.  This appears to be a purely reflection nebula, with no significant excess seen when observed with the H$_{2}$ S(1) filter at $2.12~\mu$m or the [FeII] filter at $1.64~\mu$m.  The coordinate given in Table 2 is for the K-band peak. \label{fig2}}
\end{figure}

\clearpage

\begin{figure}
\plotone{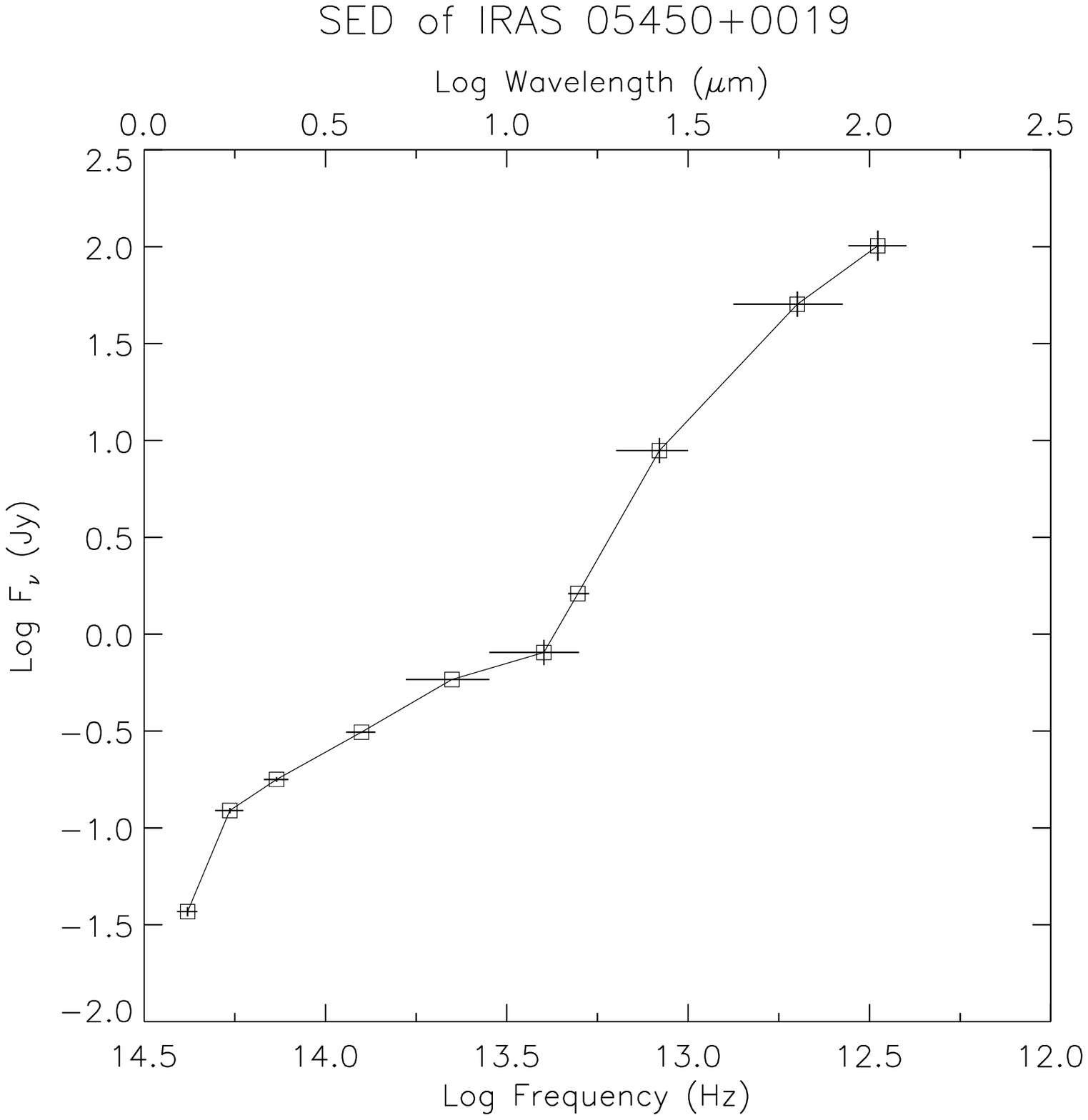}
\caption{SED for IRAS 05450+0019, which combines IRAS ($12~\mu$m, $25~\mu$m, $60~\mu$m, and $100~\mu$m), ISO ($6.7~\mu$m and $14.3~\mu$m), IRTF ($3.8~\mu$m), and 2MASS ($1.2~\mu$m, $1.6~\mu$m, $2.2~\mu$m) data.  An SED that rises from near-IR through $100~\mu$m is characteristic of Class I sources. The J, H, K, L', $6.7~\mu$m and $14.3~\mu$m flux error bars are smaller than the plot symbol size.  The horizontal error bars represent the filter bandpasses. \label{fig2}}
\end{figure}

\clearpage

\begin{figure}
\plotone{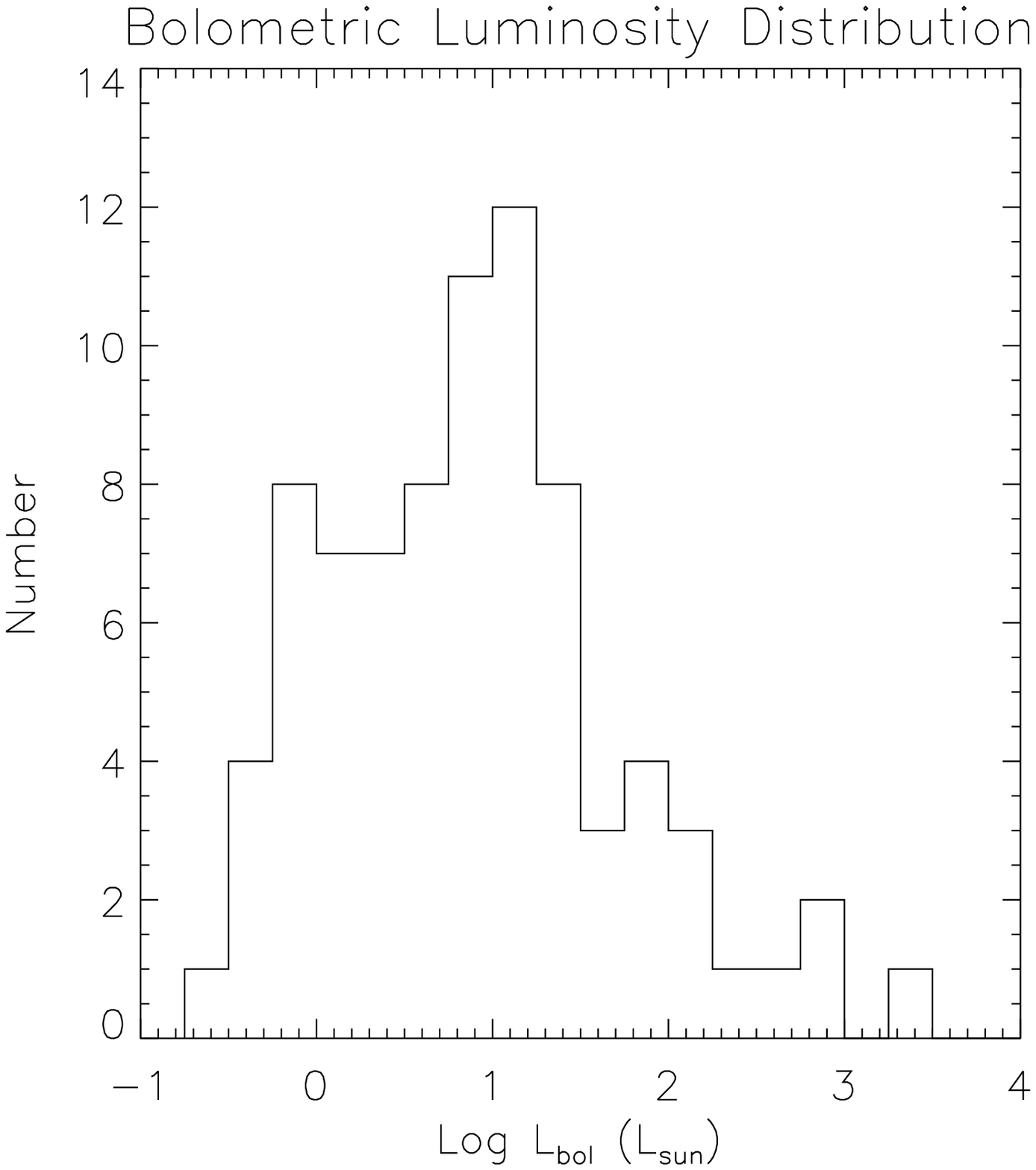}
\caption{The bolometric luminosity distribution for our sample of nebulae, with a median value of 5.8~L$_{\odot}$.\label{fig2}}
\end{figure}

\clearpage





\clearpage

\begin{figure}
\plotone{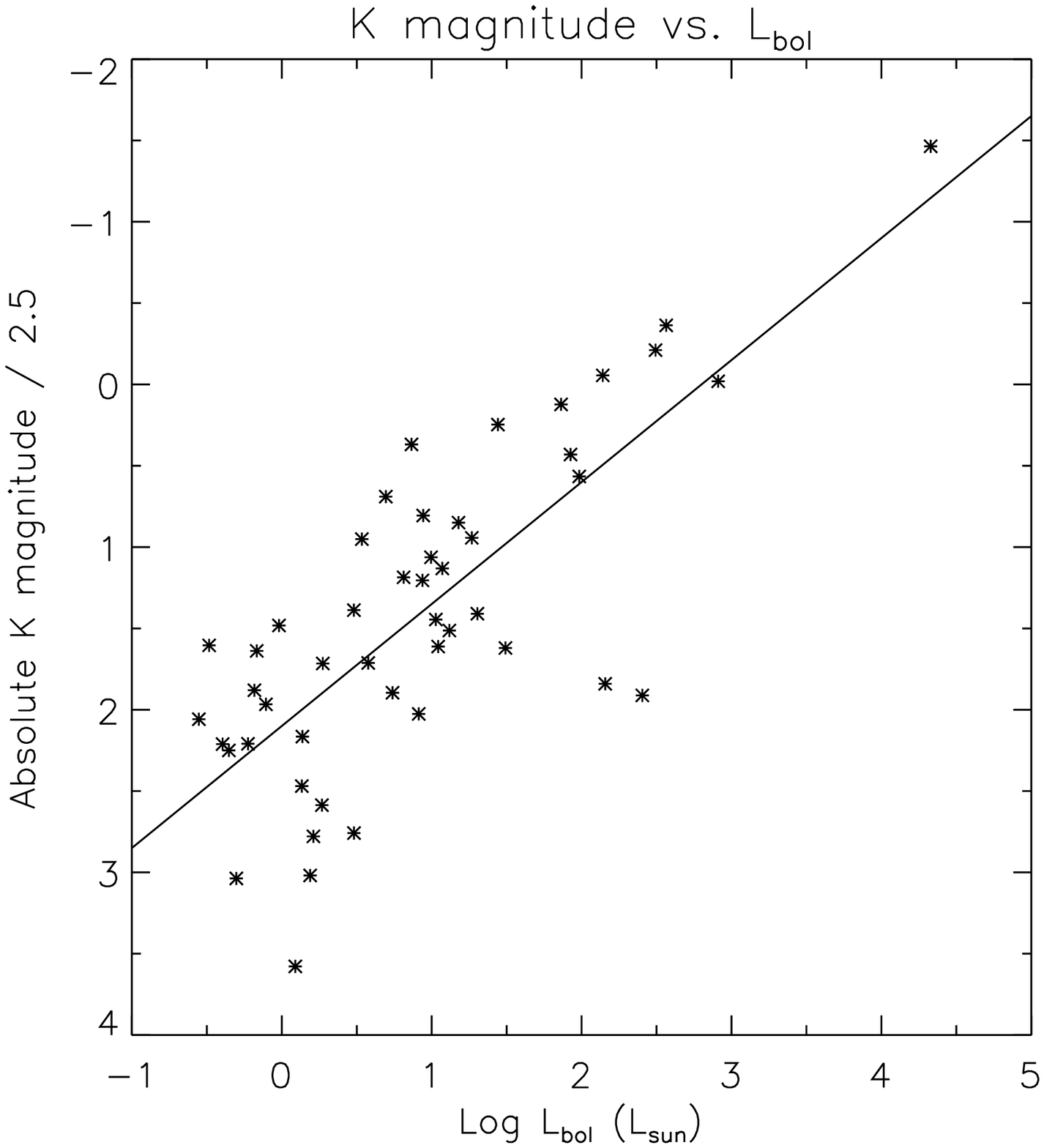}
\caption{K-band absolute magnitude (divided by 2.5) vs. bolometric luminosity.  The slope of the regression line is 0.75, with the data points having a standard deviation from the regression line of 1.63 magnitudes (a factor of 4.5). \label{fig2}}
\end{figure}

\clearpage

\begin{figure}
\plotone{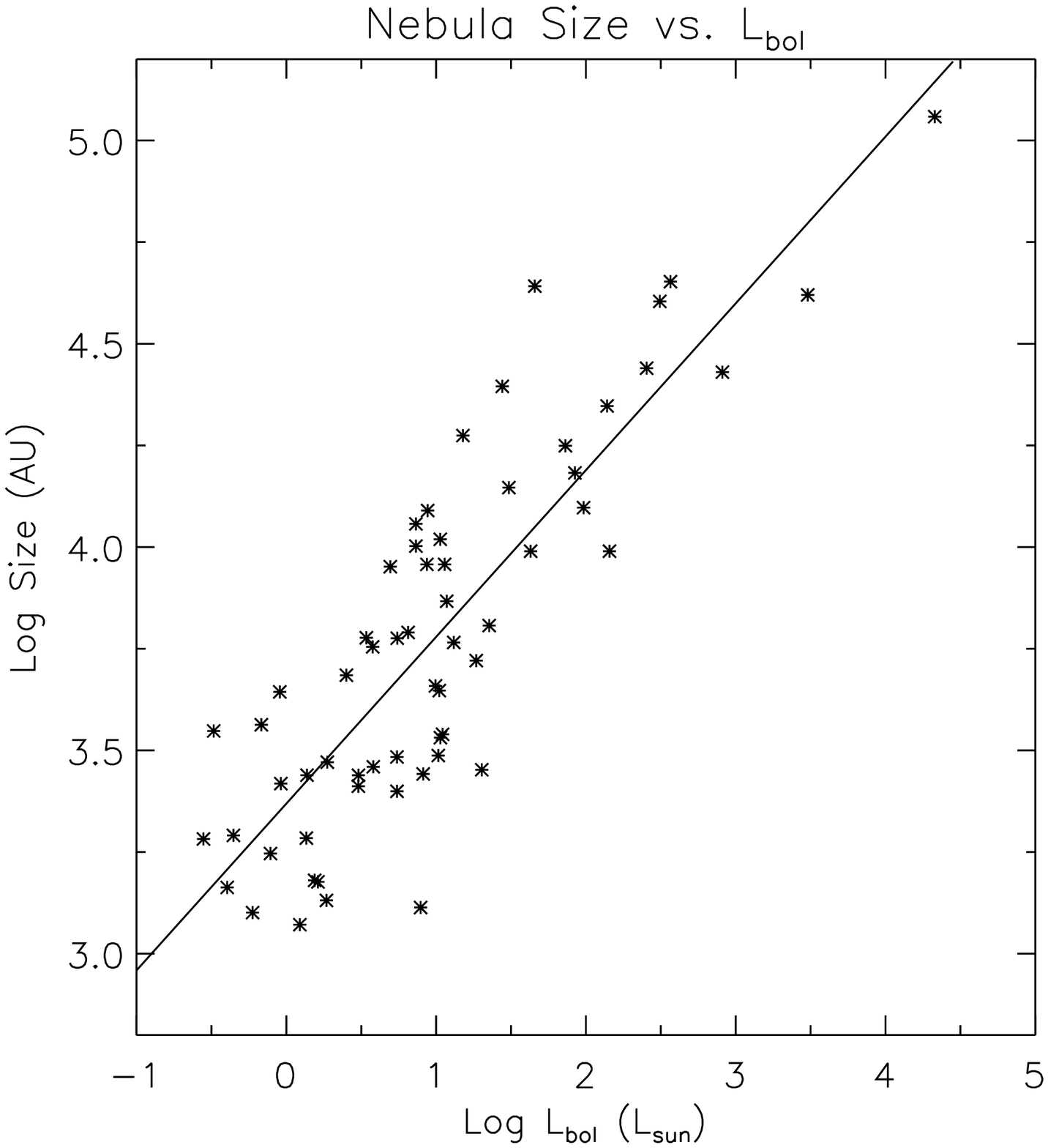}
\caption{Nebula size vs. bolometric luminosity.  The slope of the regression line is 0.41, with the data points having a standard deviation from the regression line of 0.25 (a factor of 1.8). \label{fig2}}
\end{figure}

\clearpage

\begin{figure}
\plotone{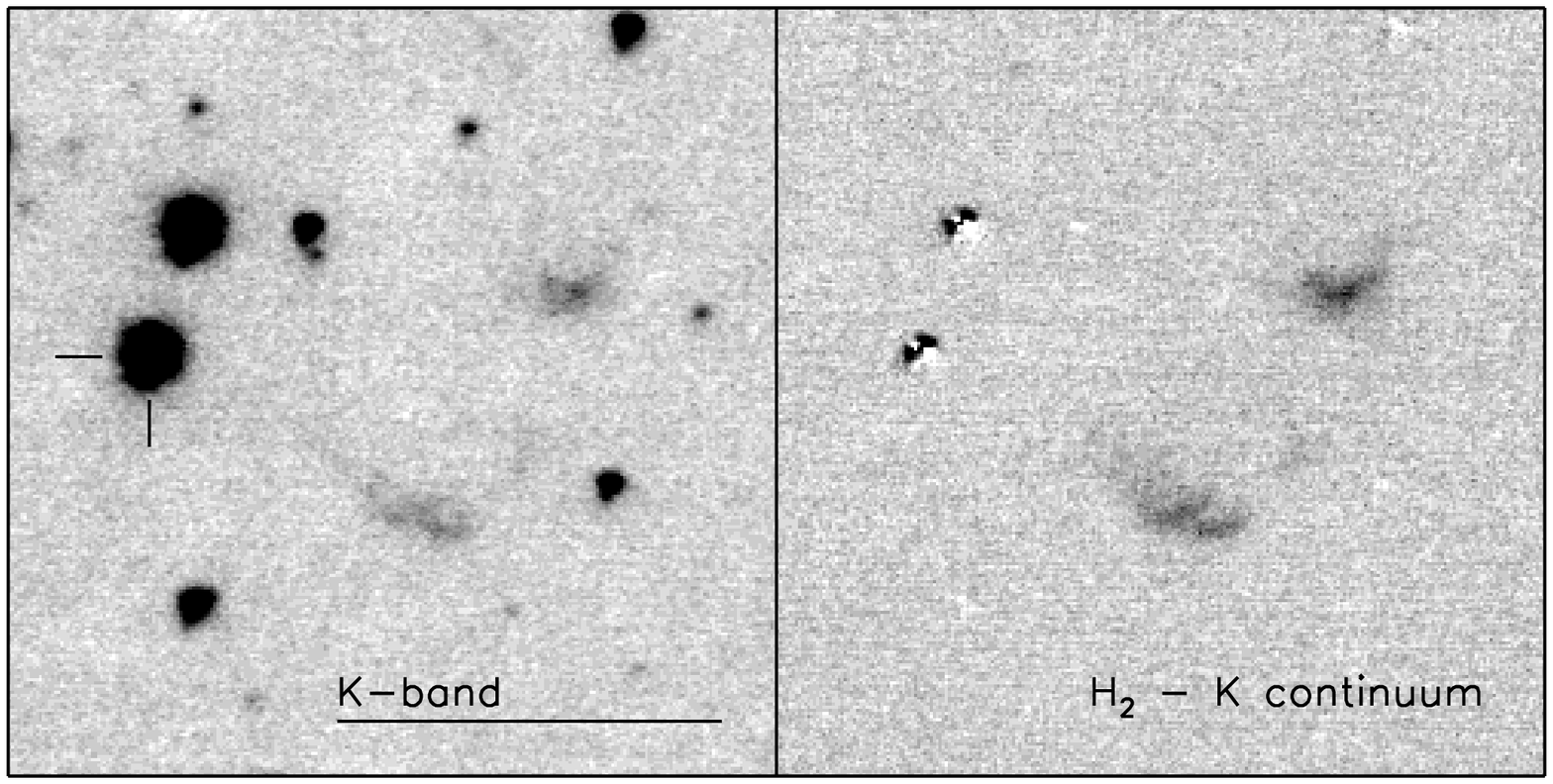}
\caption{K-band and H$_{2}$ images of IRAS 00182+6223}
\end{figure}

\clearpage

\begin{figure}
\plotone{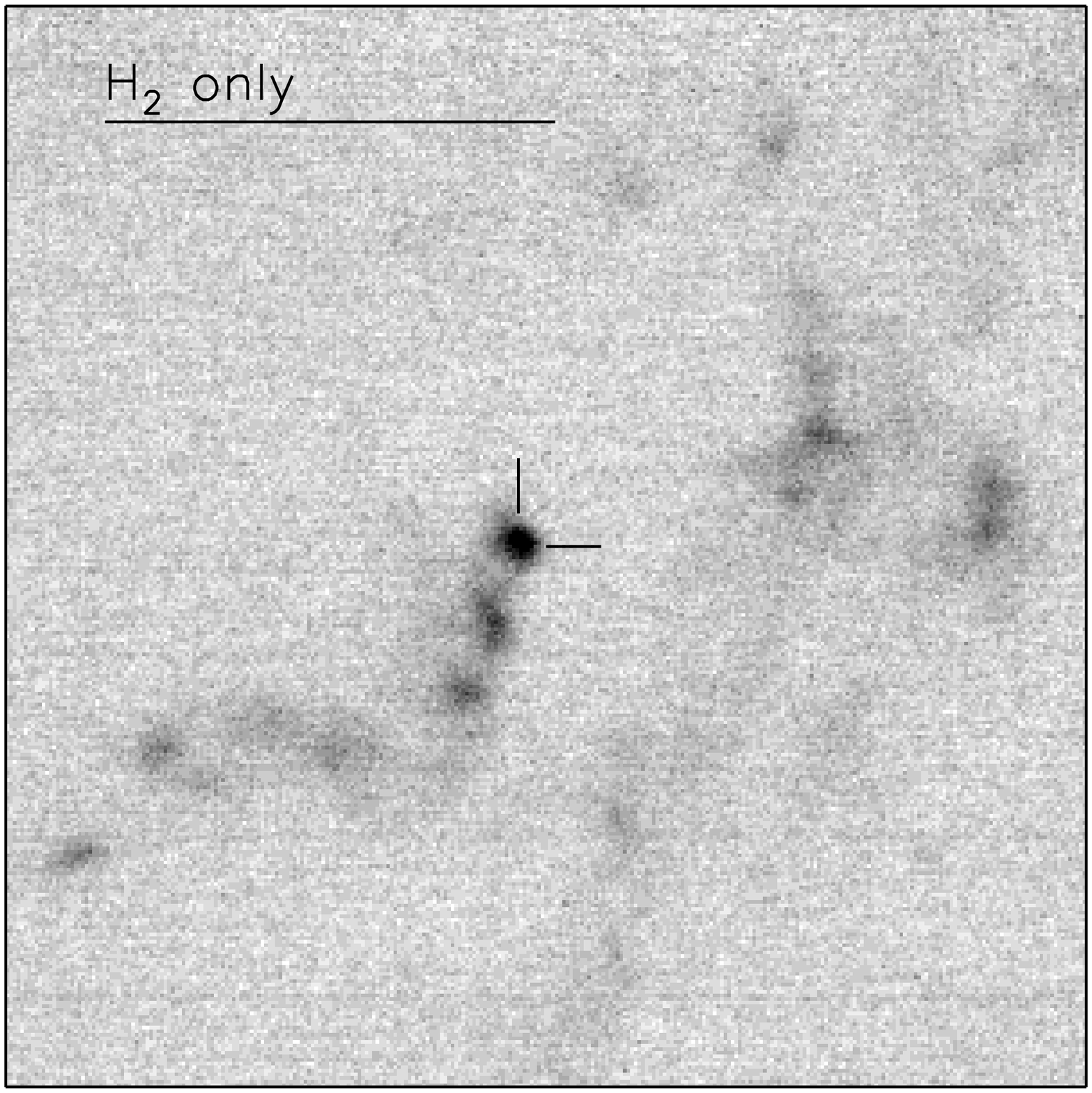}
\caption{H$_{2}$ image of IRAS 03220+3035.  This field is $\sim1.5\arcmin$ to the NE of the IRAS source coordinates, which are 03 25 09.43 +30 46 21.7.  Since no K continuum flux was detected, the K-band image is not presented.}
\end{figure}

\clearpage

\begin{figure}
\plotone{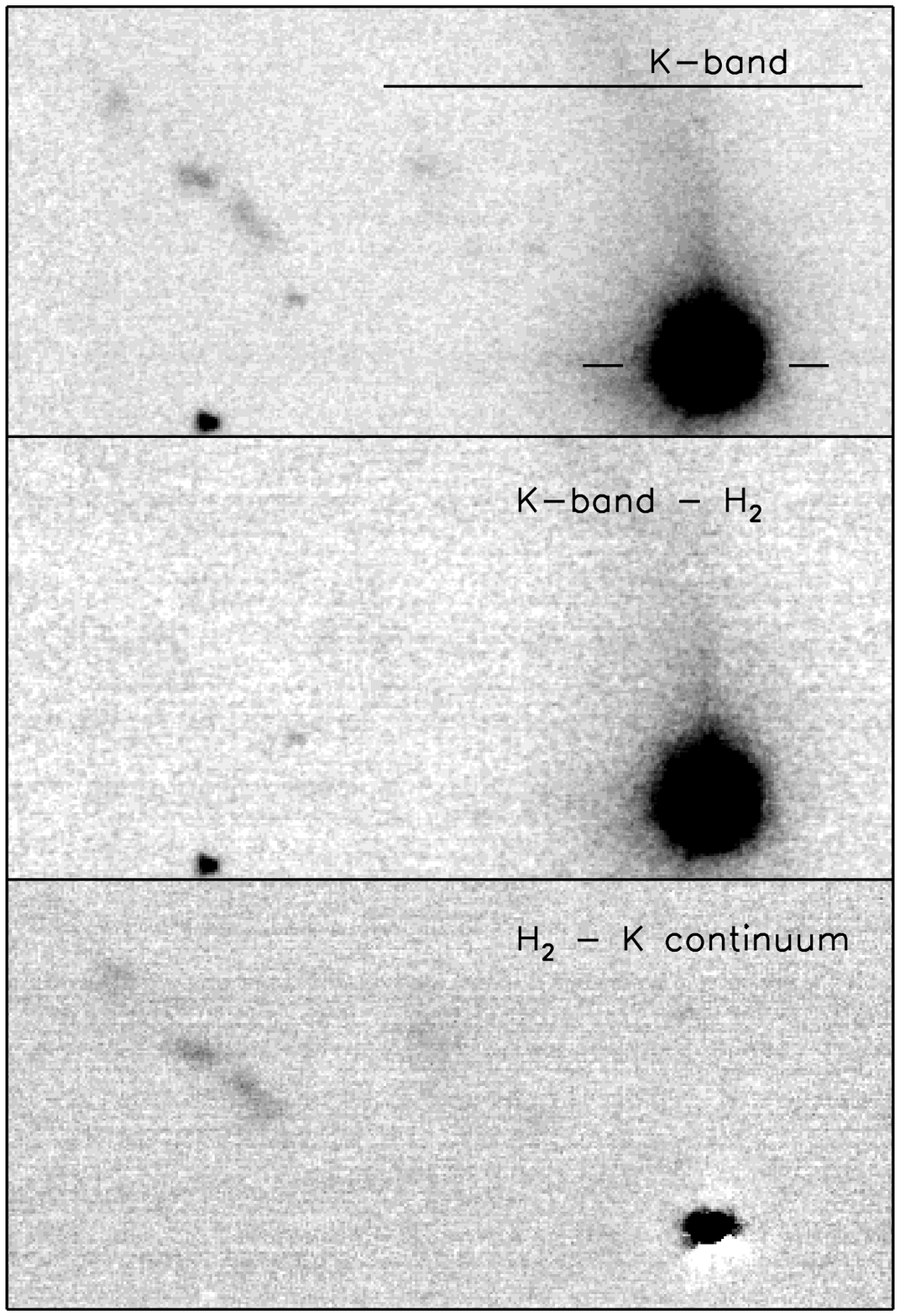}
\caption{K-band, K continuum, and H$_{2}$ images of IRAS 03445+3242}
\end{figure}

\clearpage

\begin{figure}
\plotone{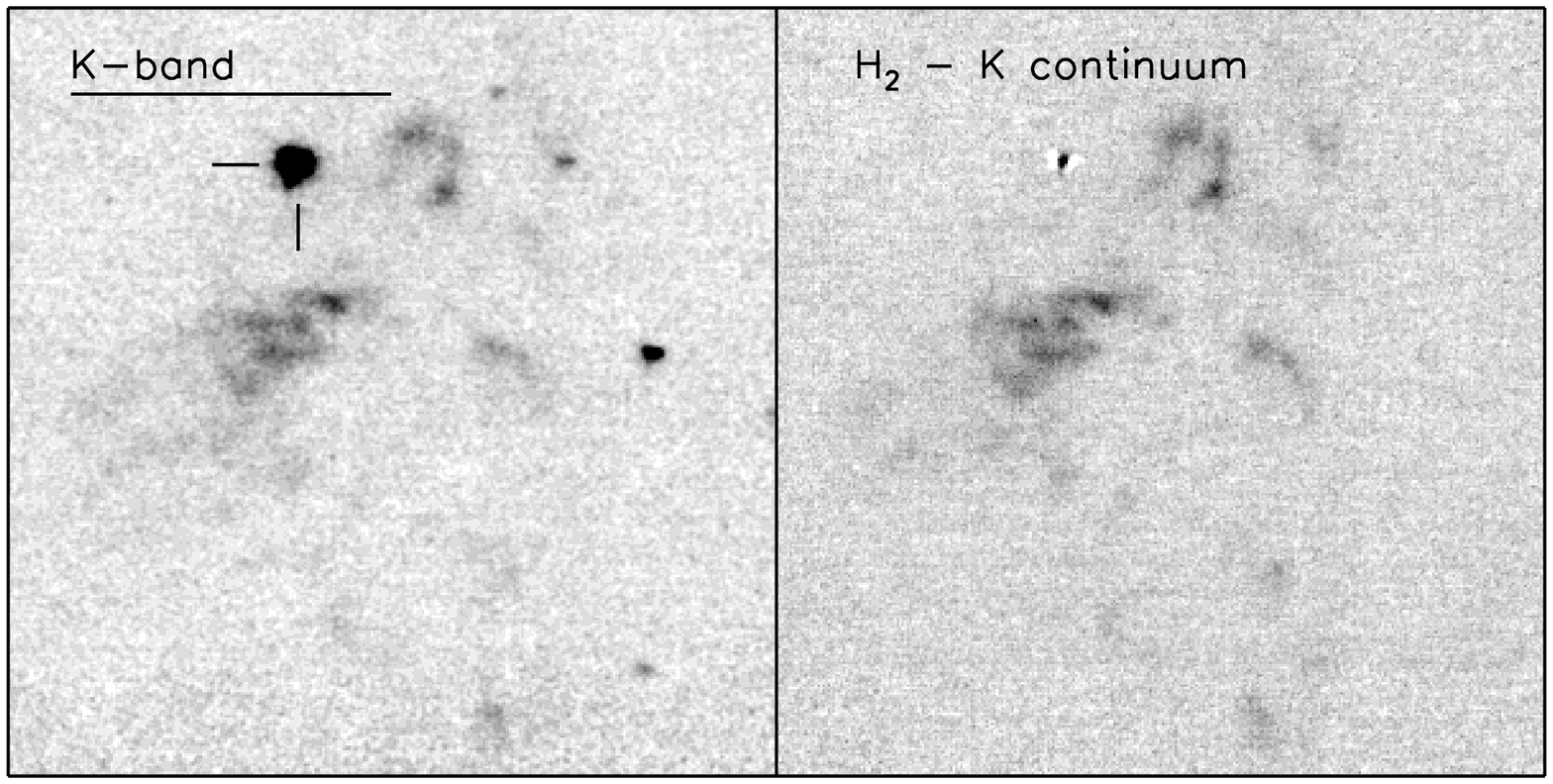}
\caption{K-band and H$_{2}$ images of IRAS 04327+5432.  This field is $\sim1\arcmin$ to the NNW of the IRAS source coordinates.}
\end{figure}

\clearpage

\begin{figure}
\plotone{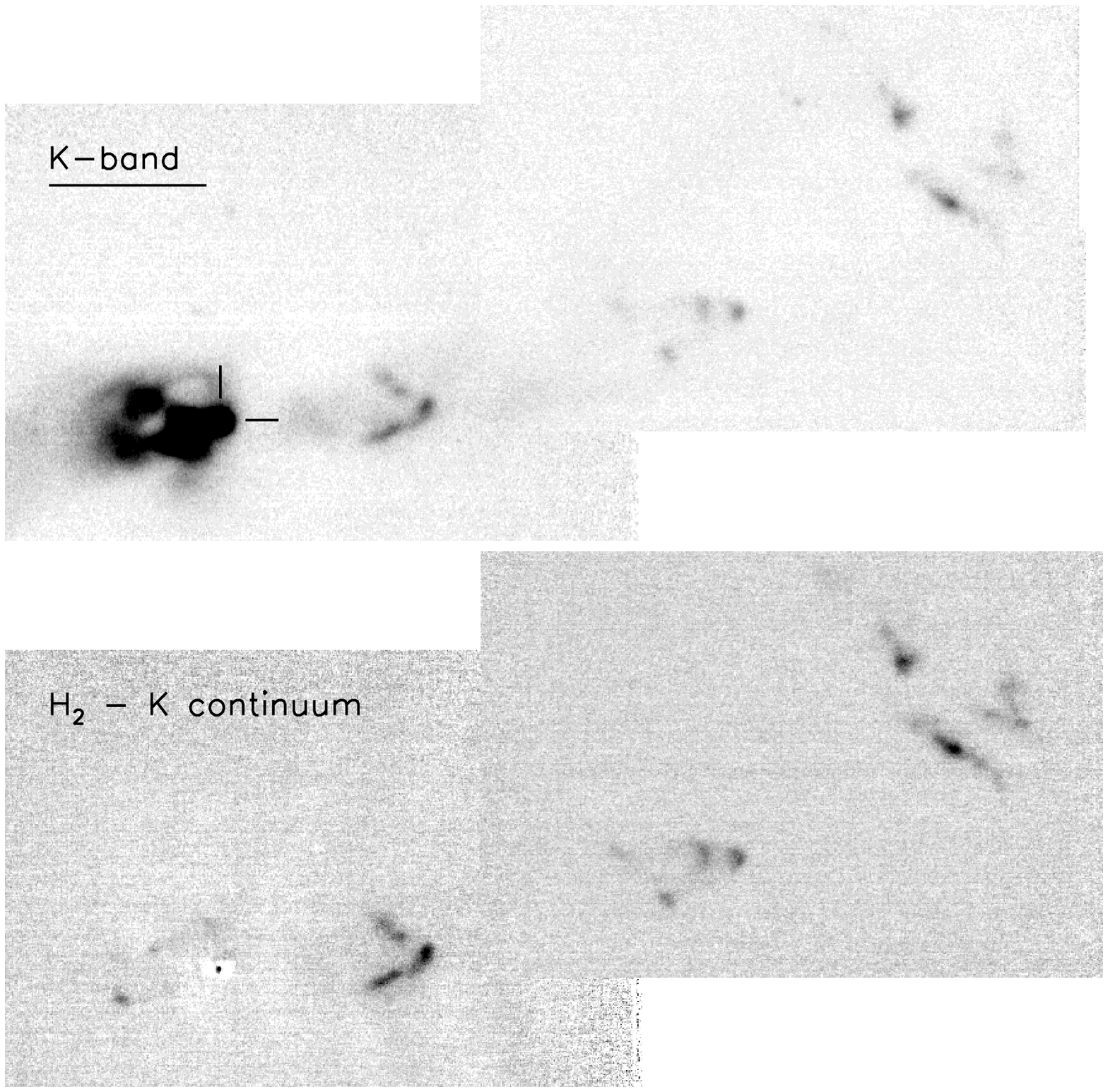}
\caption{K-band and H$_{2}$ images of IRAS 05155+0707}
\end{figure}

\clearpage

\begin{figure}
\plotone{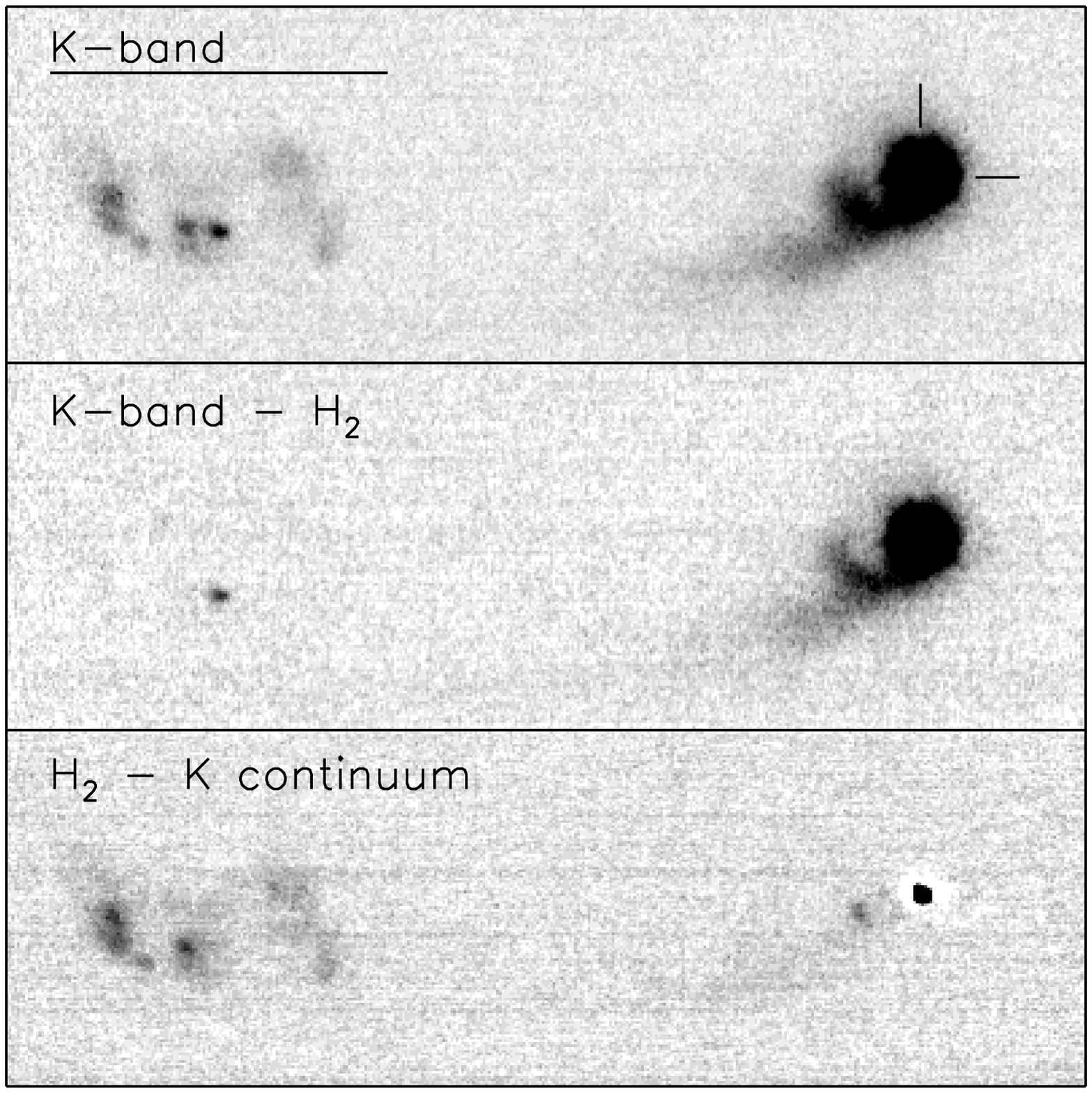}
\caption{K-band, K continuum, and H$_{2}$ images of IRAS 05403$-$0818}
\end{figure}

\clearpage

\begin{figure}
\plotone{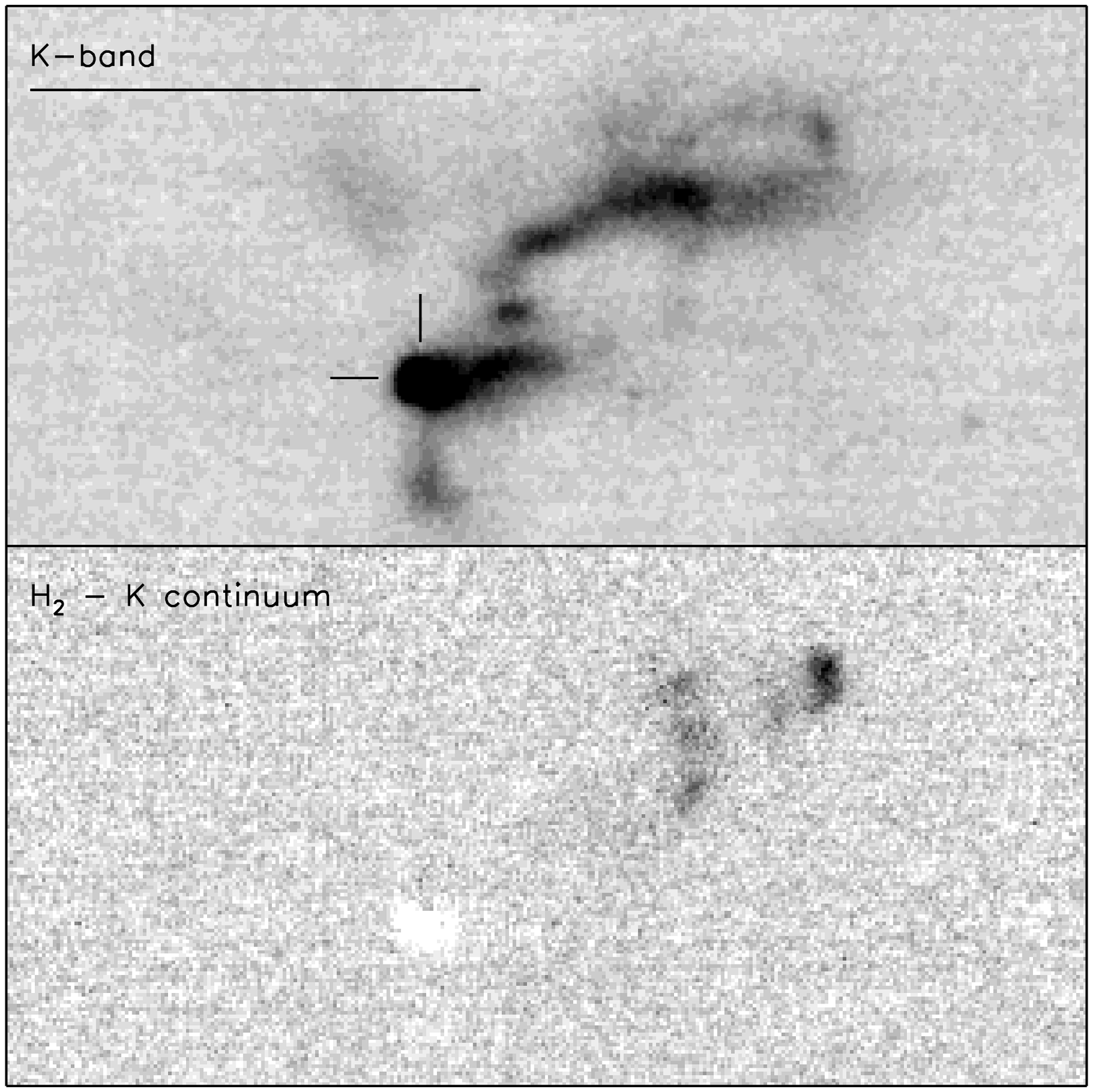}
\caption{K-band, and H$_{2}$ images of IRAS 18264$-$0143}
\end{figure}

\clearpage

\begin{figure}
\plotone{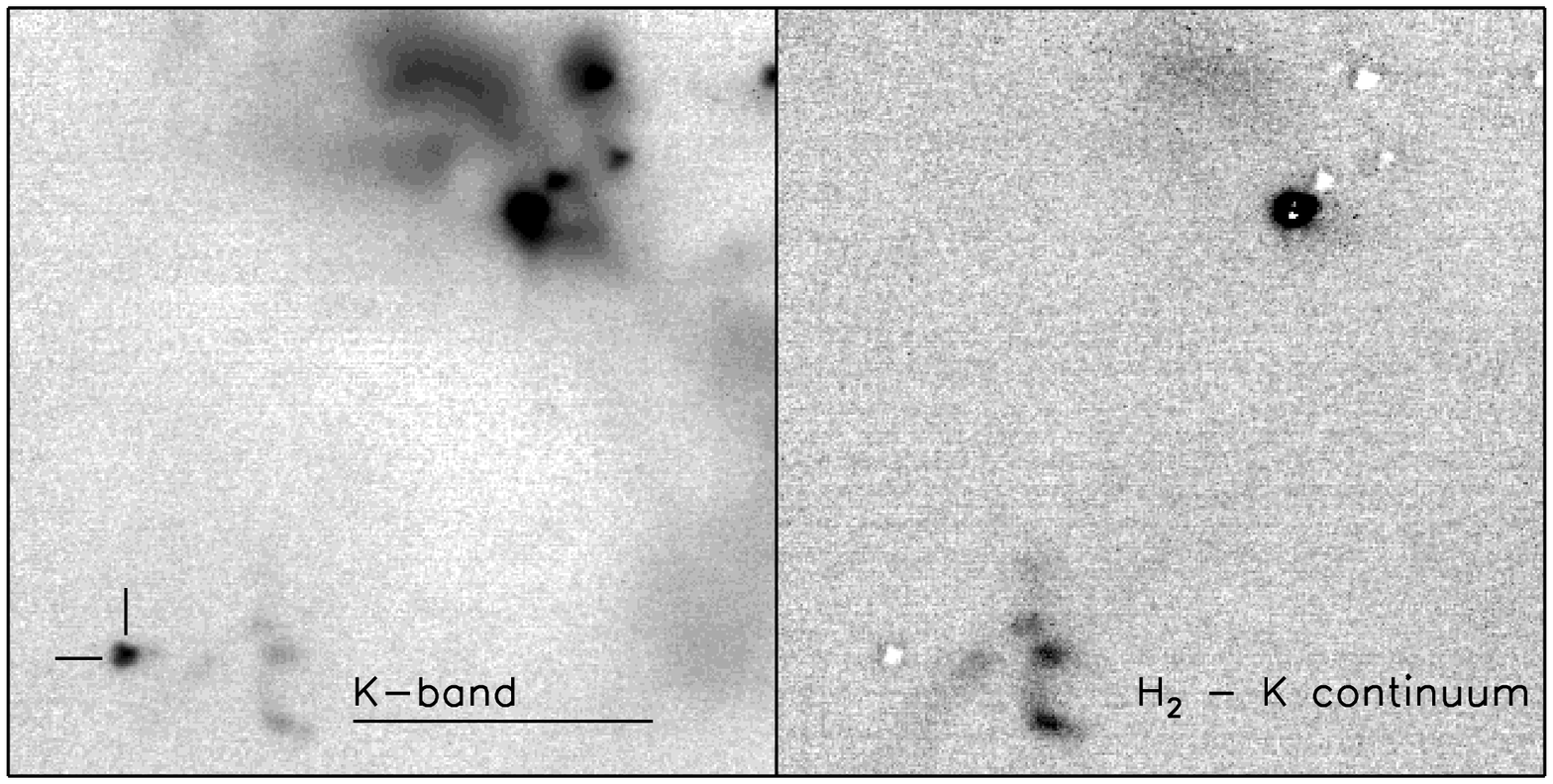}
\caption{K-band, and H$_{2}$ images of IRAS 18270$-$0153.  The stars and nebulae seen in the northern part of the field are to the east of the IRAS source coordinates}
\end{figure}

\clearpage

\begin{figure}
\plotone{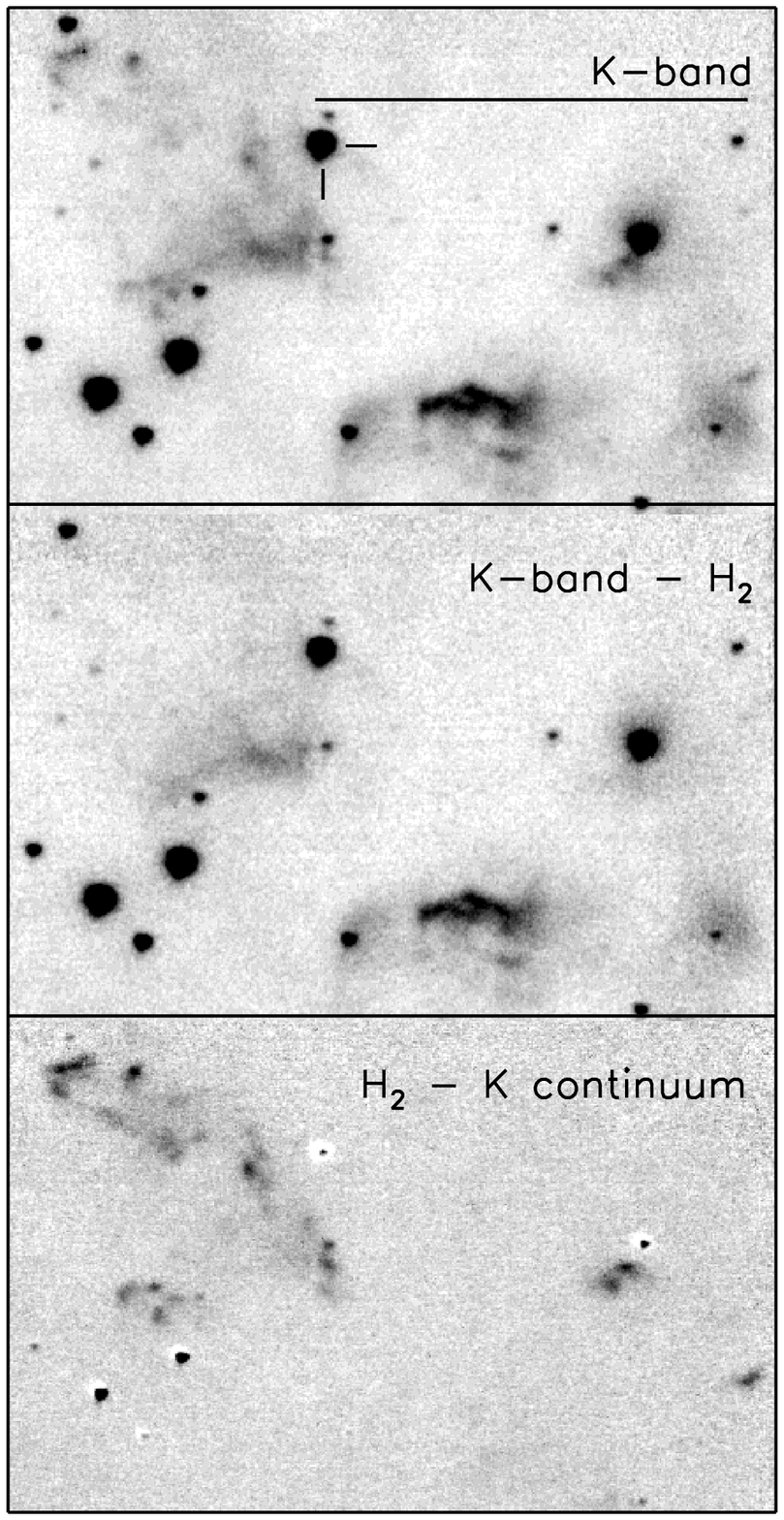}
\caption{K-band, and H$_{2}$ images of IRAS 21391+5802}
\end{figure}

\clearpage






\clearpage

\begin{deluxetable}{lrrrrccr}
\tabletypesize{\scriptsize}
\tablecaption{Observation Notes. \label{tbl-1}}
\tablewidth{0pt}
\tablehead{
\colhead{IRAS} & \colhead{$\alpha$(J2000)} & \colhead{$\delta$(J2000)} &
\colhead{t(s)\tablenotemark{a}} &
\colhead{Date (UT)}  & \colhead{FWHM($''$)\tablenotemark{b}} & \colhead{Contour\tablenotemark{c}} &
\colhead{S/N\tablenotemark{d}}
}
\startdata
00182+6223   & 00 20 58.1 &   +62 40 18 &    540 &    2003 Jul 20 &  0.80  &  19  &  2.13   \\
00465+5028   & 00 49 25.0 &   +50 44 45 &    540 &    2003 Jul 18 &  0.98  &  19  &  2.83   \\
01166+6635   & 01 20 03.7 &   +66 51 32 &    540 &    2003 Jul 17 &  0.93  &  19  &  2.15   \\
02086+7600   & 02 13 42.0 &   +76 15 02 &    540 &    2003 Sep 07 &  0.72  &  19  &  2.15   \\
03225+3034   & 03 25 36.0 &   +30 45 19 &    540 &    2003 Jul 18 &  ----  &  19  &  1.04   \\
03245+3002   & 03 27 39.0 &   +30 12 59 &    300 &    2003 Jan 10 &  0.57  &  19  &  1.83   \\
03260+3111   & 03 29 10.4 &   +31 21 59 &    540 &    2003 Sep 06 &  0.83  &  19  &  2.55   \\
03271+3013   & 03 30 14.9 &   +30 23 48 &   1620 &    2003 Sep 09 &  0.75  &  19  &  4.71   \\
03301+3057   & 03 33 16.3 &   +31 07 51 &    300 &    2003 Jan 10 &  0.68  &  19  &  1.22   \\
03331+6256   & 03 37 28.5 &   +63 06 29 &    540 &    2003 Jul 20 &  0.69  &  19  &  2.21   \\
03445+3242   & 03 47 41.3 &   +32 51 42 &   1740 &    2003 Sep 10 &  0.65  &  20  &  2.57   \\
03507+3801   & 03 54 06.3 &   +38 10 43 &   1620 &    2003 Sep 10 &  0.68  &  20  &  2.32   \\
04016+2610   & 04 04 42.9 &   +26 18 56 &    540 &    2003 Jul 18 &  1.23  &  19  &  2.95   \\
04067+3954   & 04 10 08.4 &   +40 02 25 &    300 &    2003 Jan 10 &  0.62  &  19  &  1.41   \\
04073+3800   & 04 10 41.2 &   +38 07 54 &    300 &    2003 Jan 09 &  0.55  &  19  &  1.51   \\
04169+2702   & 04 19 58.0 &   +27 09 56 &    300 &    2003 Jan 10 &  0.65  &  19  &  1.76   \\
04189+2650   & 04 22 00.6 &   +26 57 28 &    540 &    2003 Jul 20 &  0.87  &  19  &  2.23   \\
04191+1523   & 04 21 59.0 &   +15 30 17 &    300 &    2003 Jan 10 &  0.78  &  19  &  1.76   \\
04223+3700   & 04 25 38.9 &   +37 07 06 &   1620 &    2003 Sep 09 &  0.71  &  19  &  4.97   \\
04239+2436   & 04 26 55.3 &   +24 43 34 &    300 &    2003 Jan 10 &  0.74  &  18  &  3.87   \\
04248+2612   & 04 27 57.2 &   +26 19 17 &    300 &    2003 Jan 10 &  0.62  &  19  &  1.80   \\
04275+3531   & 04 30 48.2 &   +35 37 51 &   2160 &    2003 Sep 09 &  0.73  &  20  &  2.28   \\
04287+1801   & 04 31 35.8 &   +18 08 08 &    660 &    2003 Jul 20 &  ----  &  20  &  1.02   \\
04302+2247   & 04 33 16.8 &   +22 53 21 &    540 &    2003 Sep 07 &  0.70  &  19  &  2.56   \\
04325+2402   & 04 35 33.1 &   +24 08 15 &    540 &    2003 Sep 06 &  0.70  &  19  &  1.66   \\
04327+5432   & 04 36 44.2 &   +54 39 01 &    540 &    2003 Aug 09 &  1.15  &  19  &  2.38   \\
04365+2535   & 04 39 34.8 &   +25 41 43 &    300 &    2003 Jan 10 &  0.60  &  19  &  1.47   \\
04381+2540   & 04 41 11.6 &   +25 46 32 &   1620 &    2003 Sep 08 &  0.70  &  20  &  1.66   \\
04530+5126   & 04 56 56.8 &   +51 30 50 &    540 &    2003 Aug 10 &  0.75  &  19  &  2.27   \\
04591$-$0856 & 05 01 30.2 & $-$08 52 14 &    300 &    2003 Jan 10 &  0.71  &  19  &  1.51   \\
05155+0707   & 05 18 17.1 &   +07 11 01 &    540 &    2003 Aug 09 &  0.86  &  19  &  2.35   \\
05302$-$0537 & 05 32 41.7 & $-$05 35 48 &    540 &    2003 Aug 10 &  0.76  &  19  &  1.65   \\
05311$-$0631 & 05 32 41.7 & $-$05 35 48 &    300 &    2003 Jan 09 &  0.68  &  19  &  1.43   \\
05327$-$0457 & 05 35 14.4 & $-$04 55 45 &    300 &    2003 Jan 09 &  0.78  &  19  &  1.79   \\
05378$-$0750 & 05 40 16.7 & $-$07 48 33 &    540 &    2003 Sep 07 &  0.78  &  20  &  1.00   \\
05379$-$0758 & 05 40 20.5 & $-$07 56 32 &    540 &    2003 Sep 07 &  0.62  &  20  &  1.48   \\
05384$-$0808 & 05 40 48.8 & $-$08 06 51 &    300 &    2003 Jan 09 &  0.76  &  19  &  1.88   \\
05391$-$0841 & 05 41 29.8 & $-$08 40 14 &    540 &    2003 Sep 07 &  0.78  &  19  &  2.41   \\
05399$-$0121 & 05 42 27.7 & $-$01 20 02 &   1380 &    2003 Sep 10 &  0.75  &  20  &  1.38   \\
05403$-$0818 & 05 42 47.3 & $-$08 17 06 &   1080 &    2003 Sep 08 &  0.87  &  20  &  1.38   \\
05404$-$0948 & 05 42 47.7 & $-$09 47 24 &    540 &    2003 Aug 10 &  0.73  &  19  &  2.02   \\
05405$-$0117 & 05 43 02.6 & $-$01 16 23 &   1620 &    2003 Sep 08 &  0.77  &  19  &  4.28   \\
05413$-$0104 & 05 43 51.5 & $-$01 02 52 &    540 &    2003 Sep 06 &  0.65  &  20  &  0.90   \\
05417+0907   & 05 44 29.8 &   +09 08 54 &    540 &    2003 Aug 09 &  0.83  &  19  &  2.23   \\
05450+0019   & 05 47 34.6 &   +00 20 08 &    660 &    2003 Jan 10 &  0.67  &  20  &  1.52   \\
05510$-$1018 & 05 53 22.8 & $-$10 17 38 &    540 &    2003 Aug 09 &  1.12  &  19  &  1.85   \\
05513$-$1024 & 05 53 42.5 & $-$10 24 01 &    540 &    2003 Sep 06 &  0.73  &  19  &  2.77   \\
05548$-$0935 & 05 57 13.0 & $-$09 35 10 &    540 &    2003 Aug 14 &  1.16  &  19  &  1.99   \\
05564$-$1329 & 05 58 47.5 & $-$13 29 16 &    540 &    2003 Aug 10 &  0.79  &  19  &  2.00   \\
05581$-$1026 & 06 00 28.5 & $-$10 26 34 &   1620 &    2003 Sep 09 &  0.66  &  20  &  1.21   \\
05582$-$0950 & 06 00 35.9 & $-$09 50 35 &    540 &    2003 Sep 08 &  0.74  &  19  &  2.74   \\
05596$-$0903 & 06 02 01.7 & $-$09 03 06 &    540 &    2003 Sep 06 &  0.70  &  19  &  2.37   \\
05598$-$0906 & 06 02 16.3 & $-$09 06 32 &    540 &    2003 Aug 10 &  0.74  &  19  &  1.69   \\
06027$-$0714 & 06 05 08.2 & $-$07 14 42 &    300 &    2003 Jan 10 &  0.78  &  19  &  1.68   \\
06033$-$0710 & 06 05 48.2 & $-$07 10 30 &    540 &    2003 Sep 06 &  0.68  &  20  &  1.02   \\
06047$-$1117 & 06 07 08.7 & $-$11 17 51 &   1620 &    2003 Sep 10 &  0.76  &  21  &         \\
06057$-$0923 & 06 08 05.1 & $-$09 23 52 &    540 &    2003 Sep 06 &  0.72  &  19  &  2.56   \\
06249$-$0953 & 06 27 17.4 & $-$09 55 27 &    300 &    2003 Jan 09 &  0.80  &  19  &  1.60   \\
06368+0938   & 06 39 35.7 &   +09 35 35 &   1380 &    2003 Sep 10 &  0.70  &  20  &  1.63   \\
06381+1039   & 06 40 58.0 &   +10 36 49 &   1380 &    2003 Sep 09 &  0.71  &  20  &  1.79   \\
06382+1017   & 06 41 03.3 &   +10 15 01 &    540 &    2003 Sep 10 &  0.71  &  20  &  0.96   \\
06393+0913   & 06 42 06.9 &   +09 10 24 &   1140 &    2003 Sep 08 &  0.69  &  20  &  1.64   \\
07180$-$2356 & 07 20 06.7 & $-$24 02 21 &    300 &    2003 Jan 08 &  1.10  &  19  &  1.28   \\
15398$-$3359 & 15 43 02.2 & $-$34 09 06 &    540 &    2003 May 15 &  1.56  &  19  &  2.08   \\
16288$-$2450E & 16 31 56.0 & $-$24 56 28 &   540 &    2003 May 14 &  1.13  &  19  &  2.70   \\
16288$-$2450W & 16 31 56.0 & $-$24 56 28 &   540 &    2003 May 14 &  1.13  &  19  &  2.70   \\
16295$-$4452 & 16 33 08.6 & $-$44 58 23 &    540 &    2003 May 16 &  1.24  &  18  &  2.11   \\
16316$-$1540 & 16 34 29.1 & $-$15 47 00 &    540 &    2003 May 14 &  1.10  &  19  &  2.59   \\
16442$-$0930 & 16 46 57.4 & $-$09 35 19 &    540 &    2003 May 14 &  1.21  &  19  &  2.08   \\
17364$-$1946 & 17 39 23.5 & $-$19 47 51 &    540 &    2003 May 16 &  1.01  &  19  &  1.74   \\
17441$-$0433 & 17 46 50.8 & $-$04 34 34 &    540 &    2003 May 15 &  0.93  &  19  &  0.97   \\
18148$-$0440 & 18 17 29.8 & $-$04 39 38 &    540 &    2003 May 14 &  1.10  &  20  &  1.01   \\
18264$-$0143 & 18 29 05.7 & $-$01 41 57 &    540 &    2003 May 14 &  1.20  &  19  &  2.82   \\
18270$-$0153 & 18 29 37.0 & $-$01 51 02 &    540 &    2003 May 15 &  1.17  &  19  &  2.51   \\
18274$-$0212 & 18 30 01.4 & $-$02 10 26 &    540 &    2003 May 16 &  0.84  &  18  &  2.17   \\
18278$-$0212 & 18 30 28.0 & $-$02 10 48 &    540 &    2003 May 16 &  0.82  &  18  &  2.69   \\
18331$-$0035 & 18 35 42.1 & $-$00 33 18 &    540 &    2003 May 14 &  1.33  &  19  &  2.53   \\
18341$-$0113 & 18 36 45.7 & $-$01 10 30 &    540 &    2003 May 14 &  1.24  &  19  &  3.13   \\
18383+0059   & 18 40 51.7 &   +01 02 09 &    540 &    2003 May 15 &  1.17  &  19  &  3.02   \\
18595$-$3712 & 19 02 59.5 & $-$37 07 33 &    540 &    2003 May 14 &  1.24  &  20  &  0.83   \\
19266+0932   & 19 29 01.1 &   +09 38 46 &    540 &    2003 May 14 &  1.29  &  19  &  3.03   \\
19411+2306   & 19 43 18.1 &   +23 13 59 &    540 &    2003 May 16 &  0.74  &  19  &  2.32   \\
20353+6742   & 20 35 46.5 &   +67 52 59 &   1800 &    2003 Sep 06 &  0.58  &  20  &  1.55   \\
20361+5733   & 20 37 20.8 &   +57 44 13 &   2160 &    2003 Sep 06 &  0.68  &  21  &  0.89   \\
20377+5658   & 20 38 57.4 &   +57 09 34 &    540 &    2003 May 15 &  1.04  &  19  &  2.68   \\
20386+6751   & 20 39 06.6 &   +68 02 13 &    540 &    2003 Aug 09 &  0.81  &  20  &  0.92   \\
20453+6746   & 20 45 54.2 &   +67 57 39 &    540 &    2003 May 15 &  1.02  &  19  &  2.50   \\
20568+5217   & 20 58 21.4 &   +52 29 27 &    840 &    2003 May 15 &  1.03  &  19  &  2.65   \\
20582+7724   & 20 57 13.2 &   +77 35 46 &    540 &    2003 May 15 &  1.18  &  19  &  2.41   \\
21004+7811   & 20 59 14.3 &   +78 23 01 &    300 &    2003 Sep 06 &  0.63  &  19  &  1.28   \\
21007+4951   & 21 02 22.9 &   +50 03 06 &    540 &    2003 Jul 20 &  0.78  &  19  &  2.12   \\
21017+6742   & 21 02 22.6 &   +67 54 13 &    540 &    2003 Jul 18 &  1.00  &  19  &  2.38   \\
21169+6804   & 21 17 39.4 &   +68 17 32 &    540 &    2003 May 15 &  1.24  &  19  &  2.52   \\
21352+4307   & 21 37 11.3 &   +43 20 36 &    540 &    2003 Jul 17 &  1.28  &  19  &  2.61   \\
21388+5622   & 21 40 29.0 &   +56 35 58 &    540 &    2003 Jul 18 &  0.91  &  19  &  2.38   \\
21391+5802   & 21 40 42.4 &   +58 16 10 &    540 &    2003 Jul 18 &  0.97  &  19  &  2.15   \\
21445+5712   & 21 46 06.8 &   +57 26 23 &    540 &    2003 May 15 &  0.81  &  19  &  2.48   \\
21454+4718   & 21 47 21.8 &   +47 32 09 &    540 &    2003 Jul 17 &  1.37  &  19  &  2.72   \\
21569+5842   & 21 58 36.4 &   +58 57 09 &    540 &    2003 Jul 20 &  0.78  &  19  &  2.38   \\
22051+5848   & 22 06 50.7 &   +59 02 46 &    540 &    2003 May 15 &  0.81  &  19  &  2.92   \\
22176+6303   & 22 19 18.2 &   +63 18 46 &    540 &    2004 Aug 05 &  0.90  &  18  &  5.43   \\
22266+6845   & 22 28 02.9 &   +69 01 13 &    540 &    2003 Aug 10 &  0.74  &  19  &  2.22   \\
22267+6244   & 22 28 29.4 &   +62 59 44 &    540 &    2003 Sep 09 &  0.71  &  20  &  0.99   \\
22272+6358A  & 22 28 52.3 &   +64 13 43 &    540 &    2003 May 15 &  0.99  &  19  &  2.79   \\
22376+7455   & 22 38 47.2 &   +75 11 29 &    540 &    2003 Sep 07 &  0.66  &  19  &  2.32   \\
23037+6213   & 23 05 48.9 &   +62 30 02 &    540 &    2003 Jul 17 &  1.19  &  19  &  2.50   \\

\enddata
\tablecomments{Coordinates are from IRAS}
\tablenotetext{a}{t = On source integration time}
\tablenotetext{b}{Median seeing = 0.78"}
\tablenotetext{c}{This is the first (outer most) surface brightness contour, in units of K magnitudes per square arcsecond.}
\tablenotetext{d}{This is the pixel value under the first surface brightness contour divided by the standard deviation of the pixel counts on the sky}
\end{deluxetable}

\clearpage

\begin{deluxetable}{lrrrrrrrrcrcc}
\tabletypesize{\scriptsize}
\tablecaption{Source Characteristics}
\tablewidth{0pt}
\tablehead{
\colhead{IRAS} &
\colhead{Associations} & 
\colhead{D(pc)\tablenotemark{a}} & 
\colhead{$L_{bol}$} & 
\colhead{$\alpha$(J2000)\tablenotemark{b}} &
\colhead{$\delta$(J2000)\tablenotemark{b}} & 
\colhead{J\tablenotemark{c}} & 
\colhead{H\tablenotemark{c}} & 
\colhead{K$_s$\tablenotemark{c}} & 
\colhead{$\alpha$\tablenotemark{d}} &
\colhead{Size(")\tablenotemark{e}} 

}
\startdata
00182+6223   &  L1280                      & 4680(4)  &  366.93 & 00 20 56.79 &   +62 40 21.0  & 15.688 & 13.982 & 12.443 &  1.53       &  9.6 \\
00465+5028   & CB6, LBN 613 	           &  800(6)  &  8.7792 & 00 49 24.50 &   +50 44 43.6  & 13.822 & 12.297 & 11.531 &  1.01       & 15.4 \\
01166+6635   &                             &  249(4)  &  0.4437 & 01 20 03.93 &   +66 51 35.9  & 16.029 & 14.037 & 12.606 &  0.38       &  7.8 \\
02086+7600   & L1333                       &  180(5)  &  0.7839 & 02 13 43.61 &   +76 15 06.0  & 13.715 & 12.254 & 11.193 &  0.88       &  9.8 \\
03225+3034   & L1448 IRS 3, RNO 14         &  290(1)  &  13.124 &             &                & 13.745 & 12.363 & 11.095 &  1.52       & 20.1 \\
03245+3002   & L1455 IRS 1,                &  260(1)  &  7.8807 & 03 27 38.83 &   +30 13 25.0  &   ...  &   ...  &  ...   &  1.91       &  5.0 \\
             & RNO 15 FIR                  &          &         &             &                &        &        &        &             &      \\
03260+3111   & L1450, SVS 3                &  290(1)  &  138.43 & 03 29 10.38 &   +31 21 59.2  &  9.368 &  7.987 &  7.173 &  0.59       & 76.6 \\
03271+3013   & in NGC 1333                 &  290(2)  &  1.6255 & 03 30 15.16 &   +30 23 49.4  &   ...  &   ...  & 14.259 &  0.86       &  5.2 \\
03301+3057   & Barnard 1 IRS               &  290(2)  &  3.0314 & 03 33 16.68 &   +31 07 54.9  &   ...  &   ...  & 14.208 &  1.52       &  9.5 \\
03331+6256   &                             & 1560(4)  &         & 03 37 28.45 &   +63 06 31.2  &   ...  &   ...  & 14.590 &  0.45       &  3.9 \\
03445+3242   & L1471, HH 366 VLA 1,        &  280(1)  &  3.8062 & 03 47 41.60 &   +32 51 43.8  &   ...  & 14.047 & 11.214 &  0.16       & 10.3 \\
             & Barnard 5 IRS 1             &          &         &             &                &        &        &        &             &      \\
03507+3801   & HH 462                      &  350(1)  &  2.5178 & 03 54 06.19 &   +38 10 42.5  & 12.474 & 10.863 & 10.098 &  0.22       & 13.8 \\
04016+2610   & L1489 IRS, HH 360           &  140(1)  &  3.0280 & 04 04 43.05 &   +26 18 56.2  & 12.655 & 10.861 &  9.199 &  0.31       & 18.4 \\
04067+3954   & L1459                       &  350(1)  &  15.105 & 04 10 08.40 &   +40 02 24.6  & 13.767 & 11.478 &  9.844 &  1.17       & 53.7 \\
04073+3800   & L1473,  HH 463              &  350(1)  &  22.600 & 04 10 41.09 &   +38 07 54.0  & 15.339 & 13.552 & 10.500 &  0.07       & 18.3 \\
04169+2702   & L1495, near HH 391          &  140(1)  &  0.9190 & 04 19 58.45 &   +27 09 57.1  & 16.528 & 12.554 & 10.428 &  0.53       & 18.7 \\
04189+2650   & FS Tau B, HH 157,           &  140(3)  &  0.6454 & 04 22 00.70 &   +26 57 32.5  & 15.082 & 13.351 & 11.753 & -0.04       & 21.8$^{*}$ \\
             & Haro 6-5B                   &          &         &             &                &        &        &        &             &      \\
04191+1523   & --                          &  140(7)  &  0.4031 & 04 22 00.44 &   +15 30 21.2  & 16.592 & 12.354 & 11.259 &  0.97       & 10.4 \\
04223+3700   & L1478                       &  350(1)  &  2.7410 & 04 25 39.80 &   +37 07 08.2  &   ...  & 13.170 & 10.271 &  0.47       & 10.4$^{*}$ \\
04239+2436   & L1524, HH 300 VLA 1         &  140(1)  &  1.1028 & 04 26 56.30 &   +24 43 35.3  & 14.323 & 11.530 &  9.764 &  0.09       &  8.5$^{*}$ \\
04248+2612   & L1521D, HH 31 IRS2,         &  140(1)  &  0.3276 & 04 27 57.30 &   +26 19 18.4  & 11.619 & 10.270 &  9.741 &  0.52       & 25.2 \\
             & Barnard 217                 &          &         &             &                &        &        &        &             &      \\
04275+3531   &                             &  350(1)  &  1.5456 & 04 30 48.52 &   +35 37 53.2  &   ...  &   ...  & 15.268 &  0.51       &  4.3 \\
04287+1801   & L1551 IRS 5B, HH 154,       &  140(1)  &  20.179 & 04 31 34.08 &   +18 08 04.9  & 12.230 & 10.550 &  9.255 &  0.76       & 20.2 \\
04302+2247   & HH 394, near L1536          &  140(1)  &  0.2797 & 04 33 16.50 &   +22 53 20.2  & 13.489 & 11.772 & 10.876 &  1.34       & 13.7 \\
04325+2402   & L1535 IRS, Barnard 18I      &  140(1)  &  0.6805 & 04 35 35.39 &   +24 08 19.4  & 16.122 & 11.504 &  9.826 &  1.71       & 26.1 \\
04327+5432   & L1400, HH 378               &  170(1)  &  1.8530 & 04 36 45.50 &   +54 39 04.5  & 16.437 & 13.974 & 12.618 &  0.84       &  8.0 \\
04365+2535   & TMC-1A, L1534               &  140(1)  &  1.8774 & 04 39 35.19 &   +25 41 44.7  & 16.389 & 12.062 & 10.020 &  0.68       & 21.1 \\
04381+2540   & TMC-1, L1534                &  140(1)  &  0.5949 & 04 41 12.68 &   +25 46 35.4  & 16.076 & 12.954 & 11.254 &  0.64       &  9.0 \\
04530+5126   & L1438, V347 Aur, 	   &   none   &         & 04 56 57.02 &   +51 30 50.9  &  9.990 &  8.825 &  8.062 &  0.05       & 19.3$^{*}$ \\
             & RNO 33                      &          &         &             &                &        &        &        &             &      \\
04591$-$0856 & HHL 17, IC 2118             &  210(8)  &  0.9043 & 05 01 29.64 & $-$08 52 16.9  & 11.359 & 10.341 &  9.933 &  0.62       & 21.0 \\
05155+0707   & HH 114                      &  460(17) &  11.773 & 05 18 17.30 &   +07 10 59.9  &   ...  & 12.567 & 10.214 &  1.55       & 24.5 \\
05302$-$0537 & Haro 4-145		   &  470(2)  &  42.749 & 05 32 41.65 & $-$05 35 46.1  &   ...  & 15.116 & 11.389 &  0.38       & 20.8 \\
05311$-$0631 & L1641, HH 83 VLA 1          &  470(3)  &  7.3329 & 05 33 32.52 & $-$06 29 44.2  & 13.358 & 11.487 &  9.749 &  0.23       & 21.4 \\
05327$-$0457 & Ced 55e                     &  450(9)  &  920.15 & 05 35 13.10 & $-$04 55 52.5  & 13.166 & 10.886 &  9.360 &  1.76       & 22.0$^{*}$ \\
05378$-$0750 & L1641                       &  480(1)  &  8.1978 & 05 40 14.95 & $-$07 48 48.5  &   ...  & 15.392 & 13.470 &  0.25       &  5.8 \\
05379$-$0758 & L1641                       &  480(1)  &  6.3836 & 05 40 20.55 & $-$07 56 39.9  & 12.851 & 10.678 &  9.399 &  0.19       &  9.3$^{*}$ \\
05384$-$0808 & L1641 S4, S85               &  480(1)  &  10.809 & 05 40 50.59 & $-$08 05 48.7  & 13.134 & 11.349 & 10.276 &  1.03       & 28.3$^{*}$ \\
05391$-$0841 & L1641                       &  480(1)  &  3.5872 & 05 41 30.05 & $-$08 40 09.2  &   ...  & 14.729 & 11.855 &  0.77       & 5.80$^{*}$ \\
05399$-$0121 & L1630, HH 92,               &  430(1)  &  10.686 & 05 42 27.64 & $-$01 19 57.0  &   ...  &   ...  &  ...   &  1.53       &  7.9 \\
05403$-$0818 & L1641 S2                    &  480(1)  &  9.8998 & 05 42 47.07 & $-$08 17 06.9  & 15.671 & 13.155 & 11.063 &  0.40       &  9.5 \\
05404$-$0948 & L1647                       &  480(1)  &  49.781 & 05 42 47.67 & $-$09 47 22.5  & 10.818 &  9.810 &  9.232 &  0.76       & 21.4$^{*}$ \\
             &                             &          &         &             &                & 15.981 & 13.592 & 12.068 &             &      \\
05405$-$0117 & L1630                       &  430(1)  &  4.3714 & 05 43 03.06 & $-$01 16 29.2  & 14.467 & 11.877 & 10.300 &  0.71       & 10.7$^{*}$ \\
05413$-$0104 & L1630, HH 212               &  430(1)  &  10.500 & 05 43 51.50 & $-$01 02 58.5  &   ...  &    ... &   ...  &  2.91       & 10.3 \\
05417+0907   & L1594, HH 175,              &  465(1)  &  18.380 & 05 44 30.01 &   +09 08 57.1  &   ...  & 15.913 & 12.400 &  1.68       & 17.8$^{*}$ \\
             & Barnard 35A                 &          &         &             &                &        &        &        &             &      \\
05450+0019   & L1630                       &  430(1)  &  27.648 & 05 47 36.55 &   +00 20 06.3  & 11.406 &  9.604 &  8.784 &  1.26       & 57.8 \\
05510$-$1018 & --                          &   none   &         & 05 53 23.71 & $-$10 17 27.6  & 16.267 & 15.085 & 12.787 &  0.93       & 11.0 \\
05513$-$1024 & --                          &   none   &         & 05 53 42.55 & $-$10 24 00.7  &  9.803 &  7.635 &  5.956 &  0.18       & 41.7$^{*}$ \\
05548$-$0935 & --                          &   none   &         & 05 57 13.23 & $-$09 35 10.9  & 14.573 & 13.357 & 12.544 &  0.72       & 10.0$^{*}$ \\
05564$-$1329 & --                          &   none   &         & 05 58 46.91 & $-$13 29 18.8  & 14.021 & 12.061 & 10.762 &  0.38       & 12.8$^{*}$ \\
05581$-$1026 & --                          &   none   &         & 06 00 28.64 & $-$10 26 31.9  & 17.464 &   ...  & 14.701 &  0.47       &  4.1 \\
05582$-$0950 & RNO 60                      &   none   &         & 06 00 36.26 & $-$09 51 11.9  & 14.031 & 12.202 & 11.218 &  1.26       & 12.0 \\
05596$-$0903 & --                          &   none   &         & 06 02 02.47 & $-$09 03 13.3  &   ...  &   ...  &   ...  &  1.11       &  5.4 \\
05598$-$0906 & HHL 34,GGD 10               &   none   &         & 06 02 16.20 & $-$09 06 29.0  & 14.553 & 11.876 &  9.813 &  0.43       & 29.2$^{*}$ \\
06027$-$0714 &                             &  830(1)  &  8.6831 & 06 05 07.90 & $-$07 14 42.6  & 16.226 & 13.473 & 12.607 &  1.08       & 10.9 \\
06033$-$0710 &                             &  830(1)  &  10.333 & 06 05 48.61 & $-$07 10 31.2  &   ...  &   ...  &   ...  &  1.28       &  3.7 \\
06047$-$1117 & --                          &  500(10) &  4.9455 & 06 07 08.50 & $-$11 17 51.0  & 14.119 & 12.222 & 10.220 &  0.64       & 17.9 \\
06057$-$0923 & --                          &   none   &         & 06 08 05.29 & $-$09 23 47.3  &   ...  &   ...  &   ...  &  0.97       &  5.8 \\
06249$-$0953 & L1652                       &  830(1)  &  6.4991 & 06 27 17.34 & $-$09 55 27.4  & 15.034 & 13.652 & 12.559 &  1.04       &  7.4 \\
06368+0938   & L1613                       &  790(11) &  6.5302 & 06 39 32.09 &   +09 35 41.5  &   ...  &   ...  &   ...  &  0.93       &      \\
06381+1039   &                             &  960(4)  &  143.60 & 06 40 58.15 &   +10 36 52.1  &   ...  &   ...  & 14.513 &  1.93       & 10.2 \\
06382+1017   & HH 124, NGC 2264,           &  800(3)  &  84.413 & 06 41 02.64 &   +10 15 02.1  & 13.362 & 12.218 & 10.592 &  1.00       & 19.0 \\
             & L1610/1613                  &          &         &             &                &        &        &        &             &      \\
06393+0913   &                             &  950(4)  &  28.887 & 06 42 08.13 &   +09 10 30.0  & 15.243 & 12.048 & 10.593 &  1.42       & 6.78$^{*}$ \\
07180-2356   & L1660, HH 72 IRS            & 1500(17) &  185.95 & 07 20 08.36 & $-$24 02 23.0  &   ...  & 14.176 & 11.648 &  0.81       & 6.94$^{*}$ \\
15398$-$3359 & HH 185, Lupus 1,            &  170(3)  &  1.3606 & 15 43 01.32 & $-$34 09 15.3  & 15.963 & 13.992 & 12.326 &  1.59       & 11.3 \\
             & Barnard 228                 &          &         &             &                &        &        &        &             &      \\
16288-2450(E)& L1689 IRS 5, $\rho$ Oph S   &  160(1)  &         & 16 32 02.22 & $-$24 56 16.8  &   ...  & 13.813 & 10.726 &  0.70       & 15.6 \\
16288-2450(W)& $\rho$ Oph S                &  160(1)  &   5.479 & 16 31 52.98 & $-$24 56 24.6  & 11.783 &  9.391 &  7.557 &  0.70       & 19.0 \\
16295$-$4452 & --		   	   &  600(12) &  30.602 & 16 33 07.73 & $-$44 58 24.7  &   ...  & 15.086 & 12.270 &  0.79       & 23.3 \\
16316$-$1540 & L43, RNO 91                 &  160(1)  &  11.398 & 16 34 29.29 & $-$15 47 01.9  & 10.994 &  9.635 &  8.464 &  0.84       & 27.5 \\
16442$-$0930 & L260                        &  160(1)  &  0.6570 & 16 46 58.27 & $-$09 35 19.7  & 14.316 & 12.339 & 10.721 &  0.22       & 10.5$^{*}$ \\
17364$-$1946 & L219			   &   none   &         & 17 39 23.25 & $-$19 47 54.7  &   ...  &   ...  & 13.757 &  0.96       & 11.2 \\
17441$-$0433 & L425			   &   none   &         & 17 46 50.89 & $-$04 34 33.7  & 16.700 & 15.270 & 13.325 &  0.50       & 10.8 \\
18148$-$0440 & L483 FIR                    &  225(1)  &  11.050 & 18 17 29.56 & $-$04 39 35.7  & 16.188 & 12.640 & 10.790 &  1.36       & 15.4 \\
18264$-$0143 &                             &   none   &         & 18 29 05.31 & $-$01 41 56.9  &   ...  &   ...  & 13.968 &  1.39       & 13.3 \\
18270$-$0153 &                             &   none   &         & 18 29 36.69 & $-$01 50 59.1  & 16.356 & 12.956 & 10.772 &  0.49       & 24.5 \\
             &                             &          &         &             &                & 13.700 & 11.797 & 10.711 &             &      \\
18274$-$0212 &                             &   none   &         & 18 30 01.36 & $-$02 10 25.6  &   ...  & 15.145 & 11.489 &  0.12       & 6.64$^{*}$ \\
18278$-$0212 &                             &   none   &         & 18 30 27.28 & $-$02 11 00.2  &   ...  &   ...  & 14.550 &  1.62       & 14.7 \\
18331$-$0035 & L588, HH 109,               &  310(3)  &  3.7700 & 18 35 42.00 & $-$00 33 22.1  & 16.347 & 13.911 & 11.738 &  2.02       & 18.3 \\
             & HH 108 IRAS                 &          &         &             &                &        &        &        &             &      \\
18341$-$0113 & L564                        &   none   &         & 18 36 46.33 & $-$01 10 29.5  & 14.849 & 11.974 & 10.229 &  0.91       & 19.5 \\
18383+0059   & --                          &   none   &         & 18 40 51.87 &   +01 02 12.9  & 14.892 & 11.748 &  9.602 &  0.50       & 15.3$^{*}$ \\
18595$-$3712 & ISO-CrA 182                 &  129(13) &  1.2313 & 19 02 58.70 & $-$37 07 34.1  &   ...  & 15.881 & 14.498 &  1.83       &  9.1 \\
19266+0932   & HH 221, Parsamian 21        &  300(3)  &  3.4202 & 19 29 00.86 &   +09 38 42.9  & 11.205 & 10.485 &  9.763 &  0.37       & 19.9 \\
19411+2306   & --                          & 2100(14) &  3026.1 & 19 43 17.94 &   +23 14 01.6  & 13.946 & 11.548 &  9.596 &  1.11       & 19.9 \\
20353+6742   & L1152,  HH 376              &  370(1)  &  1.3738 & 20 35 46.33 &   +67 53 02.0  & 15.263 & 14.230 & 13.254 &  1.41       &  7.4 \\
20361+5733   & L1041                       &   none   &         & 20 37 20.36 &   +57 44 14.8  &   ...  &   ...  &   ...  &  1.91       &  2.5 \\
20377+5658   & L1036                       &  440(1)  &  4.7624 & 20 38 57.48 &   +57 09 37.6  & 13.925 & 11.226 &  9.507 &  0.32       & 17.6$^{*}$ \\
20386+6751   & L1157 IRS,                  &  370(1)  &  5.5027 & 20 39 06.28 &   +68 02 13.8  &   ...  &   ...  &   ...  &  2.23       & 16.1 \\
             & HH 375 VLA 1                &          &         &             &                &        &        &        &             &      \\
20453+6746   & PV Cep, HH 215,             &  500(3)  &  63.697 & 20 45 53.94 &   +67 57 38.7  & 12.453 &  9.497 &  7.291 & -0.32       & 25.2$^{*}$ \\
             & L1158                       &          &         &             &                &        &        &        &             &      \\
20568+5217   & L1002, HH 381 IRS           & 1270(4)  &  45.55  & 20 58 21.09 &   +52 29 27.7  & 11.544 &  9.813 &  8.305 &  0.62       & 34.5 \\
20582+7724   & L1228, HH 199               &  175(1)  &  1.2112 & 20 57 12.94 &   +77 35 43.7  & 13.024 & 10.608 &  9.171 &  0.31       & 16.0$^{*}$ \\
21004+7811   & HHL 66, HH 198,             &  300(3)  &  13.538 & 20 59 14.03 &   +78 23 04.1  &  9.437 &  7.530 &  6.319 &  0.20       & 25.5$^{*}$ \\
             & RNO 129                     &          &         &             &                &        &        &        &             &      \\
21007+4951   & L988                        &  700(1)  &  31.059 & 21 02 23.85 &   +50 03 06.8  & 16.368 & 14.818 & 13.276 &  0.69       & 10.1$^{*}$ \\
21017+6742   & L1172                       &  288(15) &  0.4974 & 21 02 21.27 &   +67 54 20.1  &   ...  &   ...  & 14.890 &  0.66       & 20.1 \\
21169+6804   & L1177, CB 230               &  450(6)  &  7.3301 & 21 17 38.69 &   +68 17 33.4  & 11.562 &  9.898 &  9.188 &  1.75       & 25.3 \\
21352+4307   & Barnard 158                 &  600(6)  &  11.721 & 21 37 11.39 &   +43 20 38.4  &   ...  & 15.877 & 12.915 &  0.17       &  7.6$^{*}$ \\
21388+5622   & HH 588                      &  750(16) &  96.504 & 21 40 28.98 &   +56 35 55.7  & 12.801 & 11.620 & 10.789 &  0.59       & 16.7 \\
21391+5802   & L1121, IC 1396N             &  750(16) &  254.17 & 21 40 42.80 &   +58 16 01.1  &   ...  & 15.642 & 14.155 &  2.15       & 36.7 \\
21445+5712   &                             &  360(4)  &  18.527 & 21 46 07.12 &   +57 26 31.8  & 13.950 & 11.965 & 10.139 &  0.54       & 14.6 \\
21454+4718   & L1013B/1031B,               &  900(1)  &  106.66 & 21 47 20.66 &   +47 32 03.6  &  9.889 &  8.087 &  7.040 &  0.70       & 33.0$^{*}$ \\
             & V1735 Cyg                   &          &         &             &                &        &        &        &             &      \\
21569+5842   & L1143                       &  250(4)  &  0.9577 & 21 58 35.90 &   +58 57 22.8  & 15.457 & 12.936 & 10.695 &  0.08       & 12.5$^{*}$ \\
22051+5848   & L1165, HH 354 IRS           &  750(3)  &  72.956 & 22 06 50.37 &   +59 02 45.9  & 11.370 & 10.248 &  9.682 &  1.15       & 23.7 \\
22176+6303   & L1240,  RAFGL 2884,         &  910(1)  &  21313. & 22 19 20.39 &   +63 19 38.5  & 12.304 &  9.298 &  6.135 &  0.87       & 125.7 \\
             & S 140 IRS1-3                &          &         &             &                &        &        &        &             &      \\
22266+6845   & L1221, HH 363               &  200(1)  &  1.8195 & 22 28 02.99 &   +69 01 16.7  & 16.575 & 13.544 & 11.465 &  0.53       & 14.3$^{*}$ \\
22267+6244   &                             &  900(1)  &  311.24 & 22 28 31.29 &   +63 00 28.2  & 15.826 & 11.799 &  9.244 &  1.45       & 44.7 \\
22272+6358A  & L1206                       &  950(1)  &  815.52 & 22 28 52.60 &   +64 13 41.0  & 16.690 & 12.197 &  9.840 &  1.76       & 28.3 \\
22376+7455   & L1251B 3, HH 189            &  330(1)  &  10.659 & 22 38 42.49 &   +75 11 45.6  & 13.977 & 12.073 & 11.206 &  1.09       & 31.7 \\
             &                             &          &         &             &                & 16.162 & 14.644 & 10.611 &             &      \\
23037+6213   & Cep C                       &          &  188.79 & 23 05 48.07 &   +62 30 09.9  &        &        &        &  1.23       & 34.3$^{*}$ \\
\enddata

\tablenotetext{a}{ The estimated distance to each source in parsecs.  The citation for the distances estimate is designated by the number in the parentheses, and are as follows: 1) Hilton, J., \& Lahulla, J. 1995; 2) Educated guess based on proximity to nearby objects; 3) Reipurth, A General Catalog of HH Objects, 1999 (http:casa.colorado.edu/hhcat/); 4) Wouterloot, J., \& Brand, J. 1989; 5) Obayashi et al. 1998; 6) Launhardt \& Henning 1997; 7) Andr\'{e} et al. 1999; 8) Kun et al. 2001; 9) Mookerjea et al. 2000; 10) Yun et al. 2001; 11) Sagar et al. 1983; 12) Moreira et al. 2000; 13) Marraco \& Rydgren 1981; 14) Guetter, H. 1992,; 15) Straizys et al. 1992; 16) Battinelli \& Capuzzo-Dolcetta 1991; 17) Reipurth \&  Aspin 1997; none = Searched for and could not find a distance estimate  }

\tablenotetext{b}{ 2MASS coordinate for selected object in field}
\tablenotetext{c}{ Magnitudes from the 2MASS extended source catalog, in the 2MASS photometric system.}
\tablenotetext{d}{ The spectral index of the source}
\tablenotetext{e}{ The square root of the area within the K=19 magnitudes per square arcsecond surface brightness contour. $^{*}$ denotes sources where the stellar image dominates the measured size.}
\tablecomments{RNO designates objects in "Red and Nebulous Objects in Dark Clouds: a Survey". Cohen,  M. 1980}
\end{deluxetable}

\clearpage

\begin{deluxetable}{lrr}
\tabletypesize{\scriptsize}
\tablecaption{Nebulae observed in H$_{2}$}
\tablewidth{0pt}
\tablehead{
\colhead{IRAS} &
\colhead{$\alpha$(J2000)\tablenotemark{a}} &
\colhead{$\delta$(J2000)\tablenotemark{a}} 
}

\startdata

00182+6223   &  00 21 01.47 &   +62 39 39.1 \\
03220+3035   &  03 25 14.34 &   +30 46 55.6 \\
03445+3242   &  03 47 41.60 &   +32 51 43.8 \\
04327+5432   &  04 36 41.65 &   +54 40 00.4 \\
05155+0707   &  05 18 17.30 &   +07 10 59.9 \\
05403$-$0818 &  05 42 47.07 & $-$08 17 06.9 \\
18264$-$0143 &  18 29 05.31 & $-$01 41 56.9 \\
18270$-$0153 &  18 29 40.26 & $-$01 51 27.8 \\
21391+5802   &  21 40 43.65 &   +58 16 19.1 \\

\enddata

\tablenotetext{a}{The RA and declination of the star or nebula marked in each figure.}

\end{deluxetable}



\end{document}